\documentclass[aps,prb,twocolumn,groupedaddress,showpacs,floatfix,altaffilletter]{revtex4-1}
\usepackage{graphicx}
\usepackage{grffile}
\usepackage{amsmath}
\usepackage{amssymb}
\usepackage{bm}
\usepackage{color}
\usepackage[dvipsnames]{xcolor}
\usepackage{amsmath,amssymb,amsfonts}
\usepackage{epsfig}
\usepackage{times}
\usepackage[colorlinks,bookmarks=false,citecolor=blue,linkcolor=red,urlcolor=blue]{hyperref}

\newcommand{\fref}[1]{Fig.~\ref{#1}}
\newcommand{\eref}[1]{Eq.~(\ref{#1})}

\begin{document}

\title{Charge-fluctuations in lightly hole-doped cuprates: effect of vertex corrections}

\author{R.~Nourafkan$^{1}$, M.~C{\^o}t{\'e}$^{2}$, and A.-M.S.~Tremblay$^{1,3}$}
\affiliation{$^1$D{\'e}partement de Physique and Institut quantique, Universit{\'e} de Sherbrooke, Sherbrooke, Qu{\'e}bec, Canada  J1K 2R1}
\affiliation{$^2$D{\'e}partement de Physique, Universit{\'e} de Montr{\'e}al, Montr{\'e}al, Qu{\'e}bec, Canada H3C 3J7}
\affiliation{$^3$Canadian Institute for Advanced Research, Toronto, Ontario, Canada M5G 1Z8}
%
%
\begin{abstract}
Identification of the electronic state that appears upon doping a Mott insulator is important to understand the physics of cuprate high-temperature superconductors.
Recent scanning tunneling microscopy of cuprates provides evidence that a charge-ordered state
emerges before the superconducting state upon doping the parent compound. 
We study this phenomenon by computing the charge response function of the Hubbard model including frequency-dependent local vertex corrections that satisfy the compressibility sum-rule. We find that upon approaching the Mott phase from the overdoped side, the charge fluctuations at wave vectors connecting hot spots are suppressed much faster than at the other wave-vectors. It leads to a momentum dependence of the dressed charge susceptibility that is very different from either the bare susceptibility or from the susceptibility obtained from the random phase approximation. We also find that the paramagnetic lightly hole-doped Mott phase at finite-temperature is unstable to charge ordering only at zero wave-vector, confirming the results previously obtained from the compressibility. Charge order is driven by the frequency-dependent scattering  processes that induce an attractive particle-hole interaction at large interaction strength and small doping. 
\end{abstract}

\maketitle
\section{Introduction}
Immediately following the discovery of cuprate high-temperature superconductors, and the suggestion by Anderson of the importance of the proximate Mott insulating phase~\cite{Anderson:1987}, the study of hole-doped Mott insulators became a central theme of research. 
Early Hartree-Fock studies of the hole-doped Mott insulators provide evidence of charge order, often accompanied by spin order, so-called stripes,~\cite{MachidaCharge:1989,SchulzCharge:1989,ZaanenCharge:1989,PoilblancCharge:1989} which survives even in presence of frustrating long-range interactions~\cite{EmeryKivelson:1993,LowEmeryChargeCoulomb:1994}. The experimental neutron scattering discovery of striped phases in La$_{1.6-x}$Nd$_{0.4}$Sr$_x$CuO$_4$~\cite{TranquadaStripe:1993} gave credence to these theoretical results. Further theoretical studies using Monte Carlo and density-matrix renormalization group~\cite{HellbergCharge:1997,WhiteCharge:1998,HellbergCharge:2000,WhiteScalapinoCharge:2000,PhysRevB.70.220506}, slave-boson~\cite{SeiboldCharge:1998,LorenzanaCharge:2002,HanCharge:2001} and variational studies~\cite{IvanovCharge:2004, PhysRevLett.113.046402} of the Hubbard and $t-J$ models confirmed that these models contain stripe phases. These are particularly evident at 
doping $p=1/8$ (for reviews see~\cite{KivelsonHowToRMP:2003,KivelsonIntertwined:2015}).  
Charge order, mostly in BCSSO systems, has been visualized with scanning tunneling microscopy (STM)~\cite{Mesaros08112016, 10.1038/ncomms1440, Hoffman466, Fujita612, Vershinin1995, He608, daSilvaNeto393, 10.1038/nphys1021}. Lately, charge order without spin order has been found in YBCO with quantum oscillations~\cite{10.1038/nature05872, 10.1038/nature06332},
NMR~\cite{10.1038/nature10345,10.1038/ncomms3113,10.1038/ncomms7438},
and X-ray studies~\cite{10.1038/ncomms11494}, hard~\cite{10.1038/nphys2456, PhysRevLett.110.137004} and soft~\cite{Ghiringhelli821, PhysRevLett.110.187001, PhysRevB.90.054513, 10.1038/nphys2041}, generating a large theoretical literature~\cite{PhysRevB.90.155114, 1367-2630-17-1-013025, PhysRevB.91.104509, PhysRevB.88.020506, PhysRevB.89.195115, PhysRevB.90.054511, PhysRevB.91.054502, PhysRevB.90.035149, PhysRevB.91.115103, PhysRevLett.114.197001, PhysRevB.91.195113, PhysRevB.90.195207, PhysRevB.92.045132, PhysRevB.95.224511, Caprara2017, PhysRevLett.114.257001, PhysRevB.93.155148} that finds charge order mostly around $1/8$ filling.

By contrast, two STM studies have focused on lightly hole-doped compounds ($p \ll 1/8$) Ca$_{2-x}$Na$_x$CuO$_2$Cl$_2$~\cite{kohsaka_visualization_2012} and Bi$_2$Sr$_{2-x}$La$_x$CuO$_{6+\delta}$~\cite{Nat.Phys.10.1038/nphys3840} in order to investigate how the Mott state evolves upon doping. The results suggest that a checkerboard charge order with a wavelength equal to four times the Cu-Cu bond length emerges first on doping. The charge order phase is present in both an antiferromagnetic or a superconducting background implying that it is not primarily driven by a Fermi surface or magnetic instabilities~\cite{Nat.Phys.10.1038/nphys3840}. Hence, the very-lightly doped Mott insulator by itself should show an instability towards this phase. This is the problem we study here.

Dynamical mean-field theory (DMFT)~\cite{RevModPhys.68.13}, its cluster~\cite{RevModPhys.77.1027,RevModPhys.78.865,LTP:2006} and diagrammatic extensions~\cite{PhysRevB.92.115109, PhysRevB.93.235124,PhysRevB.94.075159,PhysRevB.96.035152} are methods of choice to study the Mott transition.
The latter two methods include non-local correlations that are missed by DMFT and have been used to address the charge ordering in interacting systems. In small cluster calculations, uniform charge separation between a pseudogap phase originating from super-exchange and a correlated metal has been found~\cite{PhysRevLett.104.226402, PhysRevB.84.075161, Sordi:2012, FayeCharge:2017, CivelliCharge:2017}. In the electron-doped Mott insulator the variational cluster approximation~\cite{PhysRevB.74.024508} and the dynamical cluster approximation,~\cite{PhysRevB.74.085104,Galanakis1670} suggest phase coexistence between the Mott insulator and a correlated metal in analogy with single-site DMFT~\cite{PhysRevLett.89.046401,PhysRevX.5.021007}. However, cluster calculations  are unable to capture any ordering that extends beyond the cluster size, except for some limited special geometries~\cite{PhysRevB.95.115127,vanhala_dynamical_2017,MercureMSc:2015}. 

An alternate approach to charge instabilities goes beyond uniform phase separation and searches for finite-wave vector divergences or other anomalies of the density-density response function calculated in the normal phase. The response function is given by the second derivative of the free energy with respect to a conjugate field (scalar potential for density-density response function) and it is positive, due to the convexity of the free energy,  for a thermodynamically stable system. Therefore, a sign change of the response function indicates that the normal phase is unstable. 
In an interacting system, the response function takes into account that propagating particles and holes interact not only with the medium through their self-energy cloud, but also with each other with an amplitude called the full vertex function. The retarded part of that vertex originates from the exchange of other excitations, such as particle-hole (p-h) excitations.  This vertex function is a complex function with, in general, non-trivial momentum/frequency dependence. It is not taken into account in any mean-field theory, while it plays a crucial role in driving charge instabilities~\cite{PhysRevLett.114.257001, PhysRevB.93.155148}.

Starting from the prototypical Hubbard model for the cuprates, we show that in the hole-doped system, the spatially local part of the irreducible vertex in the charge channel becomes attractive for a range of low-frequencies.  This behavior starts at interaction values comparable with the electronic bandwidth of the system and becomes more prominent at larger interactions. A real charge instability occurs only for interactions larger than what is required to drive a Mott transition at half-filling. This indicates that the charge instability is an instability of the doped Mott insulator, and not of the Slater antiferromagnet. 

We present the model and method in section~\ref{II}. The single-site DMFT results for the compressibility are recovered in section~\ref{III}. Section~\ref{IV} describes the unusual structure of the frequency-dependent vertex in the charge channel, in particular that it can become attractive. This sets the stage for the calculation of the physically observable susceptibilities in section~\ref{V}. After concluding remarks, details of the calculation for the susceptibilities in various approximations are given in appendix~\ref{Response}. Appendix~\ref{Com} shows the results for the compressibility at lower interaction strength and higher temperature. It is shown in detail in appendix~\ref{AD} that the compressibility obtained from single-site DMFT is identical to that deduced from the zero-wave vector density-density response function at zero Matsubara frequency, in other words that the compressibility sum-rule is satisfied.  

\begin{figure}
\begin{center}
\begin{tabular}{c}
\includegraphics[width=0.95\columnwidth]{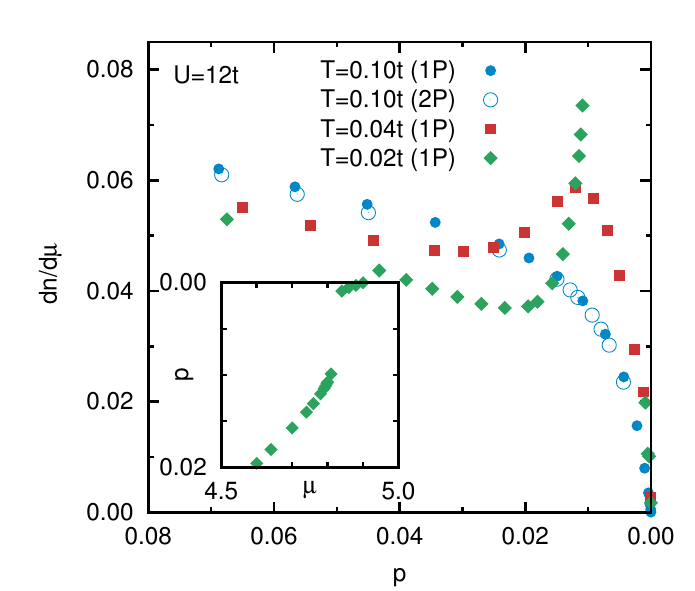} 
\end{tabular} 
\caption{ Compressibility, $\partial n/\partial \mu$, as a function of hole density, $p$, for  $U=12t$, $t^{\prime}=-0.3t$, $t^{\prime \prime}=0.2t$ at various temperatures, either computed directly within the DMFT framework, labeled 1P in the figure and discussed in section~\ref{III}, or computed from the lattice density-density correlation function, indicated as 2P and discussed in section~\ref{V}. Inset : The hole density as a function of chemical potential for $T=0.02t$ has a jump for a chemical potential $\mu_c\simeq 4.8$ with a discontinuous first-order derivative at that chemical potential, implying the discontinous compressibility seen in the main panel. }\label{fig:CompressibilityDMFT}
\end{center}
\end{figure}

\section{Model and method}\label{II}
We consider the Hubbard model on the square lattice
%
%
with $U$ the onsite Coulomb interaction and appropriate hopping parameters for cuprates, i.e, $t^{\prime}=-0.3t$, $t^{\prime \prime}=0.2t$ the second and third nearest-neighbor hopping amplitudes, $t$ being the nearest-neighbor hopping~\cite{PhysRevLett.87.047003}. The Hamiltonian is solved using DMFT
and the exact diagonalization (ED) method~\cite{PhysRevLett.72.1545}. The DMFT(ED) algorithm is also used to compute the local part of the irreducible vertex function~\cite{RevModPhys.68.13,PhysRevB.86.125114, PhysRevB.75.045118}.   The lattice response function can then be computed as described in appendix~\ref{Response}. Namely, the DMFT self-energy is included in the propagators and the vertices are obtained from four point functions on the self-consistent impurity~\cite{RevModPhys.68.13}. The resulting correlation functions are the building blocks of the ladder dynamical vertex approximation~\cite{PhysRevB.75.045118, arXiv:1705.00024}.  We focus on the charge channel because the charge and magnetic channels are independent in this approach. As in experiment, we see the charge order as an instability of the Mott insulator that appears independently of magnetic or superconducting long-range order.  Appendix~\ref{Response} contains more details on the method. 

\section{DMFT compressibility}\label{III} 
We first present the DMFT results, which are in qualitative agreement with previous DMFT results obtained at $t'=t''=0$~\cite{PhysRevLett.89.046401}.  
In the DMFT framework, the  possibility of a first order phase transition can be assessed by direct calculation of the charge compressibility, defined as $\kappa = (1/n^2)\partial n/\partial \mu $, as a function of hole-doping $p$.   A vanishing compressibility characterizes the Mott insulator, while a divergence indicates a second-order transition and a discontinuity a first-order transition. Let $U_c$ denote the DMFT critical interaction beyond which a Mott insulating phase appears at half-filling. The main panel of \fref{fig:CompressibilityDMFT} illustrates the compressibility as a function of hole density for  $U=12t > U_c$ at various temperatures. The compressibility data labeled with $1$P are calculated by a numerical derivative of the density with respect to the chemical potential. 
At $T=0.1t$ (in blue), $\partial n/\partial \mu $ continuously decreases upon approaching half-filling indicating the suppression of charge fluctuations. However, at a lower temperature, $T=0.04t$ (in red), after an initial decrease, the compressibility increases and exhibits a peak around $p\simeq 0.01$ before dropping to zero at half-filling. This increase suggests the proximity to the critical end-point of a first-order transition at a locus $(T_c,p_c)$. That first-order transition \cite{PhysRevLett.89.046401}
becomes visible at a lower temperature, $T=0.02t$ (in green) where the compressibility is discontinuous. The inset in  \fref{fig:CompressibilityDMFT} shows the hole density as a function of chemical potential at $T=0.02t$. This illustrates clearly that there is a value of the chemical potential for which there are two possible values of the hole doping, as expected in the coexistence region of a first-order transition. The critical doping, $p_c$, does not depend on $U>U_c$ sensitively but $T_c$ decreases upon increasing $U$. The phase transition is interpreted as a uniform phase separation (${\bf q}=0$) since it occurs in the compressibility. The question arises whether this actually occurs at ${\bf q}=0$ or if a more sophisticated calculation could reveal a finite wave-vector instability at a higher temperature. The answer to this question demands a calculation of the density-density response function at finite wave vector.

\section{Dynamical vertex}\label{IV} 
No charge instability is predicted by approximations with a static irreducible vertex, such as the random phase approximation (see appendix~\ref{RPA}). Hence we go beyond these approximations to obtain a reliable calculation for $U>U_c$. 
In a diagrammatic approach, the full vertex function can be decomposed into the irreducible vertex function and the generalized bubble susceptibility (see appendix~\ref{Response}).  One can diagonalize the irreducible vertex in spin space to exploit the conservation of spin in two-body scattering processes and rewrite it in the charge and magnetic channels, defined as ${ \Gamma}^{c(m),irr}={ \Gamma}^{\uparrow \uparrow,irr}+(-){ \Gamma}^{\uparrow \downarrow,irr}$. In a normal system,  there is a range and a characteristic relaxation time, beyond which ${ \Gamma}^{c/m,irr}$ becomes negligible. Hence,  the spatially local part of the  irreducible vertex function, $\big[{ \Gamma}^{irr}_{loc}(\nu_n)\big]_{\omega_m\omega_{m'}}$, is the dominant part. DMFT gives an accurate estimate of its dependence on three frequencies, $\nu_n$ the center of mass bosonic frequency and $\omega_m,\omega_{m'}$ the fermionic frequencies.

\begin{figure}
\begin{center}
\begin{tabular}{cc}
\includegraphics[width=0.49\columnwidth]{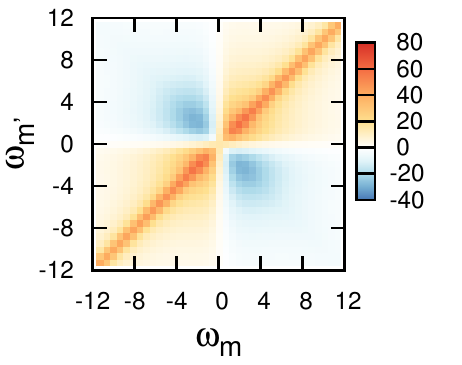}& 
\includegraphics[width=0.49\columnwidth]{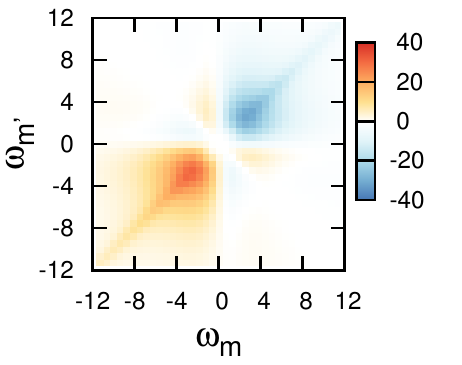}
\end{tabular}
\caption{ Real (left) and imaginary (right) parts of  $\big[{ \Gamma}^{c, irr}_{loc}(\nu_n=0)\big]_{\omega_m\omega_{m'}}$ as a function of two fermionic frequencies $\omega_m$ and $\omega_{m'}$ for hole doping level $p= 0.11$ and $T=0.1t$, $U=6t$. The static contribution, $U$, is substracted from the real part.  }\label{fig:GammaIrr_d1}
\end{center}
\end{figure}

In an interacting system, the electron spectral function demonstrates a  coherent (quasi-particle) peak and high energy incoherent (Hubbard) satellites. Upon increasing $U$, the spectral weight is transferred from the coherent peak to incoherent satellites. Thus, the irreducible vertex function contains two sets of scattering processes: (i) the scattering of the coherent part with itself (ii) the scattering of the incoherent part with itself and with the coherent part. The latter contribution is smooth and possibly featureless at low $U$, but increases and requires frequency dependence upon increasing interaction strength, leading to the nontrivial frequency dependence of ${ \Gamma}^{c, irr}$ for $U \geq W$.

Figure~\ref{fig:GammaIrr_d1} displays
the real and imaginary parts of the irreducible local  vertex in the density channel at zero bosonic frequency, $\nu_n$, for  doping level $p=0.11$ and $U=6t<W$ where $W$ is the bandwidth. The $\nu_n=0$ component measures the amplitude of scattering processes that occur at all time scales.
For $U=6t$, the system is in the perturbative regime and the behavior of $\big[{ \Gamma}^{c, irr}_{loc}(\nu_n=0)\big]_{\omega_m\omega_{m'}}$ can be understood from low-order diagrams~\cite{PhysRevB.86.125114}. The real part of the vertex function for $|\omega_m|=\omega_0$ and large $|\omega_{m'}|$ or vice versa   approaches its static limit, which is $U$ in the Hubbard model. Here, $\omega_0$ denotes the lowest fermionic Matsubara frequency.
Furthermore, the real part of the $\big[{ \Gamma}^{c, irr}_{loc}(0)\big]_{\omega_m\omega_{m'}}$ is repulsive and is large on the primary diagonal, i.e, when $\omega_m=\omega_{m'}$. This enhancement is caused by p-h scattering processes involving  the emission and reabsorption of pairs of fluctuations when the induced p-h excitations live on the Fermi surface~\cite{PhysRevB.86.125114}.
On the other hand, the scattering rate along the secondary  diagonal $\omega_m=-\omega_{m'}$ is due to particle-particle (p-p) scattering processes with energies $\omega+\nu$ and $\omega'$ and opposite spin in the Hubbard model.  These scattering processes have a large amplitude for total energies at the Fermi level, i.e., $\omega'=-\omega-\nu$ with a maximum at $\nu=0$~\cite{PhysRevB.86.125114}. The amplitude of the p-p scattering is smaller than the p-h scattering in the repulsive Hubbard model and large $U$. 
The right-hand panel of \fref{fig:GammaIrr_d1} shows the imaginary part of the irreducible vertex function. The absolute value of the imaginary part is proportional to the p-h asymmetry. It increases  along the primary diagonal and changes sign between positive and negative frequencies. The secondary diagonal peak is missing in the imaginary part. 
For interaction strengths smaller than the bandwidth, $U < W$, these characteristics  of the $\big[{ \Gamma}^{c, irr}_{loc}(\nu_n=0)\big]_{\omega_m\omega_{m'}}$ are common for all doping levels.

\begin{figure}
\begin{center}
\begin{tabular}{cc}
\includegraphics[width=0.49\columnwidth]{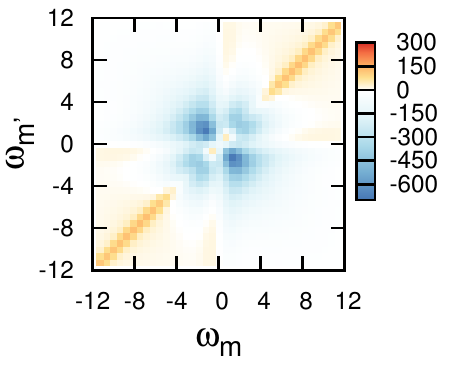} &
\includegraphics[width=0.49\columnwidth]{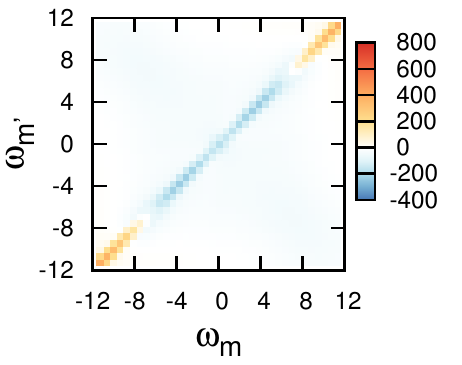}
\end{tabular}
\caption{ Real part of  $\big[{ \Gamma}^{c, irr}_{loc}(\nu_n=0)\big]_{\omega_m\omega_{m'}}$ for hole dpoing level $p=0.11$ as a function of two fermionic frequencies $\omega_m$ and $\omega_{m'}$  and $T=0.1t$. On the left panel, $U=8t$ while $U=12t $ on the right panel.  The static contribution, $U$, is subtracted. }\label{fig:GammaIrr_d2}
\end{center}
\end{figure}

In the non-perturbative regime at larger interaction strengths, $U \geq W$, the irreducible vertex in the charge channel gradually changes, as illustrated in \fref{fig:GammaIrr_d2}.  
Although the previously mentioned characteristics of $\big[{ \Gamma}^{c, irr}_{loc}(\nu_n=0)\big]_{\omega_m\omega_{m'}}$ are generally maintained in the high energy region, the low-frequency behavior  strongly depends on the interaction strength and doping level.  In the low doping region,  the real part of $\big[{ \Gamma}^{c, irr}_{\nu_n=0}\big]_{\omega_m\omega_{m'}}$  around the primary diagonal begins to show at intermediate fermionic frequencies a sign change between low and high frequencies: this means that, surprisingly, the emission and reabsorption of p-h pairs causes an effective interaction which is attractive for intermediate frequencies (see left-hand panel of the \fref{fig:GammaIrr_d2} for $U=8t$).

It is worth mentioning that, for $(U, T)$ considered here and finite doping where there is no particle-hole (p-h) symmetry, the irreducible charge vertex does not show the divergence due to a vanishing eigenvalue of the impurity susceptibility~\cite  {PhysRevB.94.235108, PhysRevB.98.235107}. By contrast with a system having p-h symmetry, the generalized impurity susceptibility here is a complex matrix with eigenvalues that are either purely real or appear in complex conjugate pairs. Depending on the interaction strength and doping level, the generalized impurity susceptibility may have egivenvalues with negative real part but their zero-crossing occurs at finite imaginary part, leading to a smooth irreducible charge vertex. For small $U$, the eigenvalues remain real for a small range of doping~\cite{PhysRevB.97.125141}.  

At larger $U=12t$, the sign change of the irreducible charge vertex includes the very low frequency region, as can be seen from the right-hand panel of \fref{fig:GammaIrr_d2}.  Indeed, a behavior different from the perturbation theory prediction occurs on an energy scale of order $U$ at low temperatures. Upon increasing $T$, the non-perturbative low-frequency region shrinks.  
The change in ${ \Gamma}^{c, irr}_{loc}$ for $U \geq W$ could be a manifestation of the replacement of the perturbative branch of the self-energy with the non-perturbative one at the physical self-energy\cite{PhysRevLett.114.156402}.

\section{Charge susceptibility}\label{V} 
From the irreducible vertex in the charge channel, we can compute the charge susceptibility. Then, instead of differentiating $n(\mu)$,  we compute the compressibility from the density-density correlation function, ${\chi}_{ph}^{c}$, using the compressibility sum rule (with $K\equiv ({\bf k},i\omega_m)$) %
\begin{equation}
\partial n/\partial \mu =2\lim_{{\bf q}\rightarrow 0,\nu\rightarrow 0}(1/N\beta)^2\sum_{KK'}[{\chi}_{ph}^{c}({\bf q},\nu_n)]_{K,K'}.\label{Compressibility}
\end{equation}
In general sum rules in approximate theories can be violated. For instance, this happens~\cite{PhysRevB.93.155162, PhysRevB.94.125144} in a different context than our analysis, for the potential energy computed by means of DMFT in finite dimensional systems. Here, instead, due to the $\Phi$-derivability of DMFT, the sum rule in \eref{Compressibility} must be necessarily fulfilled (see Appendix B and Refs.~\onlinecite{PhysRevB.92.085106, PhysRevB.90.235105}).

The density of the self-consistent impurity depends on $\mu$ explicitly and implicitly through the hybridization function, $\Delta(\mu)$, hence, the impurity compressibility is given by $(\partial n/\partial \mu)_{\Delta}+(\partial n /\partial \Delta)_{\mu}(\partial \Delta/\partial \mu)$~\cite{PhysRevB.93.155162}. The equation of motion for $(\partial n/\partial \mu)_{\Delta}$ and $(\partial n /\partial \Delta)_{\mu}$ includes, respectively, $\partial \Sigma_{\sigma}(i\omega_m)/\partial g_{\sigma}(i\omega_{m'})$ and  $\partial \Sigma_{\sigma}(i\omega_m)/\partial \Delta_{\sigma'}(i\omega_{m'})$ where $\Sigma$ is the impurity self-energy and $g$ denotes local Green's function (see appendix~\ref{AD}).  However, at strong coupling, multiple branches appear in the physical self-energy~\cite{PhysRevLett.114.156402, PhysRevLett.119.056402, PhysRevB.94.235108}. 
This makes self-energy functional derivatives  ill-defined   when  the perturbative branch crosses the non-perturbative branch. Nevertheless, our numerical results show that the compressibility calculated either directly from the chemical potential dependence of the density or from the density-density correlation function yield the same results, confirming that a perturbation expansion for functional derivatives of the self-energy remains valid even at strong interaction. This is shown in \fref{fig:CompressibilityDMFT} for $T=0.1t$ (see appendix~\ref{Com} for other interaction values and temperatures). The frequency summation at the \eref{Compressibility} converges very slowly, in particular, in the vicinity of the phase transition, hence, here we only show charge susceptibilities at $T=0.1t$~\cite{Note1} .
This is not too restrictive because a phase transition and the wave vector at which it happens is usually seen as a softening of susceptibilities at temperatures well above the transition temperature.

Before we continue, we comment on how the negative eigenvalues of the impurity suscptibility influence the charge fluctuations. In previous studies of the half-filled p-h symmetric system ($t^{\prime}=t^{\prime\prime}=0$), the appearance of negative eigenvalues of the impurity susceptibility upon increasing $U$ was interpreted as an indication of the suppression of charge fluctuation~\cite{PhysRevLett.119.056402, PhysRevB.97.245136}. The argument goes as follows: setting the oscillator matrix elements equal to one, the observable  impurity susceptibility can be expressed in terms of the generalized susceptibility as
\begin{equation}
\chi^c_{imp}(i\nu_n)=\frac{1}{\beta^2}\sum_{\omega_m\omega_{m^{\prime}}}[\chi^c_{imp}(i\nu_n)]_{\omega_m\omega_{m^{\prime}}}.\label{ImpSus}
\end{equation}
Using an expression of the generalized susceptibility in terms of its eigenvalues ($\epsilon_i$) and its eigenvectors ($|i\rangle$), the above equation can be rewritten as
\begin{align}
\chi^c_{imp}(i\nu_n)&=\sum_i\frac{1}{\beta^2}\sum_{\omega_m\omega_{m^{\prime}}}\langle m | i\rangle \epsilon_i \langle i | m^{\prime}\rangle \nonumber \\&=
\sum_i \epsilon_i \big|  \frac{1}{\beta}\sum_{\omega_m} \langle m | i\rangle\big|^2 .
\end{align}
It is argued that $\big|  \frac{1}{\beta}\sum_{\omega_m} \langle m | i\rangle\big|^2$ is in general not small, hence, obtaining a decreasing charge susceptibility upon increasing $U$ requires that some eigenvalues become negative~\cite{PhysRevB.93.245102}. Note that for weak to intermediate couplings where all eigenvalues are positive, the suppression of charge fluctuations is obtained by a reduction of the eigenvalues instead.

The dependence of eigenvalues or of the overlaps on doping or temperature is more complex than their dependence on $U$ and they do not necessarily change monotonically. For example, for $U=8t$ and $T=0.1t$, the positive  eigenvalues originally decrease upon approaching the half-filling while the absolute value of negative eigenvalues increase. However, in close vicinity to half-filling, this trend stops and reverses, leading to an enhancement of the charge fluctuations, as can be seen from \fref{fig:Compressibility2}, top panel. Hence, while the decrease of the local charge susceptibility at large doping and large $U$ is associated with the appearance of negative eigenvalues, as one approaches half-filling charge fluctuations can begin to increase again at low temperature even if negative eigenvalues are still present, albeit with a smaller absolute value.

\begin{figure}
\begin{center}
\begin{tabular}{cc}
\includegraphics[width=0.49\columnwidth]{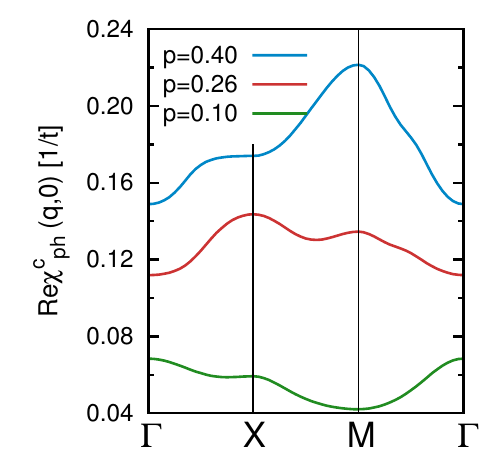} &
\includegraphics[width=0.49\columnwidth]{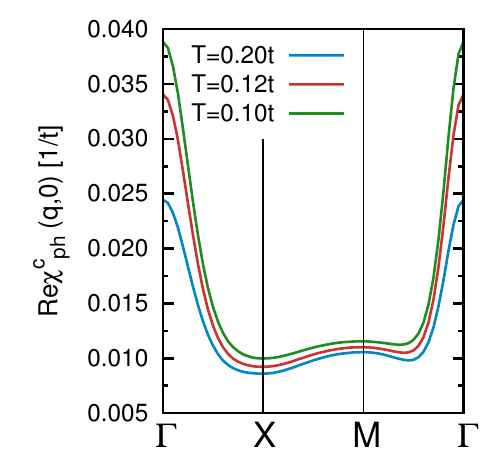}
\end{tabular} 
\caption{ Left panel: Dressed susceptibility in the charge channel for $U=12t$ and $T=0.1t$, for several hole doping values. Right panel: Dressed susceptibility in the charge channel for $U=12t$, for several temperatures at hole doping slightly larger than the critical doping}\label{fig:Compressibility}
\end{center}
\end{figure}

Figure~\ref{fig:Compressibility} left panel, shows the dressed charge susceptibility along a path as a function of hole density. At large doping 
the dressed charge susceptibility peaks at the same momenta as the bare or RPA susceptibilities. Although, the RPA susceptibility maintains this peak structure for lower doping values,  the peak structure of ${\chi}_{ph}^{c}$ changes; the peak momentum moves to $X (Y)$ point and eventually to the $\Gamma$ point upon reducing $p$. Indeed, the charge susceptibility at $(\pi,\pi)$ decreases much faster than the other wave-vectors, as can be seen from this figure. This behavior is induced by the frequency dependence of the vertex function and cannot be captured by the RPA approximation (see appendix~\ref{RPA}).

Going back to our original question regarding whether or not the system undergoes  a uniform phase separation, the momentum dependence of the dressed susceptibility in the charge channel  at a hole doping slightly larger than the critical doping is shown in the right-hand panel of \fref{fig:Compressibility} for $U=12t$ at several temperatures. As can be seen, upon reducing $T$  the charge susceptibility at ${\bf q}=0$ grows faster than at the other wave-vectors, indicating a charge instability at lower $T$ at this momentum.  This confirms previous results from compressibility studies\cite{PhysRevLett.114.156402}. 

Finally, it is worth mentioning that the largest eigenvalue of the (dimensionless) matrix  $-(1/\beta^2)\sum_{\omega_{m''}}\big[{ \Gamma}^{c/m, irr}_{loc}(0)\big]_{\omega_m\omega_{m''}}[\tilde{{\chi}}^{0}_{ph}({\bf q},0)]_{\omega_{m''},\omega_{m'}}$ with $[\tilde{{\chi}}^{0}_{ph}(Q)]_{\omega_m,\omega_{m'}}\equiv(1/N)^2 \sum_{\bf k, \bf k'}[{\chi}^{0}_{ph}(Q)]_{K,K'}$ is called the Stoner factor. When it is a real number,  it measures  the  distance from a continuous phase transition.  With a static vertex function, as in the RPA approximation, the Stoner factors in the magnetic and charge channels are purely real.  However, the eigenvalue problem that needs to be solved to search for an instability is a non-Hermitian eigenvalue problem in general. Nevertheless, the susceptibilities are always real. While our numerical results show that the Stoner factor in the magnetic channel is always real and increases with increasing $U$, eventually approaching unity, in the charge channel it is real only for $U<W$. For $U>W$,  the eigenvalues appear mostly in complex conjugate pairs, with large real and imaginary parts, in particular at low doping. Therefore, in this regime, the Stoner factor is not well defined and cannot serve as a probe of an instability. The appearance of the complex conjugate pairs might indicate a tendency towards short-range charge orders, but further investigation is required to be conclusive. 

\section{Concluding remarks}\label{VI} 
In summary, we have calculated the dressed charge susceptibility of strongly correlated metals to investigate possible charge instabilities. We verified that even for strong interactions, the compressibility sum-rule is satisfied consistently with the $\Phi$-derivability of DMFT. Our results showed that the momentum-dependent dressed charge susceptibility of a doped Mott insulator has a completely different peak structure than what an RPA analysis would predict. This is a consequence of the complicated frequency dependence of the irreducible vertex function. Indeed, the charge susceptibilities with wave vectors connecting hot spot decrease faster than at the other wave vectors upon approaching the Mott phase. Furthermore, the phase transition of a lightly hole-doped Mott insulator occurs at the ${\bf q}=0$ wave-vector in agreement with the DMFT compressibility studies. Although the single-band Hubbard model does not show any charge instability with a finite momentum, including more degrees of freedom, such as Oxygen $p$ orbitals, may change the physics. In Ref.~\onlinecite{PhysRevLett.114.257001}, finite momentum instabilities were found. It may also be that the ${\bf q}=0$ instability corresponds, in a more realistic model to an intra unit cell charge pattern~\cite{lawler_intra-unit-cell_2010}.

\begin{acknowledgments}
R.~N and A.-M.S.~T. are grateful to A.~Toschi  for critically reading the manuscript and for his helpful comments. We are thankful
to A.~Georges, M.~Ferrero, J.~Gukelberger, O.~Parcollet, and Wenhu Xu for useful discussions. R.~N is also thankful to G.~Kotliar for helpful discussions.  This work has been supported by the Canada First Research Excellence Fund, the Natural Sciences and Engineering Research Council of Canada (NSERC) under grants RGPIN-2014-04584 and RGPIN-2016-06666, and by the Research Chair in the Theory of Quantum Materials. Simulations were performed on computers provided by the Canadian Foundation for Innovation, the Minist\`ere de l'\'Education des Loisirs et du Sport (Qu\'ebec), Calcul Qu\'ebec, and Compute Canada.
\end{acknowledgments}

\appendix
\section{Response functions}\label{Response}
The response function of an interacting system can be decomposed  into the bubble response and vertex corrections.  The bubble response describes independent, but interaction-renormalized,  propagation of a particle-hole (p-h) excitation created by the field, while the vertex corrections introduce changes in the response functions due to scattering processes in which propagating particle and hole exchange multiple real or virtual p-h excitations.  Denoting the amplitude of all these scattering events by the full vertex function, defined as ${ \Gamma}^{c(m),f}={ \Gamma}^{\uparrow \uparrow,f}+(-){ \Gamma}^{\uparrow \downarrow,f}$ for charge and magnetic channels, the dressed susceptibility is given by~\cite{Bickers2004, PhysRevLett.117.137001}
\begin{align}
[{\chi}^{c/m}(Q)]_{K,K'} &=[{ \chi}^{0}_{ph}(Q)]_{K,K'}-\frac{1}{N^2\beta^2}\sum_{K_1,K_2}\nonumber \\
[{ \chi}_{ph}^{0}(Q)&]_{K,K_1}
[{ \Gamma}^{c/m,f}(Q)]_{K_1,K_2}[{\chi}^{0}_{ph}(Q)]_{K_2,K'}.
\label{eq:BSSus}
\end{align}
where the bubble susceptibility is 
\begin{equation}
[{\chi}^{0}_{ph}(Q)]_{K,K'} =-(N\beta) G(K+Q) G(K)\delta_{K,K'}.  
\end{equation}
Here, $G(K)$ is the dressed particle propagator, $K\equiv ({\bf k},i\omega_m)$ denotes momentum/energy three-vectors (the lattice is two-dimensional), $N$ is number of ${\bf k}$-points and $\beta = 1/(k_BT)$.  In \eref{eq:BSSus}  the bosonic variable $Q\equiv ({\bf q},i\nu_n)$ is always inactive in the multiplication, or conserved during the scatterings within each channel.
\eref{eq:BSSus} is the common part of the response to an external field and solely depends on the electronic structure of the system. 
An observable response function, on the other hand, is obtained by closing the external legs of \eref{eq:BSSus} using appropriate oscillator matrix elements, $O(Q)$ and  $O(-Q)$, which depends on the field wave-vector and frequency.\cite{PhysRevB.96.125140} The oscillator matrix elements for the charge channel are given by ($SU(2)$ symmetric case)
\begin{align}
{ O}_{{\bf R}_1{\bf R}_2} ({\bf q})&\equiv\frac{1}{\sqrt{V}}\int d{\bf r} 
e^{-i{\bf q}\cdot {\bf r}} 
\phi^*_{{\bf R}_1}({\bf r}) \phi_{{\bf R}_2 }({\bf r}),\label{OSReal}
\end{align}
where $\phi_{{\bf R}_1}({\bf r})$ the atomic orbital residing at the lattice point ${\bf R}_1$.

In a correlation-driven phase transition, it is the full vertex function, $[{ \Gamma}^{c(m),f}(Q)]_{K,K'}$, that causes a singular or discontinuous response. It consists of all connected diagrams. Some of these diagrams are two-particle fully irreducible. Other diagrams are reducible, i.e., cutting two Green function lines separates the diagram into two pieces. Indeed, each diagram is either fully irreducible or reducible in exactly one channel (particle-hole $ph$, particle-hole transversal $\overline{ph}$, and particle-particle $pp$), so~\cite{arXiv:1705.00024} 
\begin{align}
[{ \Gamma}^{c(m),f}&(Q)]_{K,K'} = [{ \Lambda}^{c(m)}(Q)]_{K,K'}+[{ \Phi}^{c(m)}_{ph}(Q)]_{K,K'}\nonumber \\&+[{ \Phi}^{c(m)}_{\overline{ph}}(Q)]_{K,K'}+[{ \Phi}^{c(m)}_{pp}(Q)]_{K,K'}. 
\end{align}
Here, ${ \Lambda}^{c(m)}$ and ${ \Phi}^{c(m)}_{l}$ denote, respectively, the fully irreducible vertex in all channels and reducible vertex in $l$ channel.  Moreover, one can define the irreducible diagrams in a certain channel $l$ as ${ \Gamma}^{c(m),f} = { \Gamma}^{c(m),irr}_l + { \Phi}^{c(m)}_l$.
For example, for the $ph$ channel
\begin{align}
[{ \Gamma}^{c(m),irr}_{ph}&(Q)]_{K,K'} = [{ \Lambda}^{c(m)}(Q)]_{K,K'} + 
\nonumber \\&+[{ \Phi}^{c(m)}_{\overline{ph}}(Q)]_{K,K'}+[{ \Phi}^{c(m)}_{pp}(Q)]_{K,K'}. \label{irr}
\end{align}
Having the irreducible vertex in a given channel $l$, the reducible one can be obtained from the Bethe-Salpeter equations (BSE) as~\cite{Bickers2004}
\begin{align}
[{\Gamma}^{c/m,f}(Q)]_{K,K'} &=[{ \Gamma}^{c/m, irr}_{ph}(Q)]_{K,K'}-\frac{1}{N^2\beta^2}\sum_{K_1,K_2}\nonumber \\
[{ \Gamma}^{c/m,f}(Q)&]_{K,K_1}
[{ \chi}^{0}(Q)]_{K_1,K_2}[{\Gamma}^{c,irr}_{ph}(Q)]_{K_2,K'}.
\label{eq:BSfvertex}
\end{align}
The irreducible vertex function can be evaluated using various quantum many-body approaches.

In general, $[{ \Gamma}^{c/m,irr}(Q)]_{K,K'}$ depends on the the transferred momentum/frequency in a scattering process, $Q$, and on the incoming momentum/frequency variables. The out-coming variables are determined by conservation laws.  $[{ \Gamma}^{c/m,irr}(Q)]_{K,K'}$ describes the irreducible interaction of the two elementary excitations. 
The spatially local part of the irreducible vertex function in channel $l$, $\big[{ \Gamma}^{c/m, irr}_{loc,l}(\nu_n)\big]_{\omega_m\omega_{m'}}$, can be calculated in framework of the DMFT approximation~\cite{PhysRevB.75.045118,PhysRevB.86.125114}.  
A common approximation is substituting  ${ \Gamma}^{c/m,irr}_{ph}$ by ${ \Gamma}^{c/m, irr}_{loc}(\nu_n)$ and neglecting the non-local part~\cite{RevModPhys.68.13}.  This allows to perform the summations over momentum of the internal legs in \eref{eq:BSfvertex}, leading to a full vertex function that satisfies a similar equation but with the bubble susceptibility replaced by 
\begin{equation}
[\tilde{{\chi}}^{0}_{ph}(Q)]_{\omega_m,\omega_{m'}}\equiv(\frac{1}{N})^2 \sum_{\bf k, \bf k'}[{\chi}^{0}_{ph}(Q)]_{K,K'}.\label{eq10}
\end{equation} 
Hence, the resulting full vertex depends on three frequencies but only one momentum (transferred momentum). This approximation captures the dynamics of the screening effects, which plays a significant role in correlated electron systems.~\cite{PhysRevB.86.064411} 

Focusing on the particle-hole channel, the DMFT approximation for the irreducible vertex ${ \Gamma}^{c(m),irr}_{ph}$ on the left-hand side of \eref{irr} implies the momentum dependence of the reducible vertices in the transverse particle-hole, ${ \Phi}^{c(m)}_{\overline{ph}}$,  and particle-particle channels, ${ \Phi}^{c(m)}_{pp}$, are neglected (see \eref{irr}). On the other hand, the dependence of the reducible $ph$  vertex on the transferred momentum is taken into account.

\subsection{Numerical considerations}

\begin{figure}
\begin{center}
\includegraphics[width=0.75\columnwidth]{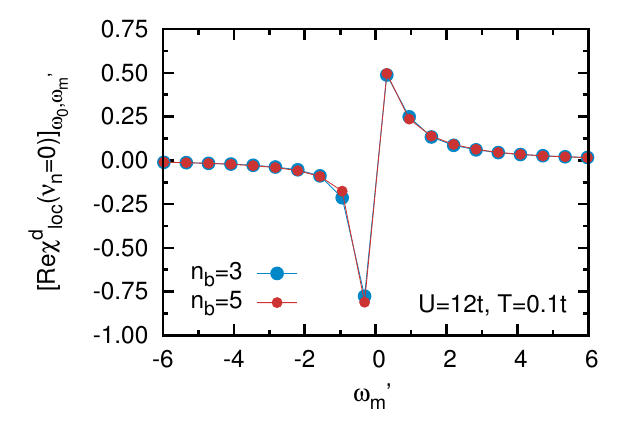}
\caption{  $\big[{ \chi}^{c}_{loc}(\nu_n=0)\big]_{\omega_m\omega_{m'}}$ with $\omega_m=\omega_0=\pi/\beta$ as a function of $\omega_{m'}$ calculated with $n_b=3$ and $n_b=5$. The doping level is $p=0.06$, $U=12t$. }\label{fig:IrrVer}
\end{center}
\end{figure}

We employed the exact diagonalization (ED) technique to solve the DMFT equations and to calculate the irreducible impurity  vertices. The latter calculation is very expensive and grows very fast with the number of bath levels in the ED method. Furthermore, calculating the dressed susceptibility requires $\big[{ \Gamma}^{c/m, irr}_{loc}(\nu_n)\big]_{\omega_m\omega_{m'}}$ calculated on a large number of fermionic Matsubara frequencies. For instance, for compressibility calculations, we took $512$ positive frequencies.  Hence, we only consider three bath levels, $n_b$, to calculate the irreducible impurity  vertices. We checked that the results do not change when increasing the number of bath level by performing calculations with five bath levels, $n_b=5$, for some interaction strengths and doping levels, albeit on a much smaller frequency range. For instance, \fref{fig:IrrVer} demonstrates the dressed impurity susceptibility in the charge channel,  $\big[{ \chi}^{c}_{loc}(\nu_n=0)\big]_{\omega_m\omega_{m'}}$, at zero bosonic frequency and $\omega_m=\omega_0=\pi/\beta$ as a function of $\omega_{m'}$ calculated with $n_b=3$ and $n_b=5$. As one can see the difference between the two calculations is negligible. The DMFT calculation of compressibility, presented in \fref{fig:CompressibilityDMFT},  on the other hand, are done using five bath levels. 

 \subsection{Random phase approximation}\label{RPA} 
Calculations with a dynamical irreducible vertex function, as in the ladder dynamical vertex approximation, differ from those done in RPA with a static vertex. In RPA, the irreducible vertex is approximated with a static, though screened, interaction which is smaller than the bare one and remains repulsive. It is parametrized by a screened Hubbard interaction $U_s$. With a static vertex, the summation over internal frequencies can be done and hence in RPA the internal bubble susceptibilities are replaced by 
\begin{equation}
{\chi}^{0,ph}_{RPA}(Q)=(\frac{1}{\beta})^2\sum_{m,m'}[\tilde{{\chi}}^{0}_{ph}(Q)]_{\omega_m,\omega_{m'}}.
\end{equation}
This makes the instability eigenvalue problem a real-symmetric one~\cite{Note2}.
Furthermore, while $[\tilde{{\chi}}^{0}_{ph}(Q)]_{\omega_m,\omega_{m}}$ has both real and imaginary parts, the RPA susceptibility, ${\chi}^{0,ph}_{RPA}(Q)$, is purely real.
\begin{figure}
	\begin{center}
		\begin{tabular}{cc}
			\includegraphics[width=0.45\columnwidth]{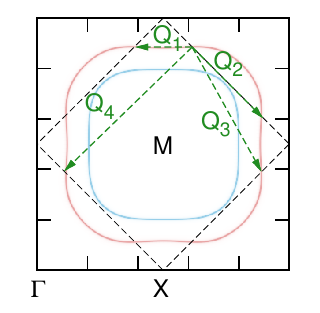}&
			\includegraphics[width=0.49\columnwidth]{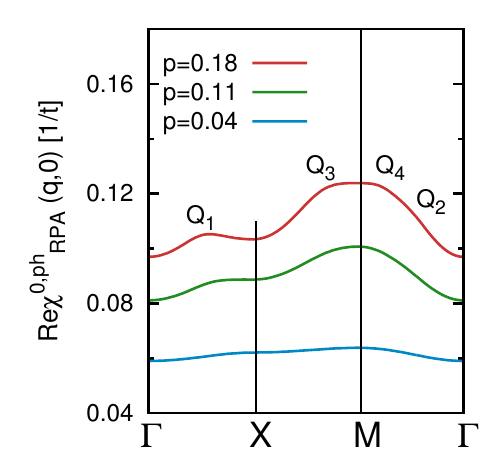} 
		\end{tabular}
		\caption{ Spectral weight at zero frequency (left panel) for hole doping levels $p= 0.18$ (red) ,  and $p= 0.04$ (blue) for  $U=12t$, $t^{\prime}=-0.3t$, $t^{\prime \prime}=0.2t$ at $T=0.1t$ on the square lattice. The dash-line shows the antiferromagnetic Brillouin zone.  The wave vectors ${\bf Q}_{1\cdots 4}$ connect hot spots when $p=0.18$.  The corresponding RPA bubble susceptibility is on the right-hand panel. 
		}\label{fig:FS}
	\end{center}
\end{figure}

To gain insight into the problem, consider the prediction of RPA when dressed propagators are used. The left panels of \fref{fig:FS} display the locus of maxima of the spectral weight at zero frequency, $A({\bf k},\omega=0)$, for two hole-doping values: one at large doping $p=0.18$ (red) and one at small doping $p=0.04$ (blue) for $U=12t$ and $T=0.1t$. Note that the depicted $A({\bf k},\omega=0)$ should not be interpretd as the Fermi surface (FS) since that concept is strictly defined only for a Fermi liquid at zero temperature.
At large doping, the spectral weights intersect the antiferromagnetic (AF) Brillouin zone  (BZ) at the so-called hot spots, i.e, regions of FS where the probability of Umklapp and $(\pi,\pi)$ scattering events is appreciable. The bubble susceptibility calculated in RPA shows peaks at the wave-vectors connecting hot spots. At smaller $U$, such as $U=8t$, $A({\bf k},\omega=0)$  crosses the AF-BZ for both doping values. At  $U=12t$, however,  the hot spots disappear at low doping due to the combination of interactions, which make the spectrum less coherent, and finite temperature. This influences the RPA bubble susceptibilities, as shown on the right-hand panel of \fref{fig:FS}. 

In RPA, the charge susceptibility is small and the leading instability occurs in the magnetic channel with  wave-vectors $(1, 1-\delta)\pi/a$ and $(1-\delta, 1)\pi/a$, where $\delta$ is small and vanishes at half-filling.

 \section{Compressibility at lower interaction strength and higher temperature}\label{Com}
\begin{figure}
\begin{center}
\begin{tabular}{c}
\includegraphics[width=0.75\columnwidth]{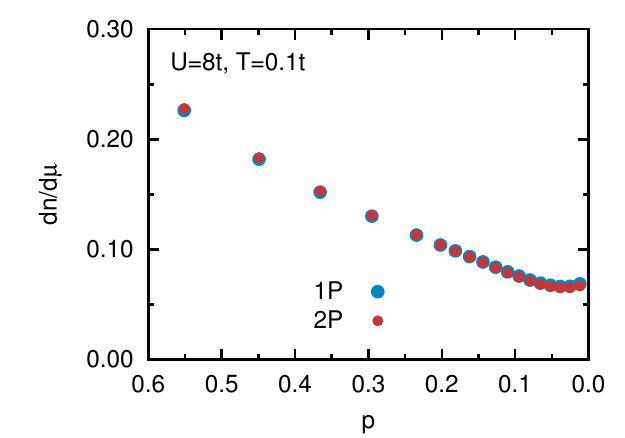} \\
\includegraphics[width=0.75\columnwidth]{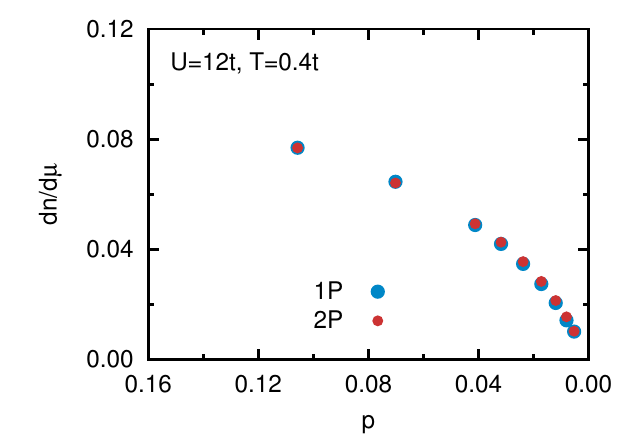}
\end{tabular}
\caption{ Compressibility $dn/d\mu$ computed from the density and from the uniform density-density correlation function, labeled by 1P and 2P respectively, as a function of hole density for $U=8t$ and $T = 0.1t$ (top panel) and  $U=12t$ and $T = 0.4t$ (bottom panel). }\label{fig:Compressibility2}
\end{center}
\end{figure}

 The transition to the charge ordered state, discussed in the main text, is absent at lower interaction strengths.  \fref{fig:Compressibility2} shows the compressibility for $U=8t$ and $T = 0.1t$ (top panel) and  $U=12t$ and $T = 0.4t$ (bottom panel) computed from the derivative of the density with respect to chemical potential and from the uniform density-density correlation function. Both methods again agree very well at all doping values considered here and they do not show a tendency towards phase separation. 

\section{Thermodynamic derivative of the density in the DMFT approximation }\label{AD}
In this section we show that the two approaches we have employed to calculate the compressibility -- namely the derivative of the density with respect to the chemical potential and the zero-frequency zero-momentum lattice charge susceptibility -- are equivalent within a local self-energy, local vertex approximation, as long as the Luttinger-Ward functional remains single-valued. For an exact calculation, the two results would obviously be equal, as required by a thermodynamic sum-rule. 

\subsection{Preliminary considerations}

The density is given by
\begin{align}
n&= \frac{1}{\beta}\sum_{\omega_m,\sigma}e^{-i\omega_m0^-} g_{\sigma}(i\omega_m)
\nonumber\\
&=\frac{1}{\beta N}\sum_{{\bf k}\omega_m,\sigma}
\frac{e^{-i\omega_m0^-}}{i\omega_m+\mu-\epsilon_{\bf k}-\Sigma_{\sigma}(i\omega_m)},
\end{align}
where the first line is for the impurity and the second line is for the lattice. Since the DMFT self-consistency equation is,
\begin{align}
g_{\sigma}(i\omega_m)=\frac{1}{N}\sum_{\bf k}G_{{\bf k}\sigma}(i\omega_m).\label{Eq:SC}
\end{align}
the density on the impurity is equal to the density on the lattice for all values of the chemical potential and $({\partial n}/{\partial \mu})_{T}$ will be identical in the two cases.

However we can also obtain $({\partial n}/{\partial \mu})_{T}$ from the susceptibility on the lattice, using the irreducible particle-hole vertex. That quantity is calculated from correlation functions on the impurity by inverting the Bethe-Salpeter equation. 
 
Remembering that the chemical potential is space-independent and imaginary-time independent, the susceptibility that is needed on the lattice to compute $({\partial n}/{\partial \mu})_{T}$ is a special case of the susceptibility considered in the previous section, namely

\begin{align}
&\frac{\partial G_{\bf k\sigma}(i\omega_m)}{\partial \mu}= - G_{{\bf k}\sigma}(i\omega_m)G_{{\bf k}\sigma}(i\omega_m)
\nonumber\\
&+G_{{\bf k}\sigma}(i\omega_m)G_{{\bf k}\sigma}(i\omega_m)\nonumber\\
&\times [\Gamma^{irr}_{\sigma\sigma,\sigma'\sigma'}(\nu_n=0)]_{\omega,\omega'}
\frac{1}{N}\sum_{{\bf k'},\sigma'}
\frac{\partial G_{\bf k'\sigma'}(i\omega_{m'})}{\partial \mu}.
\end{align}
where $[\Gamma^{irr}_{\sigma\sigma,\sigma'\sigma'}(\nu_n=0)]_{\omega,\omega'}$ is the irreducible particle-hole vertex, which we is momentum independent since it comes from a calculation on the impurity. Summing over wave vector on both sides of the equation and using the notation  
\begin{align}
G^{loc}_{\sigma}(i\omega_m)=\frac{1}{N}\sum_{\bf k}G_{{\bf k}\sigma}(i\omega_m).
\end{align}
this gives the following closed equation that can be solved by considering $\omega_m , \omega_m'$ as matrix indices and inverting:
\begin{align}
&\frac{\partial G^{loc}_{\sigma}(i\omega_m)}{\partial \mu}= - \frac{1}{N}\sum_{\bf k}G_{{\bf k}\sigma}(i\omega_m)G_{{\bf k}\sigma}(i\omega_m)
\nonumber\\
&+\frac{1}{N}\sum_{\bf k}G_{{\bf k}\sigma}(i\omega_m)G_{{\bf k}\sigma}(i\omega_m)\nonumber\\
&\times [\Gamma^{irr}_{\sigma\sigma,\sigma'\sigma'}(\nu_n=0)]_{\omega,\omega'}
\frac{\partial G^{loc}_{\sigma'}(i\omega_m')}{\partial \mu}.
\end{align}

If instead of the above approach, we compute the derivative of the impurity Green function $g_{\sigma}(i\omega_m)$ (taking into account the fact that the hybridization function depends on $\mu$ as well) and use the self-consistency equation \eref{Eq:SC}, we find the same equation as above with the following two replacements
\begin{equation}
[\Gamma^{irr}_{\sigma\sigma,\sigma'\sigma'}(\nu_n=0)]_{\omega,\omega'} \rightarrow \frac{\partial \Sigma_{\sigma}(i\omega_m)}{\partial g_{\sigma'}(i\omega_{m'})},\label{Eq:Vertex_funct_der}
\end{equation}
and
\begin{equation}
G^{loc}_{\sigma}(i\omega_m) \rightarrow g_{\sigma}(i\omega_m).
\end{equation}
$G^{loc}_{\sigma}(i\omega_m)$ and $g_{\sigma}(i\omega_m)$ are in fact equal because of the self-consistency equation \eref{Eq:SC}. By contrast, for \eref{Eq:Vertex_funct_der} to be an equality, we have to take the functional derivative with respect to $g_{\sigma'}(i\omega_{m'})$ on the physical branch when $\Sigma_{\sigma}(i\omega_m)$ is not single-valued. That $\Sigma_{\sigma}(i\omega_m)$ can be multi-valued has been documented~\cite{PhysRevLett.114.156402}. In addition there are two other possible difficulties. First, the vertex can diverge at points where two branches of the solution cross~\cite{PhysRevLett.119.056402}, a problem that is avoided for the case we consider. Second, the impurity Green function depends also on the self-consistent value of the hybridization function. That self-consistency condition can lead to phase transitions, such as the Mott transition at half-filling. In that case, precursors of the phase transition can appear in the vertex function~\cite{PhysRevLett.110.246405}. A more careful look at the separate effect of the hybridization function is given in the following section.

\subsection{Detailed derivation}
The first derivative of the free energy is related to the Green's function as follows

\begin{equation}
n = -\frac{\partial \mathcal{F}}{\beta\partial \mu} = \frac{1}{N\beta}{\rm Tr} G = \frac{1}{N\beta}\sum_{{\bf r}\sigma}G_{\sigma}({\bf r}\tau,{\bf r}\tau^+),
\end{equation}
where $ \mathcal{F}$ denotes free energy density and $\mu$ the chemical potential. The second derivative of the free energy with respect to the chemical potential gives the electron compressibility. 

To obtain an integral equation for the charge susceptibility, we apply a perturbation $\phi({\bm 1}{\bm 1}')=-\mu\delta_{{\bf r}{\bf r}'}\delta_{\tau \tau'^+}$ (where we have used the compact notation ${\bm 1} \equiv ({\bf r}, \tau)$) and calculate the response in the DMFT approximation.  The dimensionless thermodynamic derivative of interest is (with $({\bf r}'_1, \tau'_1)=({\bf r}_1, \tau^+_1)$)
\begin{equation}
\frac{\partial G_{\sigma}({\bm 1}{\bm 1}')}{\partial \phi({\bm 2}{\bm 2}')}=-G_{\sigma}({\bm 1}{\bm 3})\frac{\partial G^{-1}_{\sigma}({\bm 3}{\bm 3}';\phi)}{\partial \phi({\bm 2}{\bm 2}')}G_{\sigma}({\bm 3}'{\bm 1}'),\label{E1}
\end{equation}
where we used the identity $G_{\sigma}({\bm 1}{\bm 3};\phi)G^{-1}_{\sigma}({\bm 3}{\bm 1}';\phi)=\delta_{{\bm 1}{\bm 1}'}$ and a summation over repeated indices is assumed. 

The propagator has the following form
\begin{equation}
G_{\sigma}({\bm 1}{\bm 1}';\phi) = -[\partial_{\tau}+H_0-\phi+\Sigma_{\sigma}(G(\phi))]^{-1}_{{\bm 1}{\bm 1}'},\label{E2}
\end{equation}
where $\Sigma$ is the self-energy and the inverse should be understood as a matrix inversion in space and time coordinates. The field couples to the electron density and therefore it appears only on diagonal elements of \eref{E2}. 
The inverse propagator depends on the field explicitly and implicitly through the dependence of the self-energy on the propagator. From \eref{E1} and \eref{E2}, one can see that the explicit field dependence contribution at the derivative is given by
\begin{equation}
-G_{\sigma}({\bm 1}{\bm 2})G_{\sigma}({\bm 2}'{\bm 1}'),\label{E3}
\end{equation}
where we used the identity $\partial G^{-1}_{\sigma}(33';\phi)/\partial \phi(22') = \delta_{3,2}\delta_{3',2'}$. 

In the DMFT approximation, the self-energy is fully local, i.e., $\Sigma^{DMFT}_{\sigma}({\bm 1}{\bm 1}') = \Sigma_{\sigma}({\bm 1}{\bm 1}') \delta_{{\bf r}{\bf r}'}$. In other words, the DMFT self-energy is only a functional of the local propagator $\Sigma^{DMFT}_{\sigma}=\Sigma^{DMFT}_{\sigma}(G_{loc}(\phi))$, where $G_{loc}$ denotes the local propagator $G_{loc,\sigma}(\tau,\tau') = (1/N)\sum_{{\bf r}}G_{\sigma}({\bf r}\tau,{\bf r}\tau')$.  Employing the fact that the DMFT self-energy is a  functional of  $G_{loc}$, the derivative of the self-energy with respect to the field can be written as
\begin{widetext}
\begin{align}
G_{\sigma}({\bm 1}{\bm 3})&\frac{\delta_{{\bf r}_3{\bf r}_3'}\partial \Sigma^{ DMFT}_{\sigma}({\bm 3}{\bm 3}';\phi)}{\partial \phi({\bm 2}{\bm 2}')}G_{\sigma}({\bm 3}'{\bm 1}') = 
G_{\sigma}({\bm 1}{\bm 3})\delta_{{\bf r}_3{\bf r}_3'}\left[\frac{\partial \Sigma^{ DMFT}_{\sigma}({\bf r}_3\tau_3,{\bf r}_3\tau_{3'})}{\partial G_{loc,\sigma'}(\tau_4,\tau_{4'})}\frac{\partial G_{loc,\sigma'}(\tau_4,\tau_{4'})}{\partial \phi({\bm 2}{\bm 2}')} \right]G_{\sigma}({\bm 3}'{\bm 1}')  \nonumber \\
&=  G_{\sigma}({\bm 1}{\bm 3})\delta_{{\bf r}_3{\bf r}_3'}G_{\sigma}({\bm 3}'{\bm 1}')\Gamma_{\sigma\sigma;\sigma'\sigma'} (\tau_{3}\tau_{3'};\tau_{4}\tau_{4'}) 
\frac{\partial G_{loc,\sigma'}(\tau_4,\tau_{4'})}{\partial \phi({\bm 2}{\bm 2}')} ,
\end{align} 
\end{widetext}
where 
\begin{align}
\frac{\partial \Sigma^{DMFT}_{\sigma}({\bf r}_3\tau_3,{\bf r}_3\tau_{3'};\phi)}{\partial G_{loc,\sigma'}(\tau_4\tau_{4'})}\equiv\Gamma^{irr}_{\sigma\sigma;\sigma'\sigma'} (\tau_{3}\tau_{3'};\tau_{4}\tau_{4'}).
\end{align} 
Diagrammatically, $\Gamma_{\sigma\sigma;\sigma'\sigma'}$ is obtained by removing one internal line from the self-energy in all possible ways.

 The complete equation for the susceptibility then takes the form
\begin{widetext}
\begin{align}
\frac{\partial G_{\sigma}({\bm 1}{\bm 1}';\phi)}{\partial \phi({\bm 2}{\bm 2}')} = - G_{\sigma}({\bm 1}{\bm 2})G_{\sigma}({\bm 2}'{\bm 1}') + 
G_{\sigma}({\bm 1}{\bm 3})\delta_{{\bf r}_3{\bf r}_3'}G_{\sigma}({\bm 3}'{\bm 1}') \Gamma^{ph, irr}_{\sigma\sigma;\sigma'\sigma'} (\tau_{3}\tau_{3'};\tau_{4}\tau_{4'}) 
\frac{\partial G_{loc,\sigma'}(\tau_4,\tau_{4'})}{\partial \phi({\bm 2}{\bm 2}')}.\label{DMFT-BSE}
\end{align}
\end{widetext}

Setting ${\bf r}_1={\bf r}'_1$ on the left-hand side, the derivative of $n$ and the above derivative become the same. Then, iterating the resulting equation gives the BSE used in this study. 

So far we were working with the lattice model. Now, we follow closely the analysis of Ref.~\cite{PhysRevB.92.085106} to obtain the impurity compressibility. 
In DMFT, the lattice model is mapped on an auxiliary impurity site embedded in a non-interacting bath. The bath parameters are determined self-consistently. The impurity Green's function is 
\begin{equation}
g_{\sigma}(11';\phi) = -[\partial_{\tau}-\phi+\Delta_{\sigma}(\phi)+\Sigma_{\sigma}(g(\phi))]^{-1}_{11'},\label{E14}
\end{equation}
where $\Delta$ denotes the hybridization function determined through the self-consistency relation 
\begin{equation}
g_{\sigma}-G_{loc,\sigma}=0,\label{E14-2}
\end{equation}
which is solved in the imaginary-time independent case, i.e. for every Matsubara frequency. We leave the two imaginary times $1$ and $1^\prime$ free in \eref{E14}, but since the derivative we are interested in is for $\phi$ equal to the chemical potential, which is independent of imaginary time, the derivation goes through if the self-consistency \eref{E14-2} depends only on imaginary-time difference. Note that the inverse should be understood as a matrix inversion in imaginary time coordinates (indices are not bold anymore). Furthermore, we have $g_{\sigma}(12)g^{-1}_{\sigma}(21')=\delta_{11'}$.

The impurity Green's function depends on $\phi$ explicitly and implicitly through the hybridization function. Then the variation of the impurity density with respect to the field is given by (with $1'=1^+$)
\begin{equation}
\frac{\partial g_{\sigma}(11')}{\partial \phi(22')}\big|_{\Delta}+\frac{\partial g_{\sigma}(11')}{\partial \Delta_{\sigma'}(33')}\big|_{\phi}\frac{\partial \Delta_{\sigma'}(33')}{\partial \phi(22')}.\label{E15}
\end{equation}
Using the definition of the impurity Green's function, one can easily show that
\begin{widetext}
\begin{align}
\frac{\partial g_{\sigma}(11')}{\partial \phi(22')}\big|_{\Delta}&=
[\chi^{0,ph}_{loc,\sigma\sigma}]_{11';22'}-[\chi^{0,ph}_{loc,\sigma\sigma}]_{11';33'}
\frac{\partial \Sigma_{\sigma}(33')}{\partial \phi(22')}\big|_{\Delta}
\nonumber\\&=
[\chi^{0,ph}_{loc,\sigma\sigma}]_{11';22'}-[\chi^{0,ph}_{loc,\sigma\sigma}]_{11';33'}
\Gamma^{ph, irr}_{\sigma\sigma;\sigma'\sigma'} (33';44')
\frac{\partial g_{\sigma'}(44';\phi)}{\partial \phi(22')}\big|_{\Delta}
,\label{E16}
\end{align}
\end{widetext}
where $[\chi^{0,ph}_{loc,\sigma\sigma'}]_{11';22'}\equiv-g_{\sigma}(12)g_{\sigma}(2'1')\delta_{\sigma\sigma'}$. Note that the relevant part of the above expression is the part with $2=2'$ since $\phi$ is diagonal. 

The dependence of $g$ on the hybridization function at constant field is
\begin{widetext}
\begin{align}
\frac{\partial g_{\sigma}(11')}{\partial \Delta_{\sigma'}(33')}\big|_{\phi}&=
-[\chi^{0,ph}_{loc,\sigma\sigma'}]_{11';33'}-[\chi^{0,ph}_{loc,\sigma\sigma}]_{11';44'}
\frac{\partial \Sigma_{\sigma}(44')}{\partial \Delta_{\sigma'}(33')}\big|_{\phi}
\nonumber \\&=
-[\chi^{0,ph}_{loc,\sigma\sigma'}]_{11';33'}-[\chi^{0,ph}_{loc,\sigma\sigma}]_{11';44'}\Gamma^{ph, irr}_{\sigma\sigma;\sigma''\sigma''} (44';55') \frac{\partial g_{\sigma''}(55';\phi)}{\partial \Delta_{\sigma'}(33')}\big|_{\phi}.\label{E17}
\end{align}
\end{widetext}
Note that all imaginary time $3, 3'$ must be considered. 

 One can iterate \eref{E16} and \eref{E17} to find the corresponding dressed susceptibilities. The susceptibility obtained from \eref{E16} describes the response of a non-self consistent impurity model to a change in the chemical potential. It is a physical response and therefore positive definite at a stable state. On the other hand, the susceptibility obtained from the \eref{E17} does not describe any physical response and therefore it is not necessarily positive definite. 
 
The self-energy  and the hybridization function are functionals of the Green's function. For strong interactions, these functionals change from a non-perturbative functional at low frequency to a perturbative functional at high frequency, i. e., $\Sigma\equiv \Sigma_{\rm per}[g]$ or $\Sigma\equiv \Sigma_{\rm non-per}[g]$.  This causes an ambiguity in defining the vertex functions in  \eref{E16} and \eref{E17}  when one frequency is on the perturbative branch of the self-energy (or hybridization function)  and the other frequency is on the non-perturbative branch. Nevertheless, our numerical verification of the compressibility sum-rule confirms that \eref{E16} and \eref{E17} remain valid, at least in the range of parameters considered here.
 
Further progress requires evaluating the derivative of the hybridization function with respect to the field. This can be found from the self-consistency condition, i. e., 
\begin{align}
[F_{\sigma}(\phi,&\Delta(\phi))]_{11'}\equiv g_{\sigma}(11';\phi)+\nonumber \\ 
&\frac{1}{N}\sum_{\bf k} [ H_0({\bf k})-g_{\sigma}^{-1}(\phi)-\Delta_{\sigma}(\phi)]^{-1}_{11'}=0,\label{E18}
\end{align}
where we define $F_{\sigma}(\phi,\Delta(\phi))$ for convenience.  
The variation of the above equation with respect to the field is 
\begin{equation}
\frac{\partial F_{\sigma}(11')}{\partial \phi(22')}\big|_{\Delta}+\frac{\partial F_{\sigma}(11')}{\partial \Delta_{\sigma'}(33')}\big|_{\phi}\frac{\partial \Delta_{\sigma'}(33')}{\partial \phi(22')}=0,\label{E19}
\end{equation}
which gives
\begin{equation}
\frac{\partial \Delta_{\sigma'}(33')}{\partial \phi(22')}=-\left[\frac{\partial F}{\partial \Delta}\big|_{\phi} \right]^{-1}_{\sigma'\sigma';\sigma''\sigma''}(33';11')\frac{\partial F_{\sigma''}(11')}{\partial \phi(22')}\big|_{\Delta},\label{E20}
\end{equation}
where  we assume $\partial F/\partial \Delta$ is invert-able. One the other word, we are assuming that $\partial g/\partial \Delta$ is finite (see the discussion after \eref{E17}).
At the vicinity of a Mott phase $F(\Delta)$ may go through an  extremum with zero derivative, breaking down the above assumption.

The derivative of $F_{\sigma}(\phi,\Delta(\phi))$ with respect to the field at constant hybridization function can be written as follow, using \eref{E16},  
\begin{widetext}
\begin{align}
\frac{\partial F_{\sigma}(11')}{\partial \phi(22')}\big|_{\Delta}=&
\frac{\partial g_{\sigma}(11')}{\partial \phi(22')}\big|_{\Delta}-\frac{1}{N}\sum_{\bf k}G_{{\bf k}\sigma}(13)g^{-1}_{\sigma}(34)
\frac{\partial g_{\sigma}(44')}{\partial \phi(22')}\big|_{\Delta}g^{-1}_{\sigma}(4'3')G_{{\bf k}\sigma}(3'1')\nonumber\\
=&
\left([\chi^{0,ph}_{loc,\sigma\sigma}]_{11';33'}-[\tilde{\chi}^{0,ph}_{\sigma\sigma}({\bf q}={\bm 0})]_{11';33'}\right)[\chi^{0,ph}_{loc,\sigma\sigma}]^{-1}_{33';44'}
\frac{\partial g_{\sigma}(44')}{\partial \phi(22')}\big|_{\Delta}\nonumber\\
=&\left([\chi^{0,ph}_{loc,\sigma\sigma}]_{11';33'}-[\tilde{\chi}^{0,ph}_{\sigma\sigma}({\bf q}={\bm 0})]_{11';33'}\right)
\left( \delta_{23}\delta_{2'3'}-\Gamma^{ph, irr}_{\sigma\sigma;\sigma'\sigma'} (33';44')
\frac{\partial g_{\sigma'}(44';\phi)}{\partial \phi(22')}\big|_{\Delta}\right)
,\label{E21}
\end{align}
where
\begin{equation}
[\tilde{\chi}^{0,ph}_{\sigma\sigma'}({\bf q}={\bm 0})]_{11';33'}=(-1/N)\sum_{\bf k}G_{{\bf k}\sigma}(13)G_{{\bf k}\sigma}(3'1')\delta_{\sigma\sigma'}.
\end{equation}
The derivative of the $F_{\sigma}(\phi,\Delta(\phi))$ with respect to the hybridization function at constant field is 
%
\begin{align}
\frac{\partial F_{\sigma'}(11')}{\partial \Delta_{\sigma}(33')}\big|_{\phi}&=\frac{\partial g_{\sigma'}(11')}{\partial \Delta_{\sigma}(33')}\big|_{\phi}-[\tilde{\chi}^{0,ph}_{\sigma'\sigma}({\bf q}={\bm 0})]_{11';33'}
-\frac{1}{N}\sum_{\bf k}G_{{\bf k}\sigma'}(14)g^{-1}_{\sigma'}(45)
\frac{\partial g_{\sigma'}(55')}{\partial \Delta_{\sigma}(33')}\big|_{\phi}g^{-1}_{\sigma'}(5'4')G_{{\bf k}\sigma'}(4'1')
\nonumber\\
&=\left( \delta_{\sigma' \sigma''}\delta_{15}\delta_{1'5'}+ [\tilde{\chi}^{0,ph}_{\sigma'\sigma'}({\bf q}={\bm 0})]_{11';44'} \Gamma^{ph,irr}_{\sigma'\sigma';\sigma'' \sigma''}(44';55')\right)\frac{\partial g_{\sigma''}(55')}{\partial \Delta_{\sigma}(33')}\big|_{\phi}
,\label{E22}
\end{align}
\end{widetext}
where we have used \eref{E17}. 

To show that the \eref{E15} for $\partial n /\partial\mu$ on the impurity is identical to the $\mathbf{q}=0$ susceptibility on the lattice, note that the second term in \eref{E15} can be rewritten using \eref{E20} as
\begin{equation}
-\frac{\partial g_{\sigma}(11')}{\partial \Delta_{\sigma'}(33')}\big|_{\phi}\left[\frac{\partial F}{\partial \Delta}\big|_{\phi} \right]^{-1}_{\sigma'\sigma';\sigma''\sigma''}(33';44')\frac{\partial F_{\sigma''}(44')}{\partial \phi(22')}\big|_{\Delta}.\label{E26}
\end{equation}
%
Using the \eref{E22}, the multiplication of the first two terms at the \eref{E27} is 
\begin{widetext}
\begin{align}
\frac{\partial g_{\sigma}(11')}{\partial \Delta_{\sigma'}(33')}\big|_{\phi}\left[\frac{\partial F}{\partial \Delta}\big|_{\phi} \right]^{-1}_{\sigma'\sigma';\sigma''\sigma''}(33';44') =\left({\bm 1}+ [\tilde{\chi}^{0,ph}({\bf q}={\bm 0})] \Gamma^{ph,irr}\right)_{\sigma\sigma;\sigma'' \sigma''}^{-1}(11';44'), \label{E27}
\end{align}
\end{widetext}
where ${\bm 1}_{11'\sigma;44'\sigma''}=\delta_{\sigma\sigma''}\delta_{1,4}\delta_{1'4'}$. 
Therefore, the \eref{E15} for $\partial n /\partial\mu$ on the impurity can be rewritten as follows, using \eref{E21} and \eref{E27} 
 \begin{widetext}
 \begin{align}
 &\frac{\partial g_{\sigma}(11')}{\partial \phi(22')}\big|_{\Delta}-
\left({\bm 1}+ [\tilde{\chi}^{0,ph}({\bf q}={\bm 0})] \Gamma^{ph,irr}\right)_{\sigma\sigma;\sigma'' \sigma''}^{-1}(11';44')
\left([\chi^{0,ph}_{loc,\sigma''\sigma''}]_{44';33'}-[\tilde{\chi}^{0,ph}_{\sigma''\sigma''}({\bf q}={\bm 0})]_{44';33'}\right)\nonumber \\ &\times
\left(\delta_{32}\delta_{3'2'}-\Gamma^{ph, irr}_{\sigma''\sigma'';\sigma'\sigma'} (33';66')
\frac{\partial g_{\sigma'}(66';\phi)}{\partial \phi(22')}\big|_{\Delta}\right)= 
\left( {\bm 1}+ [\tilde{\chi}^{0,ph}({\bf q}={\bm 0})] \Gamma^{ph,irr}
\right)_{\sigma\sigma;\sigma'' \sigma''}^{-1}(11';44')[\tilde{\chi}^{0,ph}_{\sigma''\sigma''}({\bf q}={\bm 0})]_{44';22'}.
 \end{align}
 \end{widetext}
Therefore, the above expression for $\partial n /\partial\mu$ on the self-consistent impurity equals to the lattice BSE only when the self-energy is a functional of Green's function described by the perturbative branch.


\begin{thebibliography}{102}%
\makeatletter
\providecommand \@ifxundefined [1]{%
 \@ifx{#1\undefined}
}%
\providecommand \@ifnum [1]{%
 \ifnum #1\expandafter \@firstoftwo
 \else \expandafter \@secondoftwo
 \fi
}%
\providecommand \@ifx [1]{%
 \ifx #1\expandafter \@firstoftwo
 \else \expandafter \@secondoftwo
 \fi
}%
\providecommand \natexlab [1]{#1}%
\providecommand \enquote  [1]{``#1''}%
\providecommand \bibnamefont  [1]{#1}%
\providecommand \bibfnamefont [1]{#1}%
\providecommand \citenamefont [1]{#1}%
\providecommand \href@noop [0]{\@secondoftwo}%
\providecommand \href [0]{\begingroup \@sanitize@url \@href}%
\providecommand \@href[1]{\@@startlink{#1}\@@href}%
\providecommand \@@href[1]{\endgroup#1\@@endlink}%
\providecommand \@sanitize@url [0]{\catcode `\\12\catcode `\$12\catcode
  `\&12\catcode `\#12\catcode `\^12\catcode `\_12\catcode `\%12\relax}%
\providecommand \@@startlink[1]{}%
\providecommand \@@endlink[0]{}%
\providecommand \url  [0]{\begingroup\@sanitize@url \@url }%
\providecommand \@url [1]{\endgroup\@href {#1}{\urlprefix }}%
\providecommand \urlprefix  [0]{URL }%
\providecommand \Eprint [0]{\href }%
\providecommand \doibase [0]{http://dx.doi.org/}%
\providecommand \selectlanguage [0]{\@gobble}%
\providecommand \bibinfo  [0]{\@secondoftwo}%
\providecommand \bibfield  [0]{\@secondoftwo}%
\providecommand \translation [1]{[#1]}%
\providecommand \BibitemOpen [0]{}%
\providecommand \bibitemStop [0]{}%
\providecommand \bibitemNoStop [0]{.\EOS\space}%
\providecommand \EOS [0]{\spacefactor3000\relax}%
\providecommand \BibitemShut  [1]{\csname bibitem#1\endcsname}%
\let\auto@bib@innerbib\@empty
\bibitem [{\citenamefont {Anderson}(1987)}]{Anderson:1987}%
  \BibitemOpen
  \bibfield  {author} {\bibinfo {author} {\bibfnamefont {P.~W.}\ \bibnamefont
  {Anderson}},\ }\href
  {http://www.sciencemag.org/content/235/4793/1196.abstract} {\bibfield
  {journal} {\bibinfo  {journal} {Science}\ }\textbf {\bibinfo {volume}
  {235}},\ \bibinfo {pages} {1196 } (\bibinfo {year} {1987})}\BibitemShut
  {NoStop}%
\bibitem [{\citenamefont {Machida}(1989)}]{MachidaCharge:1989}%
  \BibitemOpen
  \bibfield  {author} {\bibinfo {author} {\bibfnamefont {K.}~\bibnamefont
  {Machida}},\ }\href {\doibase http://dx.doi.org/10.1016/0921-4534(89)90316-X}
  {\bibfield  {journal} {\bibinfo  {journal} {Physica C: Superconductivity}\
  }\textbf {\bibinfo {volume} {158}},\ \bibinfo {pages} {192 } (\bibinfo {year}
  {1989})}\BibitemShut {NoStop}%
\bibitem [{\citenamefont {{Schulz, H.J.}}(1989)}]{SchulzCharge:1989}%
  \BibitemOpen
  \bibfield  {author} {\bibinfo {author} {\bibnamefont {{Schulz, H.J.}}},\
  }\href {\doibase 10.1051/jphys:0198900500180283300} {\bibfield  {journal}
  {\bibinfo  {journal} {J. Phys. France}\ }\textbf {\bibinfo {volume} {50}},\
  \bibinfo {pages} {2833} (\bibinfo {year} {1989})}\BibitemShut {NoStop}%
\bibitem [{\citenamefont {Zaanen}\ and\ \citenamefont
  {Gunnarsson}(1989)}]{ZaanenCharge:1989}%
  \BibitemOpen
  \bibfield  {author} {\bibinfo {author} {\bibfnamefont {J.}~\bibnamefont
  {Zaanen}}\ and\ \bibinfo {author} {\bibfnamefont {O.}~\bibnamefont
  {Gunnarsson}},\ }\href {\doibase 10.1103/PhysRevB.40.7391} {\bibfield
  {journal} {\bibinfo  {journal} {Phys. Rev. B}\ }\textbf {\bibinfo {volume}
  {40}},\ \bibinfo {pages} {7391} (\bibinfo {year} {1989})}\BibitemShut
  {NoStop}%
\bibitem [{\citenamefont {Poilblanc}\ and\ \citenamefont
  {Rice}(1989)}]{PoilblancCharge:1989}%
  \BibitemOpen
  \bibfield  {author} {\bibinfo {author} {\bibfnamefont {D.}~\bibnamefont
  {Poilblanc}}\ and\ \bibinfo {author} {\bibfnamefont {T.~M.}\ \bibnamefont
  {Rice}},\ }\href {\doibase 10.1103/PhysRevB.39.9749} {\bibfield  {journal}
  {\bibinfo  {journal} {Phys. Rev. B}\ }\textbf {\bibinfo {volume} {39}},\
  \bibinfo {pages} {9749} (\bibinfo {year} {1989})}\BibitemShut {NoStop}%
\bibitem [{\citenamefont {Emery}\ and\ \citenamefont
  {Kivelson}(1993)}]{EmeryKivelson:1993}%
  \BibitemOpen
  \bibfield  {author} {\bibinfo {author} {\bibfnamefont {V.}~\bibnamefont
  {Emery}}\ and\ \bibinfo {author} {\bibfnamefont {S.}~\bibnamefont
  {Kivelson}},\ }\href {\doibase
  http://dx.doi.org/10.1016/0921-4534(93)90581-A} {\bibfield  {journal}
  {\bibinfo  {journal} {Physica C: Superconductivity}\ }\textbf {\bibinfo
  {volume} {209}},\ \bibinfo {pages} {597 } (\bibinfo {year}
  {1993})}\BibitemShut {NoStop}%
\bibitem [{\citenamefont {L\"ow}\ \emph {et~al.}(1994)\citenamefont {L\"ow},
  \citenamefont {Emery}, \citenamefont {Fabricius},\ and\ \citenamefont
  {Kivelson}}]{LowEmeryChargeCoulomb:1994}%
  \BibitemOpen
  \bibfield  {author} {\bibinfo {author} {\bibfnamefont {U.}~\bibnamefont
  {L\"ow}}, \bibinfo {author} {\bibfnamefont {V.~J.}\ \bibnamefont {Emery}},
  \bibinfo {author} {\bibfnamefont {K.}~\bibnamefont {Fabricius}}, \ and\
  \bibinfo {author} {\bibfnamefont {S.~A.}\ \bibnamefont {Kivelson}},\ }\href
  {\doibase 10.1103/PhysRevLett.72.1918} {\bibfield  {journal} {\bibinfo
  {journal} {Phys. Rev. Lett.}\ }\textbf {\bibinfo {volume} {72}},\ \bibinfo
  {pages} {1918} (\bibinfo {year} {1994})}\BibitemShut {NoStop}%
\bibitem [{\citenamefont {Tranquada}\ \emph {et~al.}(1995)\citenamefont
  {Tranquada}, \citenamefont {Sternlieb}, \citenamefont {Axe}, \citenamefont
  {Nakamura},\ and\ \citenamefont {Uchida}}]{TranquadaStripe:1993}%
  \BibitemOpen
  \bibfield  {author} {\bibinfo {author} {\bibfnamefont {J.~M.}\ \bibnamefont
  {Tranquada}}, \bibinfo {author} {\bibfnamefont {B.~J.}\ \bibnamefont
  {Sternlieb}}, \bibinfo {author} {\bibfnamefont {J.~D.}\ \bibnamefont {Axe}},
  \bibinfo {author} {\bibfnamefont {Y.}~\bibnamefont {Nakamura}}, \ and\
  \bibinfo {author} {\bibfnamefont {S.}~\bibnamefont {Uchida}},\ }\href
  {http://ezproxy.usherbrooke.ca/login?url=https://search.proquest.com/docview/204471163?accountid=13835}
  {\bibfield  {journal} {\bibinfo  {journal} {Nature}\ }\textbf {\bibinfo
  {volume} {375}},\ \bibinfo {pages} {561} (\bibinfo {year} {1995})},\ \bibinfo
  {note} {copyright - Copyright Macmillan Journals Ltd. Jun 15, 1995; Last
  updated - 2012-11-14; CODEN - NATUAS}\BibitemShut {NoStop}%
\bibitem [{\citenamefont {Hellberg}\ and\ \citenamefont
  {Manousakis}(1997)}]{HellbergCharge:1997}%
  \BibitemOpen
  \bibfield  {author} {\bibinfo {author} {\bibfnamefont {C.~S.}\ \bibnamefont
  {Hellberg}}\ and\ \bibinfo {author} {\bibfnamefont {E.}~\bibnamefont
  {Manousakis}},\ }\href {\doibase 10.1103/PhysRevLett.78.4609} {\bibfield
  {journal} {\bibinfo  {journal} {Phys. Rev. Lett.}\ }\textbf {\bibinfo
  {volume} {78}},\ \bibinfo {pages} {4609} (\bibinfo {year}
  {1997})}\BibitemShut {NoStop}%
\bibitem [{\citenamefont {White}\ and\ \citenamefont
  {Scalapino}(1998)}]{WhiteCharge:1998}%
  \BibitemOpen
  \bibfield  {author} {\bibinfo {author} {\bibfnamefont {S.~R.}\ \bibnamefont
  {White}}\ and\ \bibinfo {author} {\bibfnamefont {D.~J.}\ \bibnamefont
  {Scalapino}},\ }\href {\doibase 10.1103/PhysRevLett.80.1272} {\bibfield
  {journal} {\bibinfo  {journal} {Phys. Rev. Lett.}\ }\textbf {\bibinfo
  {volume} {80}},\ \bibinfo {pages} {1272} (\bibinfo {year}
  {1998})}\BibitemShut {NoStop}%
\bibitem [{\citenamefont {Hellberg}\ and\ \citenamefont
  {Manousakis}(2000)}]{HellbergCharge:2000}%
  \BibitemOpen
  \bibfield  {author} {\bibinfo {author} {\bibfnamefont {C.~S.}\ \bibnamefont
  {Hellberg}}\ and\ \bibinfo {author} {\bibfnamefont {E.}~\bibnamefont
  {Manousakis}},\ }\href {\doibase 10.1103/PhysRevB.61.11787} {\bibfield
  {journal} {\bibinfo  {journal} {Phys. Rev. B}\ }\textbf {\bibinfo {volume}
  {61}},\ \bibinfo {pages} {11787} (\bibinfo {year} {2000})}\BibitemShut
  {NoStop}%
\bibitem [{\citenamefont {White}\ and\ \citenamefont
  {Scalapino}(2000)}]{WhiteScalapinoCharge:2000}%
  \BibitemOpen
  \bibfield  {author} {\bibinfo {author} {\bibfnamefont {S.~R.}\ \bibnamefont
  {White}}\ and\ \bibinfo {author} {\bibfnamefont {D.~J.}\ \bibnamefont
  {Scalapino}},\ }\href {\doibase 10.1103/PhysRevB.61.6320} {\bibfield
  {journal} {\bibinfo  {journal} {Phys. Rev. B}\ }\textbf {\bibinfo {volume}
  {61}},\ \bibinfo {pages} {6320} (\bibinfo {year} {2000})}\BibitemShut
  {NoStop}%
\bibitem [{\citenamefont {White}\ and\ \citenamefont
  {Scalapino}(2004)}]{PhysRevB.70.220506}%
  \BibitemOpen
  \bibfield  {author} {\bibinfo {author} {\bibfnamefont {S.~R.}\ \bibnamefont
  {White}}\ and\ \bibinfo {author} {\bibfnamefont {D.~J.}\ \bibnamefont
  {Scalapino}},\ }\href {\doibase 10.1103/PhysRevB.70.220506} {\bibfield
  {journal} {\bibinfo  {journal} {Phys. Rev. B}\ }\textbf {\bibinfo {volume}
  {70}},\ \bibinfo {pages} {220506} (\bibinfo {year} {2004})}\BibitemShut
  {NoStop}%
\bibitem [{\citenamefont {Seibold}\ \emph {et~al.}(1998)\citenamefont
  {Seibold}, \citenamefont {Sigmund},\ and\ \citenamefont
  {Hizhnyakov}}]{SeiboldCharge:1998}%
  \BibitemOpen
  \bibfield  {author} {\bibinfo {author} {\bibfnamefont {G.}~\bibnamefont
  {Seibold}}, \bibinfo {author} {\bibfnamefont {E.}~\bibnamefont {Sigmund}}, \
  and\ \bibinfo {author} {\bibfnamefont {V.}~\bibnamefont {Hizhnyakov}},\
  }\href {\doibase 10.1103/PhysRevB.57.6937} {\bibfield  {journal} {\bibinfo
  {journal} {Phys. Rev. B}\ }\textbf {\bibinfo {volume} {57}},\ \bibinfo
  {pages} {6937} (\bibinfo {year} {1998})}\BibitemShut {NoStop}%
\bibitem [{\citenamefont {Lorenzana}\ and\ \citenamefont
  {Seibold}(2002)}]{LorenzanaCharge:2002}%
  \BibitemOpen
  \bibfield  {author} {\bibinfo {author} {\bibfnamefont {J.}~\bibnamefont
  {Lorenzana}}\ and\ \bibinfo {author} {\bibfnamefont {G.}~\bibnamefont
  {Seibold}},\ }\href {\doibase 10.1103/PhysRevLett.89.136401} {\bibfield
  {journal} {\bibinfo  {journal} {Phys. Rev. Lett.}\ }\textbf {\bibinfo
  {volume} {89}},\ \bibinfo {pages} {136401} (\bibinfo {year}
  {2002})}\BibitemShut {NoStop}%
\bibitem [{\citenamefont {Han}\ \emph {et~al.}(2001)\citenamefont {Han},
  \citenamefont {Wang},\ and\ \citenamefont {Lee}}]{HanCharge:2001}%
  \BibitemOpen
  \bibfield  {author} {\bibinfo {author} {\bibfnamefont {J.}~\bibnamefont
  {Han}}, \bibinfo {author} {\bibfnamefont {Q.-H.}\ \bibnamefont {Wang}}, \
  and\ \bibinfo {author} {\bibfnamefont {D.-H.}\ \bibnamefont {Lee}},\ }\href
  {\doibase 10.1142/S021797920100468X} {\bibfield  {journal} {\bibinfo
  {journal} {International Journal of Modern Physics B}\ }\textbf {\bibinfo
  {volume} {15}},\ \bibinfo {pages} {1117} (\bibinfo {year}
  {2001})}\BibitemShut {NoStop}%
\bibitem [{\citenamefont {Ivanov}(2004)}]{IvanovCharge:2004}%
  \BibitemOpen
  \bibfield  {author} {\bibinfo {author} {\bibfnamefont {D.~A.}\ \bibnamefont
  {Ivanov}},\ }\href {\doibase 10.1103/PhysRevB.70.104503} {\bibfield
  {journal} {\bibinfo  {journal} {Phys. Rev. B}\ }\textbf {\bibinfo {volume}
  {70}},\ \bibinfo {pages} {104503} (\bibinfo {year} {2004})}\BibitemShut
  {NoStop}%
\bibitem [{\citenamefont {Corboz}\ \emph {et~al.}(2014)\citenamefont {Corboz},
  \citenamefont {Rice},\ and\ \citenamefont {Troyer}}]{PhysRevLett.113.046402}%
  \BibitemOpen
  \bibfield  {author} {\bibinfo {author} {\bibfnamefont {P.}~\bibnamefont
  {Corboz}}, \bibinfo {author} {\bibfnamefont {T.~M.}\ \bibnamefont {Rice}}, \
  and\ \bibinfo {author} {\bibfnamefont {M.}~\bibnamefont {Troyer}},\ }\href
  {\doibase 10.1103/PhysRevLett.113.046402} {\bibfield  {journal} {\bibinfo
  {journal} {Phys. Rev. Lett.}\ }\textbf {\bibinfo {volume} {113}},\ \bibinfo
  {pages} {046402} (\bibinfo {year} {2014})}\BibitemShut {NoStop}%
\bibitem [{\citenamefont {Kivelson}\ \emph {et~al.}(2003)\citenamefont
  {Kivelson}, \citenamefont {Bindloss}, \citenamefont {Fradkin}, \citenamefont
  {Oganesyan}, \citenamefont {Tranquada}, \citenamefont {Kapitulnik},\ and\
  \citenamefont {Howald}}]{KivelsonHowToRMP:2003}%
  \BibitemOpen
  \bibfield  {author} {\bibinfo {author} {\bibfnamefont {S.~A.}\ \bibnamefont
  {Kivelson}}, \bibinfo {author} {\bibfnamefont {I.~P.}\ \bibnamefont
  {Bindloss}}, \bibinfo {author} {\bibfnamefont {E.}~\bibnamefont {Fradkin}},
  \bibinfo {author} {\bibfnamefont {V.}~\bibnamefont {Oganesyan}}, \bibinfo
  {author} {\bibfnamefont {J.~M.}\ \bibnamefont {Tranquada}}, \bibinfo {author}
  {\bibfnamefont {A.}~\bibnamefont {Kapitulnik}}, \ and\ \bibinfo {author}
  {\bibfnamefont {C.}~\bibnamefont {Howald}},\ }\href {\doibase
  10.1103/RevModPhys.75.1201} {\bibfield  {journal} {\bibinfo  {journal} {Rev.
  Mod. Phys.}\ }\textbf {\bibinfo {volume} {75}},\ \bibinfo {pages} {1201}
  (\bibinfo {year} {2003})}\BibitemShut {NoStop}%
\bibitem [{\citenamefont {Fradkin}\ \emph {et~al.}(2015)\citenamefont
  {Fradkin}, \citenamefont {Kivelson},\ and\ \citenamefont
  {Tranquada}}]{KivelsonIntertwined:2015}%
  \BibitemOpen
  \bibfield  {author} {\bibinfo {author} {\bibfnamefont {E.}~\bibnamefont
  {Fradkin}}, \bibinfo {author} {\bibfnamefont {S.~A.}\ \bibnamefont
  {Kivelson}}, \ and\ \bibinfo {author} {\bibfnamefont {J.~M.}\ \bibnamefont
  {Tranquada}},\ }\href {\doibase 10.1103/RevModPhys.87.457} {\bibfield
  {journal} {\bibinfo  {journal} {Rev. Mod. Phys.}\ }\textbf {\bibinfo {volume}
  {87}},\ \bibinfo {pages} {457} (\bibinfo {year} {2015})}\BibitemShut
  {NoStop}%
\bibitem [{\citenamefont {Mesaros}\ \emph {et~al.}(2016)\citenamefont
  {Mesaros}, \citenamefont {Fujita}, \citenamefont {Edkins}, \citenamefont
  {Hamidian}, \citenamefont {Eisaki}, \citenamefont {Uchida}, \citenamefont
  {Davis}, \citenamefont {Lawler},\ and\ \citenamefont
  {Kim}}]{Mesaros08112016}%
  \BibitemOpen
  \bibfield  {author} {\bibinfo {author} {\bibfnamefont {A.}~\bibnamefont
  {Mesaros}}, \bibinfo {author} {\bibfnamefont {K.}~\bibnamefont {Fujita}},
  \bibinfo {author} {\bibfnamefont {S.~D.}\ \bibnamefont {Edkins}}, \bibinfo
  {author} {\bibfnamefont {M.~H.}\ \bibnamefont {Hamidian}}, \bibinfo {author}
  {\bibfnamefont {H.}~\bibnamefont {Eisaki}}, \bibinfo {author} {\bibfnamefont
  {S.-i.}\ \bibnamefont {Uchida}}, \bibinfo {author} {\bibfnamefont {J.~C.~S.}\
  \bibnamefont {Davis}}, \bibinfo {author} {\bibfnamefont {M.~J.}\ \bibnamefont
  {Lawler}}, \ and\ \bibinfo {author} {\bibfnamefont {E.-A.}\ \bibnamefont
  {Kim}},\ }\href {\doibase 10.1073/pnas.1614247113} {\bibfield  {journal}
  {\bibinfo  {journal} {Proceedings of the National Academy of Sciences}\
  }\textbf {\bibinfo {volume} {113}},\ \bibinfo {pages} {12661} (\bibinfo
  {year} {2016})},\ \Eprint
  {http://arxiv.org/abs/http://www.pnas.org/content/113/45/12661.full.pdf}
  {http://www.pnas.org/content/113/45/12661.full.pdf} \BibitemShut {NoStop}%
\bibitem [{\citenamefont {Lalibert\'e}\ \emph {et~al.}(2011)\citenamefont
  {Lalibert\'e}, \citenamefont {Chang}, \citenamefont {Doiron-Leyraud},
  \citenamefont {Hassinger}, \citenamefont {Daou}, \citenamefont {Rondeau},
  \citenamefont {Ramshaw}, \citenamefont {Liang}, \citenamefont {Bonn},
  \citenamefont {Hardy}, \citenamefont {Pyon}, \citenamefont {Takayama},
  \citenamefont {Takagi}, \citenamefont {Sheikin}, \citenamefont {Malone},
  \citenamefont {Proust}, \citenamefont {Behnia},\ and\ \citenamefont
  {Taillefer}}]{10.1038/ncomms1440}%
  \BibitemOpen
  \bibfield  {author} {\bibinfo {author} {\bibfnamefont {F.}~\bibnamefont
  {Lalibert\'e}}, \bibinfo {author} {\bibfnamefont {J.}~\bibnamefont {Chang}},
  \bibinfo {author} {\bibfnamefont {N.}~\bibnamefont {Doiron-Leyraud}},
  \bibinfo {author} {\bibfnamefont {E.}~\bibnamefont {Hassinger}}, \bibinfo
  {author} {\bibfnamefont {R.}~\bibnamefont {Daou}}, \bibinfo {author}
  {\bibfnamefont {M.}~\bibnamefont {Rondeau}}, \bibinfo {author} {\bibfnamefont
  {B.}~\bibnamefont {Ramshaw}}, \bibinfo {author} {\bibfnamefont
  {R.}~\bibnamefont {Liang}}, \bibinfo {author} {\bibfnamefont
  {D.}~\bibnamefont {Bonn}}, \bibinfo {author} {\bibfnamefont {W.}~\bibnamefont
  {Hardy}}, \bibinfo {author} {\bibfnamefont {S.}~\bibnamefont {Pyon}},
  \bibinfo {author} {\bibfnamefont {T.}~\bibnamefont {Takayama}}, \bibinfo
  {author} {\bibfnamefont {H.}~\bibnamefont {Takagi}}, \bibinfo {author}
  {\bibfnamefont {I.}~\bibnamefont {Sheikin}}, \bibinfo {author} {\bibfnamefont
  {L.}~\bibnamefont {Malone}}, \bibinfo {author} {\bibfnamefont
  {C.}~\bibnamefont {Proust}}, \bibinfo {author} {\bibfnamefont
  {K.}~\bibnamefont {Behnia}}, \ and\ \bibinfo {author} {\bibfnamefont
  {L.}~\bibnamefont {Taillefer}},\ }\href {\doibase 10.1038/nphys3840}
  {\bibfield  {journal} {\bibinfo  {journal} {Nat. Comms.}\ }\textbf {\bibinfo
  {volume} {2}},\ \bibinfo {pages} {432} (\bibinfo {year} {2011})}\BibitemShut
  {NoStop}%
\bibitem [{\citenamefont {Hoffman}\ \emph {et~al.}(2002)\citenamefont
  {Hoffman}, \citenamefont {Hudson}, \citenamefont {Lang}, \citenamefont
  {Madhavan}, \citenamefont {Eisaki}, \citenamefont {Uchida},\ and\
  \citenamefont {Davis}}]{Hoffman466}%
  \BibitemOpen
  \bibfield  {author} {\bibinfo {author} {\bibfnamefont {J.~E.}\ \bibnamefont
  {Hoffman}}, \bibinfo {author} {\bibfnamefont {E.~W.}\ \bibnamefont {Hudson}},
  \bibinfo {author} {\bibfnamefont {K.~M.}\ \bibnamefont {Lang}}, \bibinfo
  {author} {\bibfnamefont {V.}~\bibnamefont {Madhavan}}, \bibinfo {author}
  {\bibfnamefont {H.}~\bibnamefont {Eisaki}}, \bibinfo {author} {\bibfnamefont
  {S.}~\bibnamefont {Uchida}}, \ and\ \bibinfo {author} {\bibfnamefont {J.~C.}\
  \bibnamefont {Davis}},\ }\href {\doibase 10.1126/science.1066974} {\bibfield
  {journal} {\bibinfo  {journal} {Science}\ }\textbf {\bibinfo {volume}
  {295}},\ \bibinfo {pages} {466} (\bibinfo {year} {2002})},\ \Eprint
  {http://arxiv.org/abs/http://science.sciencemag.org/content/295/5554/466.full.pdf}
  {http://science.sciencemag.org/content/295/5554/466.full.pdf} \BibitemShut
  {NoStop}%
\bibitem [{\citenamefont {Fujita}\ \emph {et~al.}(2014)\citenamefont {Fujita},
  \citenamefont {Kim}, \citenamefont {Lee}, \citenamefont {Lee}, \citenamefont
  {Hamidian}, \citenamefont {Firmo}, \citenamefont {Mukhopadhyay},
  \citenamefont {Eisaki}, \citenamefont {Uchida}, \citenamefont {Lawler},
  \citenamefont {Kim},\ and\ \citenamefont {Davis}}]{Fujita612}%
  \BibitemOpen
  \bibfield  {author} {\bibinfo {author} {\bibfnamefont {K.}~\bibnamefont
  {Fujita}}, \bibinfo {author} {\bibfnamefont {C.~K.}\ \bibnamefont {Kim}},
  \bibinfo {author} {\bibfnamefont {I.}~\bibnamefont {Lee}}, \bibinfo {author}
  {\bibfnamefont {J.}~\bibnamefont {Lee}}, \bibinfo {author} {\bibfnamefont
  {M.~H.}\ \bibnamefont {Hamidian}}, \bibinfo {author} {\bibfnamefont {I.~A.}\
  \bibnamefont {Firmo}}, \bibinfo {author} {\bibfnamefont {S.}~\bibnamefont
  {Mukhopadhyay}}, \bibinfo {author} {\bibfnamefont {H.}~\bibnamefont
  {Eisaki}}, \bibinfo {author} {\bibfnamefont {S.}~\bibnamefont {Uchida}},
  \bibinfo {author} {\bibfnamefont {M.~J.}\ \bibnamefont {Lawler}}, \bibinfo
  {author} {\bibfnamefont {E.-A.}\ \bibnamefont {Kim}}, \ and\ \bibinfo
  {author} {\bibfnamefont {J.~C.}\ \bibnamefont {Davis}},\ }\href {\doibase
  10.1126/science.1248783} {\bibfield  {journal} {\bibinfo  {journal}
  {Science}\ }\textbf {\bibinfo {volume} {344}},\ \bibinfo {pages} {612}
  (\bibinfo {year} {2014})},\ \Eprint
  {http://arxiv.org/abs/http://science.sciencemag.org/content/344/6184/612.full.pdf}
  {http://science.sciencemag.org/content/344/6184/612.full.pdf} \BibitemShut
  {NoStop}%
\bibitem [{\citenamefont {Vershinin}\ \emph {et~al.}(2004)\citenamefont
  {Vershinin}, \citenamefont {Misra}, \citenamefont {Ono}, \citenamefont {Abe},
  \citenamefont {Ando},\ and\ \citenamefont {Yazdani}}]{Vershinin1995}%
  \BibitemOpen
  \bibfield  {author} {\bibinfo {author} {\bibfnamefont {M.}~\bibnamefont
  {Vershinin}}, \bibinfo {author} {\bibfnamefont {S.}~\bibnamefont {Misra}},
  \bibinfo {author} {\bibfnamefont {S.}~\bibnamefont {Ono}}, \bibinfo {author}
  {\bibfnamefont {Y.}~\bibnamefont {Abe}}, \bibinfo {author} {\bibfnamefont
  {Y.}~\bibnamefont {Ando}}, \ and\ \bibinfo {author} {\bibfnamefont
  {A.}~\bibnamefont {Yazdani}},\ }\href {\doibase 10.1126/science.1093384}
  {\bibfield  {journal} {\bibinfo  {journal} {Science}\ }\textbf {\bibinfo
  {volume} {303}},\ \bibinfo {pages} {1995} (\bibinfo {year} {2004})},\ \Eprint
  {http://arxiv.org/abs/http://science.sciencemag.org/content/303/5666/1995.full.pdf}
  {http://science.sciencemag.org/content/303/5666/1995.full.pdf} \BibitemShut
  {NoStop}%
\bibitem [{\citenamefont {He}\ \emph {et~al.}(2014)\citenamefont {He},
  \citenamefont {Yin}, \citenamefont {Zech}, \citenamefont {Soumyanarayanan},
  \citenamefont {Yee}, \citenamefont {Williams}, \citenamefont {Boyer},
  \citenamefont {Chatterjee}, \citenamefont {Wise}, \citenamefont {Zeljkovic},
  \citenamefont {Kondo}, \citenamefont {Takeuchi}, \citenamefont {Ikuta},
  \citenamefont {Mistark}, \citenamefont {Markiewicz}, \citenamefont {Bansil},
  \citenamefont {Sachdev}, \citenamefont {Hudson},\ and\ \citenamefont
  {Hoffman}}]{He608}%
  \BibitemOpen
  \bibfield  {author} {\bibinfo {author} {\bibfnamefont {Y.}~\bibnamefont
  {He}}, \bibinfo {author} {\bibfnamefont {Y.}~\bibnamefont {Yin}}, \bibinfo
  {author} {\bibfnamefont {M.}~\bibnamefont {Zech}}, \bibinfo {author}
  {\bibfnamefont {A.}~\bibnamefont {Soumyanarayanan}}, \bibinfo {author}
  {\bibfnamefont {M.~M.}\ \bibnamefont {Yee}}, \bibinfo {author} {\bibfnamefont
  {T.}~\bibnamefont {Williams}}, \bibinfo {author} {\bibfnamefont {M.~C.}\
  \bibnamefont {Boyer}}, \bibinfo {author} {\bibfnamefont {K.}~\bibnamefont
  {Chatterjee}}, \bibinfo {author} {\bibfnamefont {W.~D.}\ \bibnamefont
  {Wise}}, \bibinfo {author} {\bibfnamefont {I.}~\bibnamefont {Zeljkovic}},
  \bibinfo {author} {\bibfnamefont {T.}~\bibnamefont {Kondo}}, \bibinfo
  {author} {\bibfnamefont {T.}~\bibnamefont {Takeuchi}}, \bibinfo {author}
  {\bibfnamefont {H.}~\bibnamefont {Ikuta}}, \bibinfo {author} {\bibfnamefont
  {P.}~\bibnamefont {Mistark}}, \bibinfo {author} {\bibfnamefont {R.~S.}\
  \bibnamefont {Markiewicz}}, \bibinfo {author} {\bibfnamefont
  {A.}~\bibnamefont {Bansil}}, \bibinfo {author} {\bibfnamefont
  {S.}~\bibnamefont {Sachdev}}, \bibinfo {author} {\bibfnamefont {E.~W.}\
  \bibnamefont {Hudson}}, \ and\ \bibinfo {author} {\bibfnamefont {J.~E.}\
  \bibnamefont {Hoffman}},\ }\href {\doibase 10.1126/science.1248221}
  {\bibfield  {journal} {\bibinfo  {journal} {Science}\ }\textbf {\bibinfo
  {volume} {344}},\ \bibinfo {pages} {608} (\bibinfo {year} {2014})},\ \Eprint
  {http://arxiv.org/abs/http://science.sciencemag.org/content/344/6184/608.full.pdf}
  {http://science.sciencemag.org/content/344/6184/608.full.pdf} \BibitemShut
  {NoStop}%
\bibitem [{\citenamefont {da~Silva~Neto}\ \emph {et~al.}(2014)\citenamefont
  {da~Silva~Neto}, \citenamefont {Aynajian}, \citenamefont {Frano},
  \citenamefont {Comin}, \citenamefont {Schierle}, \citenamefont {Weschke},
  \citenamefont {Gyenis}, \citenamefont {Wen}, \citenamefont {Schneeloch},
  \citenamefont {Xu}, \citenamefont {Ono}, \citenamefont {Gu}, \citenamefont
  {Le~Tacon},\ and\ \citenamefont {Yazdani}}]{daSilvaNeto393}%
  \BibitemOpen
  \bibfield  {author} {\bibinfo {author} {\bibfnamefont {E.~H.}\ \bibnamefont
  {da~Silva~Neto}}, \bibinfo {author} {\bibfnamefont {P.}~\bibnamefont
  {Aynajian}}, \bibinfo {author} {\bibfnamefont {A.}~\bibnamefont {Frano}},
  \bibinfo {author} {\bibfnamefont {R.}~\bibnamefont {Comin}}, \bibinfo
  {author} {\bibfnamefont {E.}~\bibnamefont {Schierle}}, \bibinfo {author}
  {\bibfnamefont {E.}~\bibnamefont {Weschke}}, \bibinfo {author} {\bibfnamefont
  {A.}~\bibnamefont {Gyenis}}, \bibinfo {author} {\bibfnamefont
  {J.}~\bibnamefont {Wen}}, \bibinfo {author} {\bibfnamefont {J.}~\bibnamefont
  {Schneeloch}}, \bibinfo {author} {\bibfnamefont {Z.}~\bibnamefont {Xu}},
  \bibinfo {author} {\bibfnamefont {S.}~\bibnamefont {Ono}}, \bibinfo {author}
  {\bibfnamefont {G.}~\bibnamefont {Gu}}, \bibinfo {author} {\bibfnamefont
  {M.}~\bibnamefont {Le~Tacon}}, \ and\ \bibinfo {author} {\bibfnamefont
  {A.}~\bibnamefont {Yazdani}},\ }\href {\doibase 10.1126/science.1243479}
  {\bibfield  {journal} {\bibinfo  {journal} {Science}\ }\textbf {\bibinfo
  {volume} {343}},\ \bibinfo {pages} {393} (\bibinfo {year} {2014})},\ \Eprint
  {http://arxiv.org/abs/http://science.sciencemag.org/content/343/6169/393.full.pdf}
  {http://science.sciencemag.org/content/343/6169/393.full.pdf} \BibitemShut
  {NoStop}%
\bibitem [{\citenamefont {Wise}\ \emph {et~al.}(2008)\citenamefont {Wise},
  \citenamefont {Boyer}, \citenamefont {Chatterjee}, \citenamefont {Kondo},
  \citenamefont {Takeuchi}, \citenamefont {Ikuta}, \citenamefont {Wang},\ and\
  \citenamefont {Hudson}}]{10.1038/nphys1021}%
  \BibitemOpen
  \bibfield  {author} {\bibinfo {author} {\bibfnamefont {W.~D.}\ \bibnamefont
  {Wise}}, \bibinfo {author} {\bibfnamefont {M.~C.}\ \bibnamefont {Boyer}},
  \bibinfo {author} {\bibfnamefont {K.}~\bibnamefont {Chatterjee}}, \bibinfo
  {author} {\bibfnamefont {T.}~\bibnamefont {Kondo}}, \bibinfo {author}
  {\bibfnamefont {T.}~\bibnamefont {Takeuchi}}, \bibinfo {author}
  {\bibfnamefont {H.}~\bibnamefont {Ikuta}}, \bibinfo {author} {\bibfnamefont
  {Y.}~\bibnamefont {Wang}}, \ and\ \bibinfo {author} {\bibfnamefont {E.~W.}\
  \bibnamefont {Hudson}},\ }\href {\doibase 10.1038/nphys1021} {\bibfield
  {journal} {\bibinfo  {journal} {Nature Physics}\ }\textbf {\bibinfo {volume}
  {4}},\ \bibinfo {pages} {565} (\bibinfo {year} {2008})}\BibitemShut {NoStop}%
\bibitem [{\citenamefont {Doiron-Leyraud}\ \emph {et~al.}(2007)\citenamefont
  {Doiron-Leyraud}, \citenamefont {Proust}, \citenamefont {LeBoeuf},
  \citenamefont {Levallois}, \citenamefont {Bonnemaison}, \citenamefont
  {Liang}, \citenamefont {Bonn}, \citenamefont {Hardy},\ and\ \citenamefont
  {Taillefer}}]{10.1038/nature05872}%
  \BibitemOpen
  \bibfield  {author} {\bibinfo {author} {\bibfnamefont {N.}~\bibnamefont
  {Doiron-Leyraud}}, \bibinfo {author} {\bibfnamefont {C.}~\bibnamefont
  {Proust}}, \bibinfo {author} {\bibfnamefont {D.}~\bibnamefont {LeBoeuf}},
  \bibinfo {author} {\bibfnamefont {J.}~\bibnamefont {Levallois}}, \bibinfo
  {author} {\bibfnamefont {J.-B.}\ \bibnamefont {Bonnemaison}}, \bibinfo
  {author} {\bibfnamefont {R.}~\bibnamefont {Liang}}, \bibinfo {author}
  {\bibfnamefont {D.~A.}\ \bibnamefont {Bonn}}, \bibinfo {author}
  {\bibfnamefont {W.~N.}\ \bibnamefont {Hardy}}, \ and\ \bibinfo {author}
  {\bibfnamefont {L.}~\bibnamefont {Taillefer}},\ }\href {\doibase
  10.1038/nature05872} {\bibfield  {journal} {\bibinfo  {journal} {Nature}\
  }\textbf {\bibinfo {volume} {447}},\ \bibinfo {pages} {565} (\bibinfo {year}
  {2007})}\BibitemShut {NoStop}%
\bibitem [{\citenamefont {LeBoeuf}\ \emph {et~al.}(2007)\citenamefont
  {LeBoeuf}, \citenamefont {Doiron-Leyraud}, \citenamefont {Levallois},
  \citenamefont {Daou}, \citenamefont {Bonnemaison}, \citenamefont {Hussey},
  \citenamefont {Balicas}, \citenamefont {Ramshaw}, \citenamefont {Liang},
  \citenamefont {Bonn}, \citenamefont {Hardy}, \citenamefont {Adachi},
  \citenamefont {Proust},\ and\ \citenamefont
  {Taillefer}}]{10.1038/nature06332}%
  \BibitemOpen
  \bibfield  {author} {\bibinfo {author} {\bibfnamefont {D.}~\bibnamefont
  {LeBoeuf}}, \bibinfo {author} {\bibfnamefont {N.}~\bibnamefont
  {Doiron-Leyraud}}, \bibinfo {author} {\bibfnamefont {J.}~\bibnamefont
  {Levallois}}, \bibinfo {author} {\bibfnamefont {R.}~\bibnamefont {Daou}},
  \bibinfo {author} {\bibfnamefont {J.-B.}\ \bibnamefont {Bonnemaison}},
  \bibinfo {author} {\bibfnamefont {N.~E.}\ \bibnamefont {Hussey}}, \bibinfo
  {author} {\bibfnamefont {L.}~\bibnamefont {Balicas}}, \bibinfo {author}
  {\bibfnamefont {B.~J.}\ \bibnamefont {Ramshaw}}, \bibinfo {author}
  {\bibfnamefont {R.}~\bibnamefont {Liang}}, \bibinfo {author} {\bibfnamefont
  {D.~A.}\ \bibnamefont {Bonn}}, \bibinfo {author} {\bibfnamefont {W.~N.}\
  \bibnamefont {Hardy}}, \bibinfo {author} {\bibfnamefont {S.}~\bibnamefont
  {Adachi}}, \bibinfo {author} {\bibfnamefont {C.}~\bibnamefont {Proust}}, \
  and\ \bibinfo {author} {\bibfnamefont {L.}~\bibnamefont {Taillefer}},\ }\href
  {\doibase 10.1038/nature06332} {\bibfield  {journal} {\bibinfo  {journal}
  {Nature}\ }\textbf {\bibinfo {volume} {450}},\ \bibinfo {pages} {533}
  (\bibinfo {year} {2007})}\BibitemShut {NoStop}%
\bibitem [{\citenamefont {Wu}\ \emph {et~al.}(2011)\citenamefont {Wu},
  \citenamefont {Mayaffre}, \citenamefont {Kramer}, \citenamefont {Horvatic},
  \citenamefont {Berthier}, \citenamefont {Hardy}, \citenamefont {Liang},
  \citenamefont {Bonn},\ and\ \citenamefont {Julien}}]{10.1038/nature10345}%
  \BibitemOpen
  \bibfield  {author} {\bibinfo {author} {\bibfnamefont {T.}~\bibnamefont
  {Wu}}, \bibinfo {author} {\bibfnamefont {H.}~\bibnamefont {Mayaffre}},
  \bibinfo {author} {\bibfnamefont {S.}~\bibnamefont {Kramer}}, \bibinfo
  {author} {\bibfnamefont {M.}~\bibnamefont {Horvatic}}, \bibinfo {author}
  {\bibfnamefont {C.}~\bibnamefont {Berthier}}, \bibinfo {author}
  {\bibfnamefont {W.~N.}\ \bibnamefont {Hardy}}, \bibinfo {author}
  {\bibfnamefont {R.}~\bibnamefont {Liang}}, \bibinfo {author} {\bibfnamefont
  {D.~A.}\ \bibnamefont {Bonn}}, \ and\ \bibinfo {author} {\bibfnamefont
  {M.-H.}\ \bibnamefont {Julien}},\ }\href {\doibase 10.1038/nphys3840}
  {\bibfield  {journal} {\bibinfo  {journal} {Nat.}\ }\textbf {\bibinfo
  {volume} {477}},\ \bibinfo {pages} {191} (\bibinfo {year}
  {2011})}\BibitemShut {NoStop}%
\bibitem [{\citenamefont {Wu}\ \emph {et~al.}(2013)\citenamefont {Wu},
  \citenamefont {Mayaffre}, \citenamefont {Kr\"{a}mer}, \citenamefont
  {Horvati\'{c}}, \citenamefont {Berthier}, \citenamefont {Kuhns},
  \citenamefont {Reyes}, \citenamefont {Liang}, \citenamefont {Hardy},
  \citenamefont {Bonn},\ and\ \citenamefont {Julien}}]{10.1038/ncomms3113}%
  \BibitemOpen
  \bibfield  {author} {\bibinfo {author} {\bibfnamefont {T.}~\bibnamefont
  {Wu}}, \bibinfo {author} {\bibfnamefont {H.}~\bibnamefont {Mayaffre}},
  \bibinfo {author} {\bibfnamefont {S.}~\bibnamefont {Kr\"{a}mer}}, \bibinfo
  {author} {\bibfnamefont {M.}~\bibnamefont {Horvati\'{c}}}, \bibinfo {author}
  {\bibfnamefont {C.}~\bibnamefont {Berthier}}, \bibinfo {author}
  {\bibfnamefont {P.~L.}\ \bibnamefont {Kuhns}}, \bibinfo {author}
  {\bibfnamefont {A.~P.}\ \bibnamefont {Reyes}}, \bibinfo {author}
  {\bibfnamefont {R.}~\bibnamefont {Liang}}, \bibinfo {author} {\bibfnamefont
  {W.~N.}\ \bibnamefont {Hardy}}, \bibinfo {author} {\bibfnamefont {D.~A.}\
  \bibnamefont {Bonn}}, \ and\ \bibinfo {author} {\bibfnamefont {M.-H.}\
  \bibnamefont {Julien}},\ }\href {\doibase 10.1038/ncomms3113} {\bibfield
  {journal} {\bibinfo  {journal} {Nature Communications}\ }\textbf {\bibinfo
  {volume} {4}},\ \bibinfo {pages} {2113} (\bibinfo {year} {2013})}\BibitemShut
  {NoStop}%
\bibitem [{\citenamefont {Wu}\ \emph {et~al.}(2015)\citenamefont {Wu},
  \citenamefont {Mayaffre}, \citenamefont {Kr\"{a}mer}, \citenamefont
  {Horvati\'{c}}, \citenamefont {Berthier}, \citenamefont {Hardy},
  \citenamefont {Liang}, \citenamefont {Bonn},\ and\ \citenamefont
  {Julien}}]{10.1038/ncomms7438}%
  \BibitemOpen
  \bibfield  {author} {\bibinfo {author} {\bibfnamefont {T.}~\bibnamefont
  {Wu}}, \bibinfo {author} {\bibfnamefont {H.}~\bibnamefont {Mayaffre}},
  \bibinfo {author} {\bibfnamefont {S.}~\bibnamefont {Kr\"{a}mer}}, \bibinfo
  {author} {\bibfnamefont {M.}~\bibnamefont {Horvati\'{c}}}, \bibinfo {author}
  {\bibfnamefont {C.}~\bibnamefont {Berthier}}, \bibinfo {author}
  {\bibfnamefont {W.}~\bibnamefont {Hardy}}, \bibinfo {author} {\bibfnamefont
  {R.}~\bibnamefont {Liang}}, \bibinfo {author} {\bibfnamefont
  {D.}~\bibnamefont {Bonn}}, \ and\ \bibinfo {author} {\bibfnamefont {M.-H.}\
  \bibnamefont {Julien}},\ }\href {\doibase 10.1038/ncomms7438} {\bibfield
  {journal} {\bibinfo  {journal} {Nature Communications}\ }\textbf {\bibinfo
  {volume} {6}},\ \bibinfo {pages} {6438} (\bibinfo {year} {2015})}\BibitemShut
  {NoStop}%
\bibitem [{\citenamefont {Chang}\ \emph {et~al.}(2016)\citenamefont {Chang},
  \citenamefont {Blackburn}, \citenamefont {Ivashko}, \citenamefont {Holmes},
  \citenamefont {Christensen}, \citenamefont {Hücker}, \citenamefont {Liang},
  \citenamefont {Hardy}, \citenamefont {Rütt}, \citenamefont {Zimmermann},
  \citenamefont {Forgan},\ and\ \citenamefont {Hayden}}]{10.1038/ncomms11494}%
  \BibitemOpen
  \bibfield  {author} {\bibinfo {author} {\bibfnamefont {J.}~\bibnamefont
  {Chang}}, \bibinfo {author} {\bibfnamefont {E.}~\bibnamefont {Blackburn}},
  \bibinfo {author} {\bibfnamefont {O.}~\bibnamefont {Ivashko}}, \bibinfo
  {author} {\bibfnamefont {A.~T.}\ \bibnamefont {Holmes}}, \bibinfo {author}
  {\bibfnamefont {N.~B.}\ \bibnamefont {Christensen}}, \bibinfo {author}
  {\bibfnamefont {M.}~\bibnamefont {Hücker}}, \bibinfo {author} {\bibfnamefont
  {D.~A.}\ \bibnamefont {Liang}, \bibfnamefont {Ruixingand~Bonn}}, \bibinfo
  {author} {\bibfnamefont {W.~N.}\ \bibnamefont {Hardy}}, \bibinfo {author}
  {\bibfnamefont {U.}~\bibnamefont {Rütt}}, \bibinfo {author} {\bibfnamefont
  {M.~v.}\ \bibnamefont {Zimmermann}}, \bibinfo {author} {\bibfnamefont
  {E.~M.}\ \bibnamefont {Forgan}}, \ and\ \bibinfo {author} {\bibfnamefont
  {S.~M.}\ \bibnamefont {Hayden}},\ }\href {\doibase 10.1038/ncomms11494}
  {\bibfield  {journal} {\bibinfo  {journal} {Nature Communications}\ }\textbf
  {\bibinfo {volume} {7}},\ \bibinfo {pages} {11494} (\bibinfo {year}
  {2016})}\BibitemShut {NoStop}%
\bibitem [{\citenamefont {Chang}\ \emph {et~al.}(2012)\citenamefont {Chang},
  \citenamefont {Blackburn}, \citenamefont {Holmes}, \citenamefont
  {Christensen}, \citenamefont {Larsen}, \citenamefont {Mesot}, \citenamefont
  {Liang}, \citenamefont {Bonn}, \citenamefont {Hardy}, \citenamefont
  {Watenphul}, \citenamefont {Zimmermann}, \citenamefont {Forgan},\ and\
  \citenamefont {Hayden}}]{10.1038/nphys2456}%
  \BibitemOpen
  \bibfield  {author} {\bibinfo {author} {\bibfnamefont {J.}~\bibnamefont
  {Chang}}, \bibinfo {author} {\bibfnamefont {E.}~\bibnamefont {Blackburn}},
  \bibinfo {author} {\bibfnamefont {A.~T.}\ \bibnamefont {Holmes}}, \bibinfo
  {author} {\bibfnamefont {N.~B.}\ \bibnamefont {Christensen}}, \bibinfo
  {author} {\bibfnamefont {J.}~\bibnamefont {Larsen}}, \bibinfo {author}
  {\bibfnamefont {J.}~\bibnamefont {Mesot}}, \bibinfo {author} {\bibfnamefont
  {R.}~\bibnamefont {Liang}}, \bibinfo {author} {\bibfnamefont {D.~A.}\
  \bibnamefont {Bonn}}, \bibinfo {author} {\bibfnamefont {W.~N.}\ \bibnamefont
  {Hardy}}, \bibinfo {author} {\bibfnamefont {A.}~\bibnamefont {Watenphul}},
  \bibinfo {author} {\bibfnamefont {M.~v.}\ \bibnamefont {Zimmermann}},
  \bibinfo {author} {\bibfnamefont {E.~M.}\ \bibnamefont {Forgan}}, \ and\
  \bibinfo {author} {\bibfnamefont {S.~M.}\ \bibnamefont {Hayden}},\ }\href
  {\doibase 10.1038/nphys2456} {\bibfield  {journal} {\bibinfo  {journal} {Nat
  Phys}\ }\textbf {\bibinfo {volume} {8}},\ \bibinfo {pages} {871} (\bibinfo
  {year} {2012})}\BibitemShut {NoStop}%
\bibitem [{\citenamefont {Blackburn}\ \emph {et~al.}(2013)\citenamefont
  {Blackburn}, \citenamefont {Chang}, \citenamefont {H\"ucker}, \citenamefont
  {Holmes}, \citenamefont {Christensen}, \citenamefont {Liang}, \citenamefont
  {Bonn}, \citenamefont {Hardy}, \citenamefont {R\"utt}, \citenamefont
  {Gutowski}, \citenamefont {Zimmermann}, \citenamefont {Forgan},\ and\
  \citenamefont {Hayden}}]{PhysRevLett.110.137004}%
  \BibitemOpen
  \bibfield  {author} {\bibinfo {author} {\bibfnamefont {E.}~\bibnamefont
  {Blackburn}}, \bibinfo {author} {\bibfnamefont {J.}~\bibnamefont {Chang}},
  \bibinfo {author} {\bibfnamefont {M.}~\bibnamefont {H\"ucker}}, \bibinfo
  {author} {\bibfnamefont {A.~T.}\ \bibnamefont {Holmes}}, \bibinfo {author}
  {\bibfnamefont {N.~B.}\ \bibnamefont {Christensen}}, \bibinfo {author}
  {\bibfnamefont {R.}~\bibnamefont {Liang}}, \bibinfo {author} {\bibfnamefont
  {D.~A.}\ \bibnamefont {Bonn}}, \bibinfo {author} {\bibfnamefont {W.~N.}\
  \bibnamefont {Hardy}}, \bibinfo {author} {\bibfnamefont {U.}~\bibnamefont
  {R\"utt}}, \bibinfo {author} {\bibfnamefont {O.}~\bibnamefont {Gutowski}},
  \bibinfo {author} {\bibfnamefont {M.~v.}\ \bibnamefont {Zimmermann}},
  \bibinfo {author} {\bibfnamefont {E.~M.}\ \bibnamefont {Forgan}}, \ and\
  \bibinfo {author} {\bibfnamefont {S.~M.}\ \bibnamefont {Hayden}},\ }\href
  {\doibase 10.1103/PhysRevLett.110.137004} {\bibfield  {journal} {\bibinfo
  {journal} {Phys. Rev. Lett.}\ }\textbf {\bibinfo {volume} {110}},\ \bibinfo
  {pages} {137004} (\bibinfo {year} {2013})}\BibitemShut {NoStop}%
\bibitem [{\citenamefont {Ghiringhelli}\ \emph {et~al.}(2012)\citenamefont
  {Ghiringhelli}, \citenamefont {Le~Tacon}, \citenamefont {Minola},
  \citenamefont {Blanco-Canosa}, \citenamefont {Mazzoli}, \citenamefont
  {Brookes}, \citenamefont {De~Luca}, \citenamefont {Frano}, \citenamefont
  {Hawthorn}, \citenamefont {He}, \citenamefont {Loew}, \citenamefont {Sala},
  \citenamefont {Peets}, \citenamefont {Salluzzo}, \citenamefont {Schierle},
  \citenamefont {Sutarto}, \citenamefont {Sawatzky}, \citenamefont {Weschke},
  \citenamefont {Keimer},\ and\ \citenamefont {Braicovich}}]{Ghiringhelli821}%
  \BibitemOpen
  \bibfield  {author} {\bibinfo {author} {\bibfnamefont {G.}~\bibnamefont
  {Ghiringhelli}}, \bibinfo {author} {\bibfnamefont {M.}~\bibnamefont
  {Le~Tacon}}, \bibinfo {author} {\bibfnamefont {M.}~\bibnamefont {Minola}},
  \bibinfo {author} {\bibfnamefont {S.}~\bibnamefont {Blanco-Canosa}}, \bibinfo
  {author} {\bibfnamefont {C.}~\bibnamefont {Mazzoli}}, \bibinfo {author}
  {\bibfnamefont {N.~B.}\ \bibnamefont {Brookes}}, \bibinfo {author}
  {\bibfnamefont {G.~M.}\ \bibnamefont {De~Luca}}, \bibinfo {author}
  {\bibfnamefont {A.}~\bibnamefont {Frano}}, \bibinfo {author} {\bibfnamefont
  {D.~G.}\ \bibnamefont {Hawthorn}}, \bibinfo {author} {\bibfnamefont
  {F.}~\bibnamefont {He}}, \bibinfo {author} {\bibfnamefont {T.}~\bibnamefont
  {Loew}}, \bibinfo {author} {\bibfnamefont {M.~M.}\ \bibnamefont {Sala}},
  \bibinfo {author} {\bibfnamefont {D.~C.}\ \bibnamefont {Peets}}, \bibinfo
  {author} {\bibfnamefont {M.}~\bibnamefont {Salluzzo}}, \bibinfo {author}
  {\bibfnamefont {E.}~\bibnamefont {Schierle}}, \bibinfo {author}
  {\bibfnamefont {R.}~\bibnamefont {Sutarto}}, \bibinfo {author} {\bibfnamefont
  {G.~A.}\ \bibnamefont {Sawatzky}}, \bibinfo {author} {\bibfnamefont
  {E.}~\bibnamefont {Weschke}}, \bibinfo {author} {\bibfnamefont
  {B.}~\bibnamefont {Keimer}}, \ and\ \bibinfo {author} {\bibfnamefont
  {L.}~\bibnamefont {Braicovich}},\ }\href {\doibase 10.1126/science.1223532}
  {\bibfield  {journal} {\bibinfo  {journal} {Science}\ }\textbf {\bibinfo
  {volume} {337}},\ \bibinfo {pages} {821} (\bibinfo {year} {2012})},\ \Eprint
  {http://arxiv.org/abs/http://science.sciencemag.org/content/337/6096/821.full.pdf}
  {http://science.sciencemag.org/content/337/6096/821.full.pdf} \BibitemShut
  {NoStop}%
\bibitem [{\citenamefont {Blanco-Canosa}\ \emph {et~al.}(2013)\citenamefont
  {Blanco-Canosa}, \citenamefont {Frano}, \citenamefont {Loew}, \citenamefont
  {Lu}, \citenamefont {Porras}, \citenamefont {Ghiringhelli}, \citenamefont
  {Minola}, \citenamefont {Mazzoli}, \citenamefont {Braicovich}, \citenamefont
  {Schierle}, \citenamefont {Weschke}, \citenamefont {Le~Tacon},\ and\
  \citenamefont {Keimer}}]{PhysRevLett.110.187001}%
  \BibitemOpen
  \bibfield  {author} {\bibinfo {author} {\bibfnamefont {S.}~\bibnamefont
  {Blanco-Canosa}}, \bibinfo {author} {\bibfnamefont {A.}~\bibnamefont
  {Frano}}, \bibinfo {author} {\bibfnamefont {T.}~\bibnamefont {Loew}},
  \bibinfo {author} {\bibfnamefont {Y.}~\bibnamefont {Lu}}, \bibinfo {author}
  {\bibfnamefont {J.}~\bibnamefont {Porras}}, \bibinfo {author} {\bibfnamefont
  {G.}~\bibnamefont {Ghiringhelli}}, \bibinfo {author} {\bibfnamefont
  {M.}~\bibnamefont {Minola}}, \bibinfo {author} {\bibfnamefont
  {C.}~\bibnamefont {Mazzoli}}, \bibinfo {author} {\bibfnamefont
  {L.}~\bibnamefont {Braicovich}}, \bibinfo {author} {\bibfnamefont
  {E.}~\bibnamefont {Schierle}}, \bibinfo {author} {\bibfnamefont
  {E.}~\bibnamefont {Weschke}}, \bibinfo {author} {\bibfnamefont
  {M.}~\bibnamefont {Le~Tacon}}, \ and\ \bibinfo {author} {\bibfnamefont
  {B.}~\bibnamefont {Keimer}},\ }\href {\doibase
  10.1103/PhysRevLett.110.187001} {\bibfield  {journal} {\bibinfo  {journal}
  {Phys. Rev. Lett.}\ }\textbf {\bibinfo {volume} {110}},\ \bibinfo {pages}
  {187001} (\bibinfo {year} {2013})}\BibitemShut {NoStop}%
\bibitem [{\citenamefont {Blanco-Canosa}\ \emph {et~al.}(2014)\citenamefont
  {Blanco-Canosa}, \citenamefont {Frano}, \citenamefont {Schierle},
  \citenamefont {Porras}, \citenamefont {Loew}, \citenamefont {Minola},
  \citenamefont {Bluschke}, \citenamefont {Weschke}, \citenamefont {Keimer},\
  and\ \citenamefont {Le~Tacon}}]{PhysRevB.90.054513}%
  \BibitemOpen
  \bibfield  {author} {\bibinfo {author} {\bibfnamefont {S.}~\bibnamefont
  {Blanco-Canosa}}, \bibinfo {author} {\bibfnamefont {A.}~\bibnamefont
  {Frano}}, \bibinfo {author} {\bibfnamefont {E.}~\bibnamefont {Schierle}},
  \bibinfo {author} {\bibfnamefont {J.}~\bibnamefont {Porras}}, \bibinfo
  {author} {\bibfnamefont {T.}~\bibnamefont {Loew}}, \bibinfo {author}
  {\bibfnamefont {M.}~\bibnamefont {Minola}}, \bibinfo {author} {\bibfnamefont
  {M.}~\bibnamefont {Bluschke}}, \bibinfo {author} {\bibfnamefont
  {E.}~\bibnamefont {Weschke}}, \bibinfo {author} {\bibfnamefont
  {B.}~\bibnamefont {Keimer}}, \ and\ \bibinfo {author} {\bibfnamefont
  {M.}~\bibnamefont {Le~Tacon}},\ }\href {\doibase 10.1103/PhysRevB.90.054513}
  {\bibfield  {journal} {\bibinfo  {journal} {Phys. Rev. B}\ }\textbf {\bibinfo
  {volume} {90}},\ \bibinfo {pages} {054513} (\bibinfo {year}
  {2014})}\BibitemShut {NoStop}%
\bibitem [{\citenamefont {Le~Tacon}\ \emph {et~al.}(2011)\citenamefont
  {Le~Tacon}, \citenamefont {Ghiringhelli}, \citenamefont {Chaloupka},
  \citenamefont {Sala}, \citenamefont {Hinkov}, \citenamefont {Haverkort},
  \citenamefont {Minola}, \citenamefont {Bakr}, \citenamefont {Zhou},
  \citenamefont {Blanco-Canosa}, \citenamefont {Monney}, \citenamefont {Song},
  \citenamefont {Sun}, \citenamefont {Lin}, \citenamefont {De~Luca},
  \citenamefont {Salluzzo}, \citenamefont {Khaliullin}, \citenamefont
  {Schmitt}, \citenamefont {Braicovich},\ and\ \citenamefont
  {Keimer}}]{10.1038/nphys2041}%
  \BibitemOpen
  \bibfield  {author} {\bibinfo {author} {\bibfnamefont {M.}~\bibnamefont
  {Le~Tacon}}, \bibinfo {author} {\bibfnamefont {G.}~\bibnamefont
  {Ghiringhelli}}, \bibinfo {author} {\bibfnamefont {J.}~\bibnamefont
  {Chaloupka}}, \bibinfo {author} {\bibfnamefont {M.~M.}\ \bibnamefont {Sala}},
  \bibinfo {author} {\bibfnamefont {V.}~\bibnamefont {Hinkov}}, \bibinfo
  {author} {\bibfnamefont {M.~W.}\ \bibnamefont {Haverkort}}, \bibinfo {author}
  {\bibfnamefont {M.}~\bibnamefont {Minola}}, \bibinfo {author} {\bibfnamefont
  {M.}~\bibnamefont {Bakr}}, \bibinfo {author} {\bibfnamefont {K.~J.}\
  \bibnamefont {Zhou}}, \bibinfo {author} {\bibfnamefont {S.}~\bibnamefont
  {Blanco-Canosa}}, \bibinfo {author} {\bibfnamefont {C.}~\bibnamefont
  {Monney}}, \bibinfo {author} {\bibfnamefont {Y.~T.}\ \bibnamefont {Song}},
  \bibinfo {author} {\bibfnamefont {G.~L.}\ \bibnamefont {Sun}}, \bibinfo
  {author} {\bibfnamefont {C.~T.}\ \bibnamefont {Lin}}, \bibinfo {author}
  {\bibfnamefont {G.~M.}\ \bibnamefont {De~Luca}}, \bibinfo {author}
  {\bibfnamefont {M.}~\bibnamefont {Salluzzo}}, \bibinfo {author}
  {\bibfnamefont {G.}~\bibnamefont {Khaliullin}}, \bibinfo {author}
  {\bibfnamefont {T.}~\bibnamefont {Schmitt}}, \bibinfo {author} {\bibfnamefont
  {L.}~\bibnamefont {Braicovich}}, \ and\ \bibinfo {author} {\bibfnamefont
  {B.}~\bibnamefont {Keimer}},\ }\href {\doibase 10.1038/nphys2041} {\bibfield
  {journal} {\bibinfo  {journal} {Nat Phys}\ }\textbf {\bibinfo {volume} {7}},\
  \bibinfo {pages} {725} (\bibinfo {year} {2011})}\BibitemShut {NoStop}%
\bibitem [{\citenamefont {Allais}\ \emph {et~al.}(2014)\citenamefont {Allais},
  \citenamefont {Bauer},\ and\ \citenamefont {Sachdev}}]{PhysRevB.90.155114}%
  \BibitemOpen
  \bibfield  {author} {\bibinfo {author} {\bibfnamefont {A.}~\bibnamefont
  {Allais}}, \bibinfo {author} {\bibfnamefont {J.}~\bibnamefont {Bauer}}, \
  and\ \bibinfo {author} {\bibfnamefont {S.}~\bibnamefont {Sachdev}},\ }\href
  {\doibase 10.1103/PhysRevB.90.155114} {\bibfield  {journal} {\bibinfo
  {journal} {Phys. Rev. B}\ }\textbf {\bibinfo {volume} {90}},\ \bibinfo
  {pages} {155114} (\bibinfo {year} {2014})}\BibitemShut {NoStop}%
\bibitem [{\citenamefont {Atkinson}\ \emph {et~al.}(2015)\citenamefont
  {Atkinson}, \citenamefont {Kampf},\ and\ \citenamefont
  {Bulut}}]{1367-2630-17-1-013025}%
  \BibitemOpen
  \bibfield  {author} {\bibinfo {author} {\bibfnamefont {W.~A.}\ \bibnamefont
  {Atkinson}}, \bibinfo {author} {\bibfnamefont {A.~P.}\ \bibnamefont {Kampf}},
  \ and\ \bibinfo {author} {\bibfnamefont {S.}~\bibnamefont {Bulut}},\ }\href
  {http://stacks.iop.org/1367-2630/17/i=1/a=013025} {\bibfield  {journal}
  {\bibinfo  {journal} {New Journal of Physics}\ }\textbf {\bibinfo {volume}
  {17}},\ \bibinfo {pages} {013025} (\bibinfo {year} {2015})}\BibitemShut
  {NoStop}%
\bibitem [{\citenamefont {Atkinson}\ and\ \citenamefont
  {Kampf}(2015)}]{PhysRevB.91.104509}%
  \BibitemOpen
  \bibfield  {author} {\bibinfo {author} {\bibfnamefont {W.~A.}\ \bibnamefont
  {Atkinson}}\ and\ \bibinfo {author} {\bibfnamefont {A.~P.}\ \bibnamefont
  {Kampf}},\ }\href {\doibase 10.1103/PhysRevB.91.104509} {\bibfield  {journal}
  {\bibinfo  {journal} {Phys. Rev. B}\ }\textbf {\bibinfo {volume} {91}},\
  \bibinfo {pages} {104509} (\bibinfo {year} {2015})}\BibitemShut {NoStop}%
\bibitem [{\citenamefont {Meier}\ \emph {et~al.}(2013)\citenamefont {Meier},
  \citenamefont {Einenkel}, \citenamefont {P\'epin},\ and\ \citenamefont
  {Efetov}}]{PhysRevB.88.020506}%
  \BibitemOpen
  \bibfield  {author} {\bibinfo {author} {\bibfnamefont {H.}~\bibnamefont
  {Meier}}, \bibinfo {author} {\bibfnamefont {M.}~\bibnamefont {Einenkel}},
  \bibinfo {author} {\bibfnamefont {C.}~\bibnamefont {P\'epin}}, \ and\
  \bibinfo {author} {\bibfnamefont {K.~B.}\ \bibnamefont {Efetov}},\ }\href
  {\doibase 10.1103/PhysRevB.88.020506} {\bibfield  {journal} {\bibinfo
  {journal} {Phys. Rev. B}\ }\textbf {\bibinfo {volume} {88}},\ \bibinfo
  {pages} {020506} (\bibinfo {year} {2013})}\BibitemShut {NoStop}%
\bibitem [{\citenamefont {Meier}\ \emph {et~al.}(2014)\citenamefont {Meier},
  \citenamefont {P\'epin}, \citenamefont {Einenkel},\ and\ \citenamefont
  {Efetov}}]{PhysRevB.89.195115}%
  \BibitemOpen
  \bibfield  {author} {\bibinfo {author} {\bibfnamefont {H.}~\bibnamefont
  {Meier}}, \bibinfo {author} {\bibfnamefont {C.}~\bibnamefont {P\'epin}},
  \bibinfo {author} {\bibfnamefont {M.}~\bibnamefont {Einenkel}}, \ and\
  \bibinfo {author} {\bibfnamefont {K.~B.}\ \bibnamefont {Efetov}},\ }\href
  {\doibase 10.1103/PhysRevB.89.195115} {\bibfield  {journal} {\bibinfo
  {journal} {Phys. Rev. B}\ }\textbf {\bibinfo {volume} {89}},\ \bibinfo
  {pages} {195115} (\bibinfo {year} {2014})}\BibitemShut {NoStop}%
\bibitem [{\citenamefont {Einenkel}\ \emph {et~al.}(2014)\citenamefont
  {Einenkel}, \citenamefont {Meier}, \citenamefont {P\'epin},\ and\
  \citenamefont {Efetov}}]{PhysRevB.90.054511}%
  \BibitemOpen
  \bibfield  {author} {\bibinfo {author} {\bibfnamefont {M.}~\bibnamefont
  {Einenkel}}, \bibinfo {author} {\bibfnamefont {H.}~\bibnamefont {Meier}},
  \bibinfo {author} {\bibfnamefont {C.}~\bibnamefont {P\'epin}}, \ and\
  \bibinfo {author} {\bibfnamefont {K.~B.}\ \bibnamefont {Efetov}},\ }\href
  {\doibase 10.1103/PhysRevB.90.054511} {\bibfield  {journal} {\bibinfo
  {journal} {Phys. Rev. B}\ }\textbf {\bibinfo {volume} {90}},\ \bibinfo
  {pages} {054511} (\bibinfo {year} {2014})}\BibitemShut {NoStop}%
\bibitem [{\citenamefont {Agterberg}\ \emph {et~al.}(2015)\citenamefont
  {Agterberg}, \citenamefont {Melchert},\ and\ \citenamefont
  {Kashyap}}]{PhysRevB.91.054502}%
  \BibitemOpen
  \bibfield  {author} {\bibinfo {author} {\bibfnamefont {D.~F.}\ \bibnamefont
  {Agterberg}}, \bibinfo {author} {\bibfnamefont {D.~S.}\ \bibnamefont
  {Melchert}}, \ and\ \bibinfo {author} {\bibfnamefont {M.~K.}\ \bibnamefont
  {Kashyap}},\ }\href {\doibase 10.1103/PhysRevB.91.054502} {\bibfield
  {journal} {\bibinfo  {journal} {Phys. Rev. B}\ }\textbf {\bibinfo {volume}
  {91}},\ \bibinfo {pages} {054502} (\bibinfo {year} {2015})}\BibitemShut
  {NoStop}%
\bibitem [{\citenamefont {Wang}\ and\ \citenamefont
  {Chubukov}(2014)}]{PhysRevB.90.035149}%
  \BibitemOpen
  \bibfield  {author} {\bibinfo {author} {\bibfnamefont {Y.}~\bibnamefont
  {Wang}}\ and\ \bibinfo {author} {\bibfnamefont {A.}~\bibnamefont
  {Chubukov}},\ }\href {\doibase 10.1103/PhysRevB.90.035149} {\bibfield
  {journal} {\bibinfo  {journal} {Phys. Rev. B}\ }\textbf {\bibinfo {volume}
  {90}},\ \bibinfo {pages} {035149} (\bibinfo {year} {2014})}\BibitemShut
  {NoStop}%
\bibitem [{\citenamefont {Wang}\ \emph
  {et~al.}(2015{\natexlab{a}})\citenamefont {Wang}, \citenamefont {Agterberg},\
  and\ \citenamefont {Chubukov}}]{PhysRevB.91.115103}%
  \BibitemOpen
  \bibfield  {author} {\bibinfo {author} {\bibfnamefont {Y.}~\bibnamefont
  {Wang}}, \bibinfo {author} {\bibfnamefont {D.~F.}\ \bibnamefont {Agterberg}},
  \ and\ \bibinfo {author} {\bibfnamefont {A.}~\bibnamefont {Chubukov}},\
  }\href {\doibase 10.1103/PhysRevB.91.115103} {\bibfield  {journal} {\bibinfo
  {journal} {Phys. Rev. B}\ }\textbf {\bibinfo {volume} {91}},\ \bibinfo
  {pages} {115103} (\bibinfo {year} {2015}{\natexlab{a}})}\BibitemShut
  {NoStop}%
\bibitem [{\citenamefont {Wang}\ \emph
  {et~al.}(2015{\natexlab{b}})\citenamefont {Wang}, \citenamefont {Agterberg},\
  and\ \citenamefont {Chubukov}}]{PhysRevLett.114.197001}%
  \BibitemOpen
  \bibfield  {author} {\bibinfo {author} {\bibfnamefont {Y.}~\bibnamefont
  {Wang}}, \bibinfo {author} {\bibfnamefont {D.~F.}\ \bibnamefont {Agterberg}},
  \ and\ \bibinfo {author} {\bibfnamefont {A.}~\bibnamefont {Chubukov}},\
  }\href {\doibase 10.1103/PhysRevLett.114.197001} {\bibfield  {journal}
  {\bibinfo  {journal} {Phys. Rev. Lett.}\ }\textbf {\bibinfo {volume} {114}},\
  \bibinfo {pages} {197001} (\bibinfo {year} {2015}{\natexlab{b}})}\BibitemShut
  {NoStop}%
\bibitem [{\citenamefont {Wang}\ and\ \citenamefont
  {Chubukov}(2015)}]{PhysRevB.91.195113}%
  \BibitemOpen
  \bibfield  {author} {\bibinfo {author} {\bibfnamefont {Y.}~\bibnamefont
  {Wang}}\ and\ \bibinfo {author} {\bibfnamefont {A.}~\bibnamefont
  {Chubukov}},\ }\href {\doibase 10.1103/PhysRevB.91.195113} {\bibfield
  {journal} {\bibinfo  {journal} {Phys. Rev. B}\ }\textbf {\bibinfo {volume}
  {91}},\ \bibinfo {pages} {195113} (\bibinfo {year} {2015})}\BibitemShut
  {NoStop}%
\bibitem [{\citenamefont {P\'epin}\ \emph {et~al.}(2014)\citenamefont
  {P\'epin}, \citenamefont {de~Carvalho}, \citenamefont {Kloss},\ and\
  \citenamefont {Montiel}}]{PhysRevB.90.195207}%
  \BibitemOpen
  \bibfield  {author} {\bibinfo {author} {\bibfnamefont {C.}~\bibnamefont
  {P\'epin}}, \bibinfo {author} {\bibfnamefont {V.~S.}\ \bibnamefont
  {de~Carvalho}}, \bibinfo {author} {\bibfnamefont {T.}~\bibnamefont {Kloss}},
  \ and\ \bibinfo {author} {\bibfnamefont {X.}~\bibnamefont {Montiel}},\ }\href
  {\doibase 10.1103/PhysRevB.90.195207} {\bibfield  {journal} {\bibinfo
  {journal} {Phys. Rev. B}\ }\textbf {\bibinfo {volume} {90}},\ \bibinfo
  {pages} {195207} (\bibinfo {year} {2014})}\BibitemShut {NoStop}%
\bibitem [{\citenamefont {Freire}\ \emph {et~al.}(2015)\citenamefont {Freire},
  \citenamefont {de~Carvalho},\ and\ \citenamefont
  {P\'epin}}]{PhysRevB.92.045132}%
  \BibitemOpen
  \bibfield  {author} {\bibinfo {author} {\bibfnamefont {H.}~\bibnamefont
  {Freire}}, \bibinfo {author} {\bibfnamefont {V.~S.}\ \bibnamefont
  {de~Carvalho}}, \ and\ \bibinfo {author} {\bibfnamefont {C.}~\bibnamefont
  {P\'epin}},\ }\href {\doibase 10.1103/PhysRevB.92.045132} {\bibfield
  {journal} {\bibinfo  {journal} {Phys. Rev. B}\ }\textbf {\bibinfo {volume}
  {92}},\ \bibinfo {pages} {045132} (\bibinfo {year} {2015})}\BibitemShut
  {NoStop}%
\bibitem [{\citenamefont {Caprara}\ \emph
  {et~al.}(2017{\natexlab{a}})\citenamefont {Caprara}, \citenamefont
  {Di~Castro}, \citenamefont {Seibold},\ and\ \citenamefont
  {Grilli}}]{PhysRevB.95.224511}%
  \BibitemOpen
  \bibfield  {author} {\bibinfo {author} {\bibfnamefont {S.}~\bibnamefont
  {Caprara}}, \bibinfo {author} {\bibfnamefont {C.}~\bibnamefont {Di~Castro}},
  \bibinfo {author} {\bibfnamefont {G.}~\bibnamefont {Seibold}}, \ and\
  \bibinfo {author} {\bibfnamefont {M.}~\bibnamefont {Grilli}},\ }\href
  {\doibase 10.1103/PhysRevB.95.224511} {\bibfield  {journal} {\bibinfo
  {journal} {Phys. Rev. B}\ }\textbf {\bibinfo {volume} {95}},\ \bibinfo
  {pages} {224511} (\bibinfo {year} {2017}{\natexlab{a}})}\BibitemShut
  {NoStop}%
\bibitem [{\citenamefont {Caprara}\ \emph
  {et~al.}(2017{\natexlab{b}})\citenamefont {Caprara}, \citenamefont {Grilli},
  \citenamefont {Di~Castro},\ and\ \citenamefont {Seibold}}]{Caprara2017}%
  \BibitemOpen
  \bibfield  {author} {\bibinfo {author} {\bibfnamefont {S.}~\bibnamefont
  {Caprara}}, \bibinfo {author} {\bibfnamefont {M.}~\bibnamefont {Grilli}},
  \bibinfo {author} {\bibfnamefont {C.}~\bibnamefont {Di~Castro}}, \ and\
  \bibinfo {author} {\bibfnamefont {G.}~\bibnamefont {Seibold}},\ }\href
  {\doibase 10.1007/s10948-016-3775-9} {\bibfield  {journal} {\bibinfo
  {journal} {Journal of Superconductivity and Novel Magnetism}\ }\textbf
  {\bibinfo {volume} {30}},\ \bibinfo {pages} {25} (\bibinfo {year}
  {2017}{\natexlab{b}})}\BibitemShut {NoStop}%
\bibitem [{\citenamefont {Yamakawa}\ and\ \citenamefont
  {Kontani}(2015)}]{PhysRevLett.114.257001}%
  \BibitemOpen
  \bibfield  {author} {\bibinfo {author} {\bibfnamefont {Y.}~\bibnamefont
  {Yamakawa}}\ and\ \bibinfo {author} {\bibfnamefont {H.}~\bibnamefont
  {Kontani}},\ }\href {\doibase 10.1103/PhysRevLett.114.257001} {\bibfield
  {journal} {\bibinfo  {journal} {Phys. Rev. Lett.}\ }\textbf {\bibinfo
  {volume} {114}},\ \bibinfo {pages} {257001} (\bibinfo {year}
  {2015})}\BibitemShut {NoStop}%
\bibitem [{\citenamefont {Tsuchiizu}\ \emph {et~al.}(2016)\citenamefont
  {Tsuchiizu}, \citenamefont {Yamakawa},\ and\ \citenamefont
  {Kontani}}]{PhysRevB.93.155148}%
  \BibitemOpen
  \bibfield  {author} {\bibinfo {author} {\bibfnamefont {M.}~\bibnamefont
  {Tsuchiizu}}, \bibinfo {author} {\bibfnamefont {Y.}~\bibnamefont {Yamakawa}},
  \ and\ \bibinfo {author} {\bibfnamefont {H.}~\bibnamefont {Kontani}},\ }\href
  {\doibase 10.1103/PhysRevB.93.155148} {\bibfield  {journal} {\bibinfo
  {journal} {Phys. Rev. B}\ }\textbf {\bibinfo {volume} {93}},\ \bibinfo
  {pages} {155148} (\bibinfo {year} {2016})}\BibitemShut {NoStop}%
\bibitem [{\citenamefont {Kohsaka}\ \emph {et~al.}(2012)\citenamefont
  {Kohsaka}, \citenamefont {Hanaguri}, \citenamefont {Azuma}, \citenamefont
  {Takano}, \citenamefont {Davis},\ and\ \citenamefont
  {Takagi}}]{kohsaka_visualization_2012}%
  \BibitemOpen
  \bibfield  {author} {\bibinfo {author} {\bibfnamefont {Y.}~\bibnamefont
  {Kohsaka}}, \bibinfo {author} {\bibfnamefont {T.}~\bibnamefont {Hanaguri}},
  \bibinfo {author} {\bibfnamefont {M.}~\bibnamefont {Azuma}}, \bibinfo
  {author} {\bibfnamefont {M.}~\bibnamefont {Takano}}, \bibinfo {author}
  {\bibfnamefont {J.~C.}\ \bibnamefont {Davis}}, \ and\ \bibinfo {author}
  {\bibfnamefont {H.}~\bibnamefont {Takagi}},\ }\href {\doibase
  10.1038/nphys2321} {\bibfield  {journal} {\bibinfo  {journal} {Nature
  Physics}\ }\textbf {\bibinfo {volume} {8}},\ \bibinfo {pages} {534} (\bibinfo
  {year} {2012})}\BibitemShut {NoStop}%
\bibitem [{\citenamefont {Cai}\ \emph {et~al.}(2016)\citenamefont {Cai},
  \citenamefont {Ruan}, \citenamefont {Peng}, \citenamefont {Ye}, \citenamefont
  {Li}, \citenamefont {Hao}, \citenamefont {Zhou}, \citenamefont {Lee},\ and\
  \citenamefont {Wang}}]{Nat.Phys.10.1038/nphys3840}%
  \BibitemOpen
  \bibfield  {author} {\bibinfo {author} {\bibfnamefont {P.}~\bibnamefont
  {Cai}}, \bibinfo {author} {\bibfnamefont {W.}~\bibnamefont {Ruan}}, \bibinfo
  {author} {\bibfnamefont {Y.}~\bibnamefont {Peng}}, \bibinfo {author}
  {\bibfnamefont {C.}~\bibnamefont {Ye}}, \bibinfo {author} {\bibfnamefont
  {X.}~\bibnamefont {Li}}, \bibinfo {author} {\bibfnamefont {Z.}~\bibnamefont
  {Hao}}, \bibinfo {author} {\bibfnamefont {X.}~\bibnamefont {Zhou}}, \bibinfo
  {author} {\bibfnamefont {D.-H.}\ \bibnamefont {Lee}}, \ and\ \bibinfo
  {author} {\bibfnamefont {Y.}~\bibnamefont {Wang}},\ }\href {\doibase
  10.1038/nphys3840} {\bibfield  {journal} {\bibinfo  {journal} {Nat. Phys.}\
  }\textbf {\bibinfo {volume} {12}},\ \bibinfo {pages} {1047} (\bibinfo {year}
  {2016})}\BibitemShut {NoStop}%
\bibitem [{\citenamefont {Georges}\ \emph {et~al.}(1996)\citenamefont
  {Georges}, \citenamefont {Kotliar}, \citenamefont {Krauth},\ and\
  \citenamefont {Rozenberg}}]{RevModPhys.68.13}%
  \BibitemOpen
  \bibfield  {author} {\bibinfo {author} {\bibfnamefont {A.}~\bibnamefont
  {Georges}}, \bibinfo {author} {\bibfnamefont {G.}~\bibnamefont {Kotliar}},
  \bibinfo {author} {\bibfnamefont {W.}~\bibnamefont {Krauth}}, \ and\ \bibinfo
  {author} {\bibfnamefont {M.~J.}\ \bibnamefont {Rozenberg}},\ }\href {\doibase
  10.1103/RevModPhys.68.13} {\bibfield  {journal} {\bibinfo  {journal} {Rev.
  Mod. Phys.}\ }\textbf {\bibinfo {volume} {68}},\ \bibinfo {pages} {13}
  (\bibinfo {year} {1996})}\BibitemShut {NoStop}%
\bibitem [{\citenamefont {Maier}\ \emph {et~al.}(2005)\citenamefont {Maier},
  \citenamefont {Jarrell}, \citenamefont {Pruschke},\ and\ \citenamefont
  {Hettler}}]{RevModPhys.77.1027}%
  \BibitemOpen
  \bibfield  {author} {\bibinfo {author} {\bibfnamefont {T.}~\bibnamefont
  {Maier}}, \bibinfo {author} {\bibfnamefont {M.}~\bibnamefont {Jarrell}},
  \bibinfo {author} {\bibfnamefont {T.}~\bibnamefont {Pruschke}}, \ and\
  \bibinfo {author} {\bibfnamefont {M.~H.}\ \bibnamefont {Hettler}},\ }\href
  {\doibase 10.1103/RevModPhys.77.1027} {\bibfield  {journal} {\bibinfo
  {journal} {Rev. Mod. Phys.}\ }\textbf {\bibinfo {volume} {77}},\ \bibinfo
  {pages} {1027} (\bibinfo {year} {2005})}\BibitemShut {NoStop}%
\bibitem [{\citenamefont {Kotliar}\ \emph {et~al.}(2006)\citenamefont
  {Kotliar}, \citenamefont {Savrasov}, \citenamefont {Haule}, \citenamefont
  {Oudovenko}, \citenamefont {Parcollet},\ and\ \citenamefont
  {Marianetti}}]{RevModPhys.78.865}%
  \BibitemOpen
  \bibfield  {author} {\bibinfo {author} {\bibfnamefont {G.}~\bibnamefont
  {Kotliar}}, \bibinfo {author} {\bibfnamefont {S.}~\bibnamefont {Savrasov}},
  \bibinfo {author} {\bibfnamefont {K.}~\bibnamefont {Haule}}, \bibinfo
  {author} {\bibfnamefont {V.}~\bibnamefont {Oudovenko}}, \bibinfo {author}
  {\bibfnamefont {O.}~\bibnamefont {Parcollet}}, \ and\ \bibinfo {author}
  {\bibfnamefont {C.}~\bibnamefont {Marianetti}},\ }\href {\doibase
  10.1103/RevModPhys.78.865} {\bibfield  {journal} {\bibinfo  {journal} {Rev.
  Mod. Phys.}\ }\textbf {\bibinfo {volume} {78}},\ \bibinfo {pages} {865}
  (\bibinfo {year} {2006})}\BibitemShut {NoStop}%
\bibitem [{\citenamefont {Tremblay}\ \emph {et~al.}(2006)\citenamefont
  {Tremblay}, \citenamefont {Kyung},\ and\ \citenamefont
  {S\'en\'echal}}]{LTP:2006}%
  \BibitemOpen
  \bibfield  {author} {\bibinfo {author} {\bibfnamefont {A.~M.~S.}\
  \bibnamefont {Tremblay}}, \bibinfo {author} {\bibfnamefont {B.}~\bibnamefont
  {Kyung}}, \ and\ \bibinfo {author} {\bibfnamefont {D.}~\bibnamefont
  {S\'en\'echal}},\ }\href {http://dx.doi.org/10.1063/1.2199446} {\bibfield
  {journal} {\bibinfo  {journal} {Low Temp. Phys.}\ }\textbf {\bibinfo {volume}
  {32}},\ \bibinfo {pages} {424} (\bibinfo {year} {2006})}\BibitemShut
  {NoStop}%
\bibitem [{\citenamefont {Ayral}\ and\ \citenamefont
  {Parcollet}(2015)}]{PhysRevB.92.115109}%
  \BibitemOpen
  \bibfield  {author} {\bibinfo {author} {\bibfnamefont {T.}~\bibnamefont
  {Ayral}}\ and\ \bibinfo {author} {\bibfnamefont {O.}~\bibnamefont
  {Parcollet}},\ }\href {\doibase 10.1103/PhysRevB.92.115109} {\bibfield
  {journal} {\bibinfo  {journal} {Phys. Rev. B}\ }\textbf {\bibinfo {volume}
  {92}},\ \bibinfo {pages} {115109} (\bibinfo {year} {2015})}\BibitemShut
  {NoStop}%
\bibitem [{\citenamefont {Ayral}\ and\ \citenamefont
  {Parcollet}(2016{\natexlab{a}})}]{PhysRevB.93.235124}%
  \BibitemOpen
  \bibfield  {author} {\bibinfo {author} {\bibfnamefont {T.}~\bibnamefont
  {Ayral}}\ and\ \bibinfo {author} {\bibfnamefont {O.}~\bibnamefont
  {Parcollet}},\ }\href {\doibase 10.1103/PhysRevB.93.235124} {\bibfield
  {journal} {\bibinfo  {journal} {Phys. Rev. B}\ }\textbf {\bibinfo {volume}
  {93}},\ \bibinfo {pages} {235124} (\bibinfo {year}
  {2016}{\natexlab{a}})}\BibitemShut {NoStop}%
\bibitem [{\citenamefont {Ayral}\ and\ \citenamefont
  {Parcollet}(2016{\natexlab{b}})}]{PhysRevB.94.075159}%
  \BibitemOpen
  \bibfield  {author} {\bibinfo {author} {\bibfnamefont {T.}~\bibnamefont
  {Ayral}}\ and\ \bibinfo {author} {\bibfnamefont {O.}~\bibnamefont
  {Parcollet}},\ }\href {\doibase 10.1103/PhysRevB.94.075159} {\bibfield
  {journal} {\bibinfo  {journal} {Phys. Rev. B}\ }\textbf {\bibinfo {volume}
  {94}},\ \bibinfo {pages} {075159} (\bibinfo {year}
  {2016}{\natexlab{b}})}\BibitemShut {NoStop}%
\bibitem [{\citenamefont {Gukelberger}\ \emph {et~al.}(2017)\citenamefont
  {Gukelberger}, \citenamefont {Kozik},\ and\ \citenamefont
  {Hafermann}}]{PhysRevB.96.035152}%
  \BibitemOpen
  \bibfield  {author} {\bibinfo {author} {\bibfnamefont {J.}~\bibnamefont
  {Gukelberger}}, \bibinfo {author} {\bibfnamefont {E.}~\bibnamefont {Kozik}},
  \ and\ \bibinfo {author} {\bibfnamefont {H.}~\bibnamefont {Hafermann}},\
  }\href {\doibase 10.1103/PhysRevB.96.035152} {\bibfield  {journal} {\bibinfo
  {journal} {Phys. Rev. B}\ }\textbf {\bibinfo {volume} {96}},\ \bibinfo
  {pages} {035152} (\bibinfo {year} {2017})}\BibitemShut {NoStop}%
\bibitem [{\citenamefont {Sordi}\ \emph {et~al.}(2010)\citenamefont {Sordi},
  \citenamefont {Haule},\ and\ \citenamefont
  {Tremblay}}]{PhysRevLett.104.226402}%
  \BibitemOpen
  \bibfield  {author} {\bibinfo {author} {\bibfnamefont {G.}~\bibnamefont
  {Sordi}}, \bibinfo {author} {\bibfnamefont {K.}~\bibnamefont {Haule}}, \ and\
  \bibinfo {author} {\bibfnamefont {A.-M.~S.}\ \bibnamefont {Tremblay}},\
  }\href {\doibase 10.1103/PhysRevLett.104.226402} {\bibfield  {journal}
  {\bibinfo  {journal} {Phys. Rev. Lett.}\ }\textbf {\bibinfo {volume} {104}},\
  \bibinfo {pages} {226402} (\bibinfo {year} {2010})}\BibitemShut {NoStop}%
\bibitem [{\citenamefont {Sordi}\ \emph {et~al.}(2011)\citenamefont {Sordi},
  \citenamefont {Haule},\ and\ \citenamefont {Tremblay}}]{PhysRevB.84.075161}%
  \BibitemOpen
  \bibfield  {author} {\bibinfo {author} {\bibfnamefont {G.}~\bibnamefont
  {Sordi}}, \bibinfo {author} {\bibfnamefont {K.}~\bibnamefont {Haule}}, \ and\
  \bibinfo {author} {\bibfnamefont {A.-M.~S.}\ \bibnamefont {Tremblay}},\
  }\href {\doibase 10.1103/PhysRevB.84.075161} {\bibfield  {journal} {\bibinfo
  {journal} {Phys. Rev. B}\ }\textbf {\bibinfo {volume} {84}},\ \bibinfo
  {pages} {075161} (\bibinfo {year} {2011})}\BibitemShut {NoStop}%
\bibitem [{\citenamefont {Sordi}\ \emph {et~al.}(2012)\citenamefont {Sordi},
  \citenamefont {S{\'e}mon}, \citenamefont {Haule},\ and\ \citenamefont
  {Tremblay}}]{Sordi:2012}%
  \BibitemOpen
  \bibfield  {author} {\bibinfo {author} {\bibfnamefont {G.}~\bibnamefont
  {Sordi}}, \bibinfo {author} {\bibfnamefont {P.}~\bibnamefont {S{\'e}mon}},
  \bibinfo {author} {\bibfnamefont {K.}~\bibnamefont {Haule}}, \ and\ \bibinfo
  {author} {\bibfnamefont {A.-M.~S.}\ \bibnamefont {Tremblay}},\ }\href
  {\doibase 10.1038/srep00547} {\bibfield  {journal} {\bibinfo  {journal}
  {Scientific Reports}\ }\textbf {\bibinfo {volume} {2}} (\bibinfo {year}
  {2012}),\ 10.1038/srep00547}\BibitemShut {NoStop}%
\bibitem [{\citenamefont {Faye}\ and\ \citenamefont
  {S\'en\'echal}(2017{\natexlab{a}})}]{FayeCharge:2017}%
  \BibitemOpen
  \bibfield  {author} {\bibinfo {author} {\bibfnamefont {J.~P.~L.}\
  \bibnamefont {Faye}}\ and\ \bibinfo {author} {\bibfnamefont {D.}~\bibnamefont
  {S\'en\'echal}},\ }\href {http://arxiv.org/abs/1707.09446} {\bibfield
  {journal} {\bibinfo  {journal} {arxiv:1707.09446}\ } (\bibinfo {year}
  {2017}{\natexlab{a}})}\BibitemShut {NoStop}%
\bibitem [{\citenamefont {Braganc}\ \emph {et~al.}(2017)\citenamefont
  {Braganc}, \citenamefont {Sakai}, \citenamefont {Aguiar},\ and\ \citenamefont
  {Civelli}}]{CivelliCharge:2017}%
  \BibitemOpen
  \bibfield  {author} {\bibinfo {author} {\bibfnamefont {H.}~\bibnamefont
  {Braganc}}, \bibinfo {author} {\bibfnamefont {S.}~\bibnamefont {Sakai}},
  \bibinfo {author} {\bibfnamefont {M.~C.~O.}\ \bibnamefont {Aguiar}}, \ and\
  \bibinfo {author} {\bibfnamefont {M.}~\bibnamefont {Civelli}},\ }\href
  {http://arxiv.org/abs/1708.02084} {\bibfield  {journal} {\bibinfo  {journal}
  {arxiv:1708.02084}\ } (\bibinfo {year} {2017})}\BibitemShut {NoStop}%
\bibitem [{\citenamefont {Aichhorn}\ \emph {et~al.}(2006)\citenamefont
  {Aichhorn}, \citenamefont {Arrigoni}, \citenamefont {Potthoff},\ and\
  \citenamefont {Hanke}}]{PhysRevB.74.024508}%
  \BibitemOpen
  \bibfield  {author} {\bibinfo {author} {\bibfnamefont {M.}~\bibnamefont
  {Aichhorn}}, \bibinfo {author} {\bibfnamefont {E.}~\bibnamefont {Arrigoni}},
  \bibinfo {author} {\bibfnamefont {M.}~\bibnamefont {Potthoff}}, \ and\
  \bibinfo {author} {\bibfnamefont {W.}~\bibnamefont {Hanke}},\ }\href
  {\doibase 10.1103/PhysRevB.74.024508} {\bibfield  {journal} {\bibinfo
  {journal} {Phys. Rev. B}\ }\textbf {\bibinfo {volume} {74}},\ \bibinfo
  {pages} {024508} (\bibinfo {year} {2006})}\BibitemShut {NoStop}%
\bibitem [{\citenamefont {Macridin}\ \emph {et~al.}(2006)\citenamefont
  {Macridin}, \citenamefont {Jarrell},\ and\ \citenamefont
  {Maier}}]{PhysRevB.74.085104}%
  \BibitemOpen
  \bibfield  {author} {\bibinfo {author} {\bibfnamefont {A.}~\bibnamefont
  {Macridin}}, \bibinfo {author} {\bibfnamefont {M.}~\bibnamefont {Jarrell}}, \
  and\ \bibinfo {author} {\bibfnamefont {T.}~\bibnamefont {Maier}},\ }\href
  {\doibase 10.1103/PhysRevB.74.085104} {\bibfield  {journal} {\bibinfo
  {journal} {Phys. Rev. B}\ }\textbf {\bibinfo {volume} {74}},\ \bibinfo
  {pages} {085104} (\bibinfo {year} {2006})}\BibitemShut {NoStop}%
\bibitem [{\citenamefont {Galanakis}\ \emph {et~al.}(2011)\citenamefont
  {Galanakis}, \citenamefont {Khatami}, \citenamefont {Mikelsons},
  \citenamefont {Macridin}, \citenamefont {Moreno}, \citenamefont {Browne},\
  and\ \citenamefont {Jarrell}}]{Galanakis1670}%
  \BibitemOpen
  \bibfield  {author} {\bibinfo {author} {\bibfnamefont {D.}~\bibnamefont
  {Galanakis}}, \bibinfo {author} {\bibfnamefont {E.}~\bibnamefont {Khatami}},
  \bibinfo {author} {\bibfnamefont {K.}~\bibnamefont {Mikelsons}}, \bibinfo
  {author} {\bibfnamefont {A.}~\bibnamefont {Macridin}}, \bibinfo {author}
  {\bibfnamefont {J.}~\bibnamefont {Moreno}}, \bibinfo {author} {\bibfnamefont
  {D.~A.}\ \bibnamefont {Browne}}, \ and\ \bibinfo {author} {\bibfnamefont
  {M.}~\bibnamefont {Jarrell}},\ }\href {\doibase 10.1098/rsta.2010.0228}
  {\bibfield  {journal} {\bibinfo  {journal} {Philosophical Transactions of the
  Royal Society of London A: Mathematical, Physical and Engineering Sciences}\
  }\textbf {\bibinfo {volume} {369}},\ \bibinfo {pages} {1670} (\bibinfo {year}
  {2011})}\BibitemShut {NoStop}%
\bibitem [{\citenamefont {Kotliar}\ \emph {et~al.}(2002)\citenamefont
  {Kotliar}, \citenamefont {Murthy},\ and\ \citenamefont
  {Rozenberg}}]{PhysRevLett.89.046401}%
  \BibitemOpen
  \bibfield  {author} {\bibinfo {author} {\bibfnamefont {G.}~\bibnamefont
  {Kotliar}}, \bibinfo {author} {\bibfnamefont {S.}~\bibnamefont {Murthy}}, \
  and\ \bibinfo {author} {\bibfnamefont {M.~J.}\ \bibnamefont {Rozenberg}},\
  }\href {\doibase 10.1103/PhysRevLett.89.046401} {\bibfield  {journal}
  {\bibinfo  {journal} {Phys. Rev. Lett.}\ }\textbf {\bibinfo {volume} {89}},\
  \bibinfo {pages} {046401} (\bibinfo {year} {2002})}\BibitemShut {NoStop}%
\bibitem [{\citenamefont {Yee}\ and\ \citenamefont
  {Balents}(2015)}]{PhysRevX.5.021007}%
  \BibitemOpen
  \bibfield  {author} {\bibinfo {author} {\bibfnamefont {C.-H.}\ \bibnamefont
  {Yee}}\ and\ \bibinfo {author} {\bibfnamefont {L.}~\bibnamefont {Balents}},\
  }\href {\doibase 10.1103/PhysRevX.5.021007} {\bibfield  {journal} {\bibinfo
  {journal} {Phys. Rev. X}\ }\textbf {\bibinfo {volume} {5}},\ \bibinfo {pages}
  {021007} (\bibinfo {year} {2015})}\BibitemShut {NoStop}%
\bibitem [{\citenamefont {Faye}\ and\ \citenamefont
  {S\'en\'echal}(2017{\natexlab{b}})}]{PhysRevB.95.115127}%
  \BibitemOpen
  \bibfield  {author} {\bibinfo {author} {\bibfnamefont {J.~P.~L.}\
  \bibnamefont {Faye}}\ and\ \bibinfo {author} {\bibfnamefont {D.}~\bibnamefont
  {S\'en\'echal}},\ }\href {\doibase 10.1103/PhysRevB.95.115127} {\bibfield
  {journal} {\bibinfo  {journal} {Phys. Rev. B}\ }\textbf {\bibinfo {volume}
  {95}},\ \bibinfo {pages} {115127} (\bibinfo {year}
  {2017}{\natexlab{b}})}\BibitemShut {NoStop}%
\bibitem [{\citenamefont {Vanhala}\ and\ \citenamefont
  {Törmä}(2017)}]{vanhala_dynamical_2017}%
  \BibitemOpen
  \bibfield  {author} {\bibinfo {author} {\bibfnamefont {T.~I.}\ \bibnamefont
  {Vanhala}}\ and\ \bibinfo {author} {\bibfnamefont {P.}~\bibnamefont
  {Törmä}},\ }\href {http://arxiv.org/abs/1708.06749} {\bibfield  {journal}
  {\bibinfo  {journal} {arXiv:1708.06749 [cond-mat]}\ } (\bibinfo {year}
  {2017})},\ \bibinfo {note} {arXiv: 1708.06749}\BibitemShut {NoStop}%
\bibitem [{\citenamefont {Mercure-Boissonault}(2015)}]{MercureMSc:2015}%
  \BibitemOpen
  \bibfield  {author} {\bibinfo {author} {\bibfnamefont {P.}~\bibnamefont
  {Mercure-Boissonault}},\ }\emph {\bibinfo {title} {Ordre de charge en rayures
  et supraconductivit\'e dans le modèle de Hubbard}},\ \href
  {https://www.physique.usherbrooke.ca/pages/sites/default/files/Mercure_P_MSc_PDFA.pdf}
  {Master's thesis},\ \bibinfo  {school} {Universit\'e de Sherbrooke} (\bibinfo
  {year} {2015})\BibitemShut {NoStop}%
\bibitem [{\citenamefont {Pavarini}\ \emph {et~al.}(2001)\citenamefont
  {Pavarini}, \citenamefont {Dasgupta}, \citenamefont {Saha-Dasgupta},
  \citenamefont {Jepsen},\ and\ \citenamefont
  {Andersen}}]{PhysRevLett.87.047003}%
  \BibitemOpen
  \bibfield  {author} {\bibinfo {author} {\bibfnamefont {E.}~\bibnamefont
  {Pavarini}}, \bibinfo {author} {\bibfnamefont {I.}~\bibnamefont {Dasgupta}},
  \bibinfo {author} {\bibfnamefont {T.}~\bibnamefont {Saha-Dasgupta}}, \bibinfo
  {author} {\bibfnamefont {O.}~\bibnamefont {Jepsen}}, \ and\ \bibinfo {author}
  {\bibfnamefont {O.~K.}\ \bibnamefont {Andersen}},\ }\href {\doibase
  10.1103/PhysRevLett.87.047003} {\bibfield  {journal} {\bibinfo  {journal}
  {Phys. Rev. Lett.}\ }\textbf {\bibinfo {volume} {87}},\ \bibinfo {pages}
  {047003} (\bibinfo {year} {2001})}\BibitemShut {NoStop}%
\bibitem [{\citenamefont {Caffarel}\ and\ \citenamefont
  {Krauth}(1994)}]{PhysRevLett.72.1545}%
  \BibitemOpen
  \bibfield  {author} {\bibinfo {author} {\bibfnamefont {M.}~\bibnamefont
  {Caffarel}}\ and\ \bibinfo {author} {\bibfnamefont {W.}~\bibnamefont
  {Krauth}},\ }\href {\doibase 10.1103/PhysRevLett.72.1545} {\bibfield
  {journal} {\bibinfo  {journal} {Phys. Rev. Lett.}\ }\textbf {\bibinfo
  {volume} {72}},\ \bibinfo {pages} {1545} (\bibinfo {year}
  {1994})}\BibitemShut {NoStop}%
\bibitem [{\citenamefont {Rohringer}\ \emph {et~al.}(2012)\citenamefont
  {Rohringer}, \citenamefont {Valli},\ and\ \citenamefont
  {Toschi}}]{PhysRevB.86.125114}%
  \BibitemOpen
  \bibfield  {author} {\bibinfo {author} {\bibfnamefont {G.}~\bibnamefont
  {Rohringer}}, \bibinfo {author} {\bibfnamefont {A.}~\bibnamefont {Valli}}, \
  and\ \bibinfo {author} {\bibfnamefont {A.}~\bibnamefont {Toschi}},\ }\href
  {\doibase 10.1103/PhysRevB.86.125114} {\bibfield  {journal} {\bibinfo
  {journal} {Phys. Rev. B}\ }\textbf {\bibinfo {volume} {86}},\ \bibinfo
  {pages} {125114} (\bibinfo {year} {2012})}\BibitemShut {NoStop}%
\bibitem [{\citenamefont {Toschi}\ \emph {et~al.}(2007)\citenamefont {Toschi},
  \citenamefont {Katanin},\ and\ \citenamefont {Held}}]{PhysRevB.75.045118}%
  \BibitemOpen
  \bibfield  {author} {\bibinfo {author} {\bibfnamefont {A.}~\bibnamefont
  {Toschi}}, \bibinfo {author} {\bibfnamefont {A.}~\bibnamefont {Katanin}}, \
  and\ \bibinfo {author} {\bibfnamefont {K.}~\bibnamefont {Held}},\ }\href
  {\doibase 10.1103/PhysRevB.75.045118} {\bibfield  {journal} {\bibinfo
  {journal} {Phys. Rev. B}\ }\textbf {\bibinfo {volume} {75}},\ \bibinfo
  {pages} {045118} (\bibinfo {year} {2007})}\BibitemShut {NoStop}%
\bibitem [{\citenamefont {Rohringer}\ \emph {et~al.}(2017)\citenamefont
  {Rohringer}, \citenamefont {Hafermann}, \citenamefont {Toschi}, \citenamefont
  {Katanin}, \citenamefont {Antipov}, \citenamefont {Katsnelson}, \citenamefont
  {Lichtenstein}, \citenamefont {Rubstov},\ and\ \citenamefont
  {Held}}]{arXiv:1705.00024}%
  \BibitemOpen
  \bibfield  {author} {\bibinfo {author} {\bibfnamefont {G.}~\bibnamefont
  {Rohringer}}, \bibinfo {author} {\bibfnamefont {H.}~\bibnamefont
  {Hafermann}}, \bibinfo {author} {\bibfnamefont {A.}~\bibnamefont {Toschi}},
  \bibinfo {author} {\bibfnamefont {A.~A.}\ \bibnamefont {Katanin}}, \bibinfo
  {author} {\bibfnamefont {A.~E.}\ \bibnamefont {Antipov}}, \bibinfo {author}
  {\bibfnamefont {M.~I.}\ \bibnamefont {Katsnelson}}, \bibinfo {author}
  {\bibfnamefont {A.~I.}\ \bibnamefont {Lichtenstein}}, \bibinfo {author}
  {\bibfnamefont {A.~N.}\ \bibnamefont {Rubstov}}, \ and\ \bibinfo {author}
  {\bibfnamefont {K.}~\bibnamefont {Held}},\ }\href
  {https://arxiv.org/abs/1705.00024} {\bibfield  {journal} {\bibinfo  {journal}
  {arXiv:1705.00024}\ } (\bibinfo {year} {2017})}\BibitemShut {NoStop}%
\bibitem [{\citenamefont {Sch\"afer}\ \emph {et~al.}(2016)\citenamefont
  {Sch\"afer}, \citenamefont {Ciuchi}, \citenamefont {Wallerberger},
  \citenamefont {Thunstr\"om}, \citenamefont {Gunnarsson}, \citenamefont
  {Sangiovanni}, \citenamefont {Rohringer},\ and\ \citenamefont
  {Toschi}}]{PhysRevB.94.235108}%
  \BibitemOpen
  \bibfield  {author} {\bibinfo {author} {\bibfnamefont {T.}~\bibnamefont
  {Sch\"afer}}, \bibinfo {author} {\bibfnamefont {S.}~\bibnamefont {Ciuchi}},
  \bibinfo {author} {\bibfnamefont {M.}~\bibnamefont {Wallerberger}}, \bibinfo
  {author} {\bibfnamefont {P.}~\bibnamefont {Thunstr\"om}}, \bibinfo {author}
  {\bibfnamefont {O.}~\bibnamefont {Gunnarsson}}, \bibinfo {author}
  {\bibfnamefont {G.}~\bibnamefont {Sangiovanni}}, \bibinfo {author}
  {\bibfnamefont {G.}~\bibnamefont {Rohringer}}, \ and\ \bibinfo {author}
  {\bibfnamefont {A.}~\bibnamefont {Toschi}},\ }\href {\doibase
  10.1103/PhysRevB.94.235108} {\bibfield  {journal} {\bibinfo  {journal} {Phys.
  Rev. B}\ }\textbf {\bibinfo {volume} {94}},\ \bibinfo {pages} {235108}
  (\bibinfo {year} {2016})}\BibitemShut {NoStop}%
\bibitem [{\citenamefont {Thunstr\"om}\ \emph {et~al.}(2018)\citenamefont
  {Thunstr\"om}, \citenamefont {Gunnarsson}, \citenamefont {Ciuchi},\ and\
  \citenamefont {Rohringer}}]{PhysRevB.98.235107}%
  \BibitemOpen
  \bibfield  {author} {\bibinfo {author} {\bibfnamefont {P.}~\bibnamefont
  {Thunstr\"om}}, \bibinfo {author} {\bibfnamefont {O.}~\bibnamefont
  {Gunnarsson}}, \bibinfo {author} {\bibfnamefont {S.}~\bibnamefont {Ciuchi}},
  \ and\ \bibinfo {author} {\bibfnamefont {G.}~\bibnamefont {Rohringer}},\
  }\href {\doibase 10.1103/PhysRevB.98.235107} {\bibfield  {journal} {\bibinfo
  {journal} {Phys. Rev. B}\ }\textbf {\bibinfo {volume} {98}},\ \bibinfo
  {pages} {235107} (\bibinfo {year} {2018})}\BibitemShut {NoStop}%
\bibitem [{\citenamefont {Vu\ifmmode \check{c}\else \v{c}\fi{}i\ifmmode
  \check{c}\else \v{c}\fi{}evi\ifmmode~\acute{c}\else \'{c}\fi{}}\ \emph
  {et~al.}(2018)\citenamefont {Vu\ifmmode \check{c}\else \v{c}\fi{}i\ifmmode
  \check{c}\else \v{c}\fi{}evi\ifmmode~\acute{c}\else \'{c}\fi{}},
  \citenamefont {Wentzell}, \citenamefont {Ferrero},\ and\ \citenamefont
  {Parcollet}}]{PhysRevB.97.125141}%
  \BibitemOpen
  \bibfield  {author} {\bibinfo {author} {\bibfnamefont {J.}~\bibnamefont
  {Vu\ifmmode \check{c}\else \v{c}\fi{}i\ifmmode \check{c}\else
  \v{c}\fi{}evi\ifmmode~\acute{c}\else \'{c}\fi{}}}, \bibinfo {author}
  {\bibfnamefont {N.}~\bibnamefont {Wentzell}}, \bibinfo {author}
  {\bibfnamefont {M.}~\bibnamefont {Ferrero}}, \ and\ \bibinfo {author}
  {\bibfnamefont {O.}~\bibnamefont {Parcollet}},\ }\href {\doibase
  10.1103/PhysRevB.97.125141} {\bibfield  {journal} {\bibinfo  {journal} {Phys.
  Rev. B}\ }\textbf {\bibinfo {volume} {97}},\ \bibinfo {pages} {125141}
  (\bibinfo {year} {2018})}\BibitemShut {NoStop}%
\bibitem [{\citenamefont {Kozik}\ \emph {et~al.}(2015)\citenamefont {Kozik},
  \citenamefont {Ferrero},\ and\ \citenamefont
  {Georges}}]{PhysRevLett.114.156402}%
  \BibitemOpen
  \bibfield  {author} {\bibinfo {author} {\bibfnamefont {E.}~\bibnamefont
  {Kozik}}, \bibinfo {author} {\bibfnamefont {M.}~\bibnamefont {Ferrero}}, \
  and\ \bibinfo {author} {\bibfnamefont {A.}~\bibnamefont {Georges}},\ }\href
  {\doibase 10.1103/PhysRevLett.114.156402} {\bibfield  {journal} {\bibinfo
  {journal} {Phys. Rev. Lett.}\ }\textbf {\bibinfo {volume} {114}},\ \bibinfo
  {pages} {156402} (\bibinfo {year} {2015})}\BibitemShut {NoStop}%
\bibitem [{\citenamefont {van Loon}\ \emph {et~al.}(2016)\citenamefont {van
  Loon}, \citenamefont {Krien}, \citenamefont {Hafermann}, \citenamefont
  {Stepanov}, \citenamefont {Lichtenstein},\ and\ \citenamefont
  {Katsnelson}}]{PhysRevB.93.155162}%
  \BibitemOpen
  \bibfield  {author} {\bibinfo {author} {\bibfnamefont {E.~G. C.~P.}\
  \bibnamefont {van Loon}}, \bibinfo {author} {\bibfnamefont {F.}~\bibnamefont
  {Krien}}, \bibinfo {author} {\bibfnamefont {H.}~\bibnamefont {Hafermann}},
  \bibinfo {author} {\bibfnamefont {E.~A.}\ \bibnamefont {Stepanov}}, \bibinfo
  {author} {\bibfnamefont {A.~I.}\ \bibnamefont {Lichtenstein}}, \ and\
  \bibinfo {author} {\bibfnamefont {M.~I.}\ \bibnamefont {Katsnelson}},\ }\href
  {\doibase 10.1103/PhysRevB.93.155162} {\bibfield  {journal} {\bibinfo
  {journal} {Phys. Rev. B}\ }\textbf {\bibinfo {volume} {93}},\ \bibinfo
  {pages} {155162} (\bibinfo {year} {2016})}\BibitemShut {NoStop}%
\bibitem [{\citenamefont {Rohringer}\ and\ \citenamefont
  {Toschi}(2016)}]{PhysRevB.94.125144}%
  \BibitemOpen
  \bibfield  {author} {\bibinfo {author} {\bibfnamefont {G.}~\bibnamefont
  {Rohringer}}\ and\ \bibinfo {author} {\bibfnamefont {A.}~\bibnamefont
  {Toschi}},\ }\href {\doibase 10.1103/PhysRevB.94.125144} {\bibfield
  {journal} {\bibinfo  {journal} {Phys. Rev. B}\ }\textbf {\bibinfo {volume}
  {94}},\ \bibinfo {pages} {125144} (\bibinfo {year} {2016})}\BibitemShut
  {NoStop}%
\bibitem [{\citenamefont {van Loon}\ \emph {et~al.}(2015)\citenamefont {van
  Loon}, \citenamefont {Hafermann}, \citenamefont {Lichtenstein},\ and\
  \citenamefont {Katsnelson}}]{PhysRevB.92.085106}%
  \BibitemOpen
  \bibfield  {author} {\bibinfo {author} {\bibfnamefont {E.~G. C.~P.}\
  \bibnamefont {van Loon}}, \bibinfo {author} {\bibfnamefont {H.}~\bibnamefont
  {Hafermann}}, \bibinfo {author} {\bibfnamefont {A.~I.}\ \bibnamefont
  {Lichtenstein}}, \ and\ \bibinfo {author} {\bibfnamefont {M.~I.}\
  \bibnamefont {Katsnelson}},\ }\href {\doibase 10.1103/PhysRevB.92.085106}
  {\bibfield  {journal} {\bibinfo  {journal} {Phys. Rev. B}\ }\textbf {\bibinfo
  {volume} {92}},\ \bibinfo {pages} {085106} (\bibinfo {year}
  {2015})}\BibitemShut {NoStop}%
\bibitem [{\citenamefont {Hafermann}\ \emph {et~al.}(2014)\citenamefont
  {Hafermann}, \citenamefont {van Loon}, \citenamefont {Katsnelson},
  \citenamefont {Lichtenstein},\ and\ \citenamefont
  {Parcollet}}]{PhysRevB.90.235105}%
  \BibitemOpen
  \bibfield  {author} {\bibinfo {author} {\bibfnamefont {H.}~\bibnamefont
  {Hafermann}}, \bibinfo {author} {\bibfnamefont {E.~G. C.~P.}\ \bibnamefont
  {van Loon}}, \bibinfo {author} {\bibfnamefont {M.~I.}\ \bibnamefont
  {Katsnelson}}, \bibinfo {author} {\bibfnamefont {A.~I.}\ \bibnamefont
  {Lichtenstein}}, \ and\ \bibinfo {author} {\bibfnamefont {O.}~\bibnamefont
  {Parcollet}},\ }\href {\doibase 10.1103/PhysRevB.90.235105} {\bibfield
  {journal} {\bibinfo  {journal} {Phys. Rev. B}\ }\textbf {\bibinfo {volume}
  {90}},\ \bibinfo {pages} {235105} (\bibinfo {year} {2014})}\BibitemShut
  {NoStop}%
\bibitem [{\citenamefont {Gunnarsson}\ \emph {et~al.}(2017)\citenamefont
  {Gunnarsson}, \citenamefont {Rohringer}, \citenamefont {Sch\"afer},
  \citenamefont {Sangiovanni},\ and\ \citenamefont
  {Toschi}}]{PhysRevLett.119.056402}%
  \BibitemOpen
  \bibfield  {author} {\bibinfo {author} {\bibfnamefont {O.}~\bibnamefont
  {Gunnarsson}}, \bibinfo {author} {\bibfnamefont {G.}~\bibnamefont
  {Rohringer}}, \bibinfo {author} {\bibfnamefont {T.}~\bibnamefont
  {Sch\"afer}}, \bibinfo {author} {\bibfnamefont {G.}~\bibnamefont
  {Sangiovanni}}, \ and\ \bibinfo {author} {\bibfnamefont {A.}~\bibnamefont
  {Toschi}},\ }\href {\doibase 10.1103/PhysRevLett.119.056402} {\bibfield
  {journal} {\bibinfo  {journal} {Phys. Rev. Lett.}\ }\textbf {\bibinfo
  {volume} {119}},\ \bibinfo {pages} {056402} (\bibinfo {year}
  {2017})}\BibitemShut {NoStop}%
\bibitem [{\citenamefont {Chalupa}\ \emph {et~al.}(2018)\citenamefont
  {Chalupa}, \citenamefont {Gunacker}, \citenamefont {Sch\"afer}, \citenamefont
  {Held},\ and\ \citenamefont {Toschi}}]{PhysRevB.97.245136}%
  \BibitemOpen
  \bibfield  {author} {\bibinfo {author} {\bibfnamefont {P.}~\bibnamefont
  {Chalupa}}, \bibinfo {author} {\bibfnamefont {P.}~\bibnamefont {Gunacker}},
  \bibinfo {author} {\bibfnamefont {T.}~\bibnamefont {Sch\"afer}}, \bibinfo
  {author} {\bibfnamefont {K.}~\bibnamefont {Held}}, \ and\ \bibinfo {author}
  {\bibfnamefont {A.}~\bibnamefont {Toschi}},\ }\href {\doibase
  10.1103/PhysRevB.97.245136} {\bibfield  {journal} {\bibinfo  {journal} {Phys.
  Rev. B}\ }\textbf {\bibinfo {volume} {97}},\ \bibinfo {pages} {245136}
  (\bibinfo {year} {2018})}\BibitemShut {NoStop}%
\bibitem [{\citenamefont {Gunnarsson}\ \emph {et~al.}(2016)\citenamefont
  {Gunnarsson}, \citenamefont {Sch\"afer}, \citenamefont {LeBlanc},
  \citenamefont {Merino}, \citenamefont {Sangiovanni}, \citenamefont
  {Rohringer},\ and\ \citenamefont {Toschi}}]{PhysRevB.93.245102}%
  \BibitemOpen
  \bibfield  {author} {\bibinfo {author} {\bibfnamefont {O.}~\bibnamefont
  {Gunnarsson}}, \bibinfo {author} {\bibfnamefont {T.}~\bibnamefont
  {Sch\"afer}}, \bibinfo {author} {\bibfnamefont {J.~P.~F.}\ \bibnamefont
  {LeBlanc}}, \bibinfo {author} {\bibfnamefont {J.}~\bibnamefont {Merino}},
  \bibinfo {author} {\bibfnamefont {G.}~\bibnamefont {Sangiovanni}}, \bibinfo
  {author} {\bibfnamefont {G.}~\bibnamefont {Rohringer}}, \ and\ \bibinfo
  {author} {\bibfnamefont {A.}~\bibnamefont {Toschi}},\ }\href {\doibase
  10.1103/PhysRevB.93.245102} {\bibfield  {journal} {\bibinfo  {journal} {Phys.
  Rev. B}\ }\textbf {\bibinfo {volume} {93}},\ \bibinfo {pages} {245102}
  (\bibinfo {year} {2016})}\BibitemShut {NoStop}%
\bibitem [{\citenamefont {Lawler}\ \emph {et~al.}(2010)\citenamefont {Lawler},
  \citenamefont {Fujita}, \citenamefont {Lee}, \citenamefont {Schmidt},
  \citenamefont {Kohsaka}, \citenamefont {Kim}, \citenamefont {Eisaki},
  \citenamefont {Uchida}, \citenamefont {Davis}, \citenamefont {Sethna},\ and\
  \citenamefont {Kim}}]{lawler_intra-unit-cell_2010}%
  \BibitemOpen
  \bibfield  {author} {\bibinfo {author} {\bibfnamefont {M.~J.}\ \bibnamefont
  {Lawler}}, \bibinfo {author} {\bibfnamefont {K.}~\bibnamefont {Fujita}},
  \bibinfo {author} {\bibfnamefont {J.}~\bibnamefont {Lee}}, \bibinfo {author}
  {\bibfnamefont {A.~R.}\ \bibnamefont {Schmidt}}, \bibinfo {author}
  {\bibfnamefont {Y.}~\bibnamefont {Kohsaka}}, \bibinfo {author} {\bibfnamefont
  {C.~K.}\ \bibnamefont {Kim}}, \bibinfo {author} {\bibfnamefont
  {H.}~\bibnamefont {Eisaki}}, \bibinfo {author} {\bibfnamefont
  {S.}~\bibnamefont {Uchida}}, \bibinfo {author} {\bibfnamefont {J.~C.}\
  \bibnamefont {Davis}}, \bibinfo {author} {\bibfnamefont {J.~P.}\ \bibnamefont
  {Sethna}}, \ and\ \bibinfo {author} {\bibfnamefont {E.-A.}\ \bibnamefont
  {Kim}},\ }\href {\doibase 10.1038/nature09169} {\bibfield  {journal}
  {\bibinfo  {journal} {Nature}\ }\textbf {\bibinfo {volume} {466}},\ \bibinfo
  {pages} {347} (\bibinfo {year} {2010})}\BibitemShut {NoStop}%
\bibitem [{\citenamefont {Bickers}(2004)}]{Bickers2004}%
  \BibitemOpen
  \bibfield  {author} {\bibinfo {author} {\bibfnamefont {N.~E.}\ \bibnamefont
  {Bickers}},\ }\enquote {\bibinfo {title} {Self-consistent many-body theory
  for condensed matter systems},}\ \ (\bibinfo  {publisher} {Springer-Verlag},\
  \bibinfo {address} {New York},\ \bibinfo {year} {2004})\ Chap.~\bibinfo
  {chapter} {6}, pp.\ \bibinfo {pages} {237--296}\BibitemShut {NoStop}%
\bibitem [{\citenamefont {Nourafkan}\ \emph {et~al.}(2016)\citenamefont
  {Nourafkan}, \citenamefont {Kotliar},\ and\ \citenamefont
  {Tremblay}}]{PhysRevLett.117.137001}%
  \BibitemOpen
  \bibfield  {author} {\bibinfo {author} {\bibfnamefont {R.}~\bibnamefont
  {Nourafkan}}, \bibinfo {author} {\bibfnamefont {G.}~\bibnamefont {Kotliar}},
  \ and\ \bibinfo {author} {\bibfnamefont {A.-M.~S.}\ \bibnamefont
  {Tremblay}},\ }\href {\doibase 10.1103/PhysRevLett.117.137001} {\bibfield
  {journal} {\bibinfo  {journal} {Phys. Rev. Lett.}\ }\textbf {\bibinfo
  {volume} {117}},\ \bibinfo {pages} {137001} (\bibinfo {year}
  {2016})}\BibitemShut {NoStop}%
\bibitem [{\citenamefont {Nourafkan}\ and\ \citenamefont
  {Tremblay}(2017)}]{PhysRevB.96.125140}%
  \BibitemOpen
  \bibfield  {author} {\bibinfo {author} {\bibfnamefont {R.}~\bibnamefont
  {Nourafkan}}\ and\ \bibinfo {author} {\bibfnamefont {A.-M.~S.}\ \bibnamefont
  {Tremblay}},\ }\href {\doibase 10.1103/PhysRevB.96.125140} {\bibfield
  {journal} {\bibinfo  {journal} {Phys. Rev. B}\ }\textbf {\bibinfo {volume}
  {96}},\ \bibinfo {pages} {125140} (\bibinfo {year} {2017})}\BibitemShut
  {NoStop}%
\bibitem [{\citenamefont {Toschi}\ \emph {et~al.}(2012)\citenamefont {Toschi},
  \citenamefont {Arita}, \citenamefont {Hansmann}, \citenamefont
  {Sangiovanni},\ and\ \citenamefont {Held}}]{PhysRevB.86.064411}%
  \BibitemOpen
  \bibfield  {author} {\bibinfo {author} {\bibfnamefont {A.}~\bibnamefont
  {Toschi}}, \bibinfo {author} {\bibfnamefont {R.}~\bibnamefont {Arita}},
  \bibinfo {author} {\bibfnamefont {P.}~\bibnamefont {Hansmann}}, \bibinfo
  {author} {\bibfnamefont {G.}~\bibnamefont {Sangiovanni}}, \ and\ \bibinfo
  {author} {\bibfnamefont {K.}~\bibnamefont {Held}},\ }\href {\doibase
  10.1103/PhysRevB.86.064411} {\bibfield  {journal} {\bibinfo  {journal} {Phys.
  Rev. B}\ }\textbf {\bibinfo {volume} {86}},\ \bibinfo {pages} {064411}
  (\bibinfo {year} {2012})}\BibitemShut {NoStop}%
\bibitem [{\citenamefont {Sch\"afer}\ \emph {et~al.}(2013)\citenamefont
  {Sch\"afer}, \citenamefont {Rohringer}, \citenamefont {Gunnarsson},
  \citenamefont {Ciuchi}, \citenamefont {Sangiovanni},\ and\ \citenamefont
  {Toschi}}]{PhysRevLett.110.246405}%
  \BibitemOpen
  \bibfield  {author} {\bibinfo {author} {\bibfnamefont {T.}~\bibnamefont
  {Sch\"afer}}, \bibinfo {author} {\bibfnamefont {G.}~\bibnamefont
  {Rohringer}}, \bibinfo {author} {\bibfnamefont {O.}~\bibnamefont
  {Gunnarsson}}, \bibinfo {author} {\bibfnamefont {S.}~\bibnamefont {Ciuchi}},
  \bibinfo {author} {\bibfnamefont {G.}~\bibnamefont {Sangiovanni}}, \ and\
  \bibinfo {author} {\bibfnamefont {A.}~\bibnamefont {Toschi}},\ }\href
  {\doibase 10.1103/PhysRevLett.110.246405} {\bibfield  {journal} {\bibinfo
  {journal} {Phys. Rev. Lett.}\ }\textbf {\bibinfo {volume} {110}},\ \bibinfo
  {pages} {246405} (\bibinfo {year} {2013})}\BibitemShut {NoStop}%
\end{thebibliography}
%

\end{document}